\documentclass[aip,jcp,amsmath,amssymb,reprint]{revtex4-1}
\usepackage{graphicx}
\usepackage{dcolumn}
\usepackage{bm}
\usepackage{mathrsfs}
\usepackage{amsmath, amssymb, amsthm, amsfonts}
\usepackage[dvipsnames]{xcolor}
\usepackage{ulem}
\usepackage{caption}
\usepackage[textfont=rm]{subcaption}
\begin{document}
\title{Correlation between structure and dynamics: Role of attractive interaction}
\title{A comparative study of the correlation between the structure and the dynamics for systems interacting via attractive and repulsive potentials }
\author{Mohit Sharma}
\address{\textit{Polymer Science and Engineering Division, CSIR-National Chemical Laboratory, Pune-411008, India}}
\affiliation{\textit{Academy of Scientific and Innovative Research (AcSIR), Ghaziabad 201002, India}}
\author{Manoj Kumar Nandi}
\affiliation{\textit{Université Claude Bernard Lyon 1, Institut National de la Santé et de la Recherche Médicale, Stem Cell and Brain Research Institute, Bron 69500 France}}

\author{Sarika Maitra Bhattacharyya}
\email{mb.sarika@ncl.res.in}
\address{\textit{Polymer Science and Engineering Division, CSIR-National Chemical Laboratory, Pune-411008, India}}
\affiliation{\textit{Academy of Scientific and Innovative Research (AcSIR), Ghaziabad 201002, India}}

\begin{abstract}
We present the study of the structure-dynamics correlation for systems interacting via attractive Lennard-Jones and its repulsive counterpart, the WCA potentials. The structural order parameter (SOP) is related to the microscopic mean-field caging potential. At a particle level, the SOP shows a distribution. Although the two systems have similar pair structures, their average SOP differs. However, this difference alone is insufficient to explain the well known slowing down of the dynamics in LJ system at low temperatures. The slowing down can be explained in terms of a stronger coupling between the SOP and the dynamics. To understand the origin of this system specific coupling, we study the difference in the microscopic structure between the hard and soft particles. We find that for the LJ system, the structural differences of the hard and soft particles are more significant and have a much stronger temperature dependence compared to the WCA system. Thus the study suggests that attractive interaction creates more structurally different communities. This broader difference in the structural communities is probably responsible for stronger coupling between the structure and dynamics. Thus the system specific structure-dynamics correlation, which also leads to a faster slowing down in the dynamics, appears to have a structural origin. A comparison of the predictive power of our SOP with the local energy and two body excess entropy shows that in the LJ system, the dynamics is driven by enthalpy, whereas in the WCA system, it is driven by entropy, and our SOP can capture both these contributions. 

\end{abstract}
\maketitle
\section{Introduction}

In the supercooled liquid community, there is a long-standing debate about the role the pair structure plays in the dynamics \cite{berthier_tarjus_prl_2009, atreyee_prl,schweizer_prl,olivier_pre,tanaka_prl2020}. In the normal liquid regime, the pair structure is known to play a dominant role in determining the thermodynamic properties and also the dynamics via well known theories like mode coupling theory and density functional theory\cite{Hansen_and_McDonald}. However, the dramatic slowing down of the dynamics in the supercooled liquid is accompanied by only a marginal increase in the structure, suggesting that the role of pair structure in the supercooled liquid dynamics needs to be relooked. The observation that two glass-forming systems, namely the Kob-Andersen model \cite{kob_1994} with attractive Lennard Jones (LJ) potential and its repulsive counterpart, the WCA potential, have similar pair structures, but at low temperatures, their dynamics are orders of magnitude apart also questioned the role of thermodynamics in the supercooled liquid dynamics \cite{berthier_tarjus_prl_2009}. According to van der Waals theory \cite{vander_waal}, which was the guiding principle in developing the WCA potential \cite{wca_potential} at high density, the liquid structure is determined by the short-range repulsive potential, and the long-range attractive potential creates a cohesive background which can be treated in a perturbative manner \cite{wca_potential}. 
The difference in dynamics of the two systems definitely highlights the nonperturbative role of attractive interaction in the dynamics. Follow up studies showed that although the systems are similar at the pair correlation level, they are different in higher-order correlation functions \cite{coslovich_jcp2013, hocky_prl2012}. Studies have also confirmed that the two systems are thermodynamically different. It was shown that there is a difference in the configurational entropies of the two systems, which via the well known Adam Gibbs theory, could explain the difference in the dynamics \cite{atreyee_prl}. It was also shown that although the isomorph theory applies to both the systems, the density scaling exponent used to define the isomorphs vary weakly with density in the LJ system but shows a strong variation over two decades in the WCA system \cite{jpdyre_prl2010,jpdyre_wca}.

In a series of works involving some of us, it was shown that even at the pair level, the configuration entropies of the two systems are different \cite{atreyee_prl, atreyee_density, manoj_unravel}. According to these studies, small differences in structure can contribute to large differences in thermodynamic quantities like entropy which can drive the difference in the dynamics. Schweizer and co-workers have shown that when the attractive forces are explicitly included in the theory, it can predict the difference in the dynamics for a similar pair structure \cite{schweizer_prl}. These studies brought the focus back to the role of pair structure in the dynamics. Landes {\it et al.} used a well-developed machine learning (ML) technique \cite{liu_nature}, which describes an ML softness parameter. This parameter depends on the local pair structure of the liquid. The study has shown that this ML softness parameter of the LJ and the WCA systems are different, suggesting that this difference can explain the difference in dynamics \cite{olivier_pre}. Tong and Tanaka described a structural order parameter which is many body in nature \cite{tanaka_nature,tanaka_prx,tanaka_prl2020} and showed that this order parameter is different for the LJ and the WCA systems \cite{tanaka_prl2020}. Thus suggesting that the attractive interaction, along with the dynamics, also affects the structure in a non-perturbative manner.

Recently some of us have developed a microscopic mean-field theory where we assumed that each particle is in a mean-field caging potential where the potential is described in terms of the pair structure of the liquid \cite{manoj_prl_2017, manoj_prl2021}. Using the information of this caging potential, we have described a structural order parameter. We termed it ``softness", as it describes the softness/curvature of the mean-field caging potential and is inversely proportional to the depth of the potential. Note that this softness parameter is obtained from the microscopic theory and is different from the softness parameter described in the ML study \cite{liu_nature,olivier_pre}. We showed that this softness parameter at the macroscopic level could describe the dynamics of a large class of systems \cite{manoj_prl2021}. The master plot between dynamics and the softness parameter suggested a universal correlation between structure and dynamics. However, our study further suggested that this universal correlation is not uniform. The strength of the correlation varies between systems and plays a role in the divergence of the relaxation time and, thus, the fragility of the system. Interestingly such system specific correlations between the dynamics and other order parameters like the ML softness parameter \cite{olivier_pre} or the many body structural order parameter\cite{tanaka_prl2020} have also been observed especially when studying the correlation between the structure and dynamics for the LJ and WCA systems. 

In this work, we first study the correlation between the dynamics and our structural order parameter (SOP) for the LJ and the WCA systems. Note that our SOP is described only in terms of pair correlation. Despite the two systems having similar pair structures, we show that the SOPs of the systems are different. Similar observations were also made for other structural order parameters \cite{atreyee_prl,olivier_pre,tanaka_prl2020}. However, the most important observation is that differences in the SOPs are not enough to describe the difference in the dynamics. There appears to be a system specific correlation between the dynamics and the SOP, an observation which was reported earlier for other systems \cite{manoj_prl2021}and, although not highlighted, were present in other studies \cite{olivier_pre,tanaka_prl2020}. The primary goal of this work is to understand the origin of this system specific structure-dynamics correlation. Towards this goal, we first study the correlation between structure and dynamics at the microscopic level using the framework developed for microscopic SOP\cite{mohit_pre}. The study at the microscopic level corroborates the macroscopic observation. We then use a recently proposed technique called community information \cite{coslovich2020} to quantify the difference in the local structure of the particles which are in a well defined mean-field caging potential (hard) and particles where the caging potential is less defined (soft). We find that the difference in the local structure of the hard and soft particles is larger for the LJ system, and this larger difference is also correlated with the stronger coupling between the structure and dynamics. Thus our study suggests that this system specific coupling between the structure and dynamics may have a structural origin. Finally, we also compare the predictive power of our SOP with the local energy and the local two body excess entropy. We find that for the attractive LJ system, the dynamics is controlled by the local energy, whereas for the repulsive WCA system, the local two body excess entropy controls the dynamics. Our SOP contains both the information of energy and entropy and thus is a good predictor of the dynamics for both system.

The organization of the rest of the paper is the following. Section \ref{simulation_section} contains the simulation details. In Section \ref{methodology_section}, we discuss the calculation of structure order parameter(SOP), entropy, relaxation time, the identification method of fast particles, isoconfigurational runs, and Spearman rank correlation. In Section \ref{macroscopic_section}, we discuss the correlation between structure and dynamics at the macroscopic level. In Section \ref{microscopic_section}, we discuss the correlation between structure and dynamics at the microscopic level. In Section \ref{section_diffpara}, we compare the predictive power of our SOP with other order parameters, and the paper ends with a brief conclusion in Section \ref{conclusion_section}.
This paper also contains 4 Appendix sections at the end.

\section{Simulation Details}
\label{simulation_section}
The system we study is the 3-Dimensional Kob-Andersen model for glass-forming liquid, which is a binary mixture (80:20) of Lennard-Jones (LJ) particles \cite{kob_1994} and its repulsive counterpart Weeks-Chandler-Andersen potential (WCA)\cite{wca_potential}. The interaction between the particles i and j, where i,j = A,B (the type of the particles), is given by
\begin{equation}
 U_{ij}(r)=
\begin{cases}
 U_{ij}^{(LJ)}(r;\sigma_{ij},\epsilon_{ij})- U_{ij}^{(LJ)}(r^{(c)}_{ij};\sigma_{ij},\epsilon_{ij}),    & r\leq r^{(c)}_{ij}\\
   0,                                                                                       & r> r^{(c)}_{ij},
\end{cases}
\label{pot}
\end{equation}
\noindent
where $U_{ij}^{(LJ)}(r)$ = $4\epsilon_{ij}[(\sigma_{ij}/r)^{12}-(\sigma_{ij}/r)^6]$, $r$ is the distance between particles i and j and $\sigma_{ij} $ is effective diameter of particle and $r_{ij}^{(c)}=2.5\sigma_{ij}$ for LJ and for WCA system $r^{(c)}_{ij}$ is the position of minima of $U_{ij}^{(LJ)}(r)$. The length, temperature, and time are given in units of $\sigma_{AA}$, $\epsilon_{AA}/k_B$, $(m\sigma_{AA}^2 /\epsilon_{AA})^{1/2}$, respectively.
We use $\sigma_{AA}=1.0 $, $\sigma_{AB}=0.8 $, $\sigma_{BB}=0.88 $, $\epsilon_{AA}=1.0 $, $\epsilon_{AB}=1.5 $, $\epsilon_{BB}=0.5 $, $ m_{A}=m_{B}=1 $ and Boltzmann constant $k_{B} = 1$.
We have performed MD simulation(using LAMMPS package\cite{lammps}), we have used periodic boundary conditions and Nos\'{e}-Hoover thermostat with integration timestep 0.001$\tau$ for high and 0.005$\tau$ for low temperatures. The time constants for Nos\'{e}-Hoover thermostat are taken to be 100 timesteps. The total number density $\rho=N/V=1.2$ is fixed for both systems, where V is the system volume, and N=4000 is the total number of particles.

\section{Methodology}
\label{methodology_section}
\subsection{Structural order parameter(SOP)}
\label{sop}

In recent studies involving some of us \cite{manoj_prl_2017,manoj_prl2021,mohit_pre}, we have obtained an effective mean-field caging potential derived from the Ramakrishnan-Yussouff free energy functional\cite{ry_form}. The mean-field caging potential is a potential energy term that captures the effect of particle interactions in a simplified manner, assuming that each particle is subjected to an average interaction from the surrounding frozen background, where the background is described in terms of the static structure factor. The mean-field potential for a binary system is written as,

\begin{equation}
\begin{aligned}
\beta & \Phi_{q}(\Delta r) = \\ 
& -\int\frac{{\bf{dq}}}{(2\pi)^{3}}\sum_{uv}C_{uv}(q)\sqrt{x_{u}x_{v}}[S_{uv}(q) - \delta_{uv}]e^{\frac{-q^{2}{\Delta r}^{2}}{6}},
\end{aligned}
\label{old_softness}
\end{equation}
\noindent
where $\Delta r$ is the displacement of the central particle from its equilibrium position. $\beta=1/k_BT$ and $x_{u/v}$ represents the fraction of particle of type A/B in the binary mixture. In the above expression, the partial structure factor is expressed as $S_{uv}({\bf{q}})$ = $(1/\sqrt{N_{u}N_{v}})\sum_{i=1}^{N_{u}}\sum_{j=1}^{N_{v}}$ $exp[-i{\bf{q.}}({\bf{r}}_{i}^{u}$-${\bf{r}}_{j}^{v})]$ and the direct correlation function, ${\bf{C(q) = 1 - S^{-1}(q)}}$. ${\bf{C(q)}}$ and ${\bf{S(q)}}$ are in matrix form. The SOP is the inverse of the depth of the caging potential, {\it i.e.} the value of the potential at the equilibrium position ($\Delta r=0$). For this work, we need to obtain the SOP at the microscopic level for each particle. We cannot calculate the structure factor at a single particle level.
Thus we first rewrite the expression of the mean-field caging potential at $\Delta r=0$ in terms of the radial distribution function (rdf) and the direct correlation function expressed in real space as,
\begin{equation}
\begin{aligned}
\beta\Phi_{r}(\Delta r=0) = -\rho\int{{\bf{dr}}\sum_{uv}C_{uv}(r){x_{u}x_{v}}g_{uv}(r)} ,
\end{aligned}
\label{depth_r}
\end{equation}
\noindent
where $\rho$ is the density, $g_{uv}(r)$ is the partial rdf and $C_{uv}(r)$ is the partial direct correlation function in real space. To obtain $C_{uv}(r)$, we use the hypernetted chain (HNC) approximation \cite{Hansen_and_McDonald}, which offers an expression for the direct correlation function in terms of the rdf and the interaction potential and is given by,
\begin{equation}
\begin{aligned}
C_{uv}(r) = -\beta U_{uv}(r) + (g_{uv}(r) - 1) -ln(g_{uv}(r)),
\end{aligned}
\label{direct}
\end{equation}
\noindent
where $U_{uv}(r)$ is the interaction potential given by Eq.\ref{pot}, which is an input to the theory. Please note that we deal with the absolute magnitude of the caging potential since we view the caging potential's depth as an energy barrier (Eq.\ref{depth_r}). In the calculation of the local particle level potential, the rdf is calculated at the particle level. We employ a Gaussian approximation to estimate the particle-level radial distribution function (rdf)\cite{piaggi_PRL}, 
\begin{equation}
\begin{aligned}
 g_{uv}^{i}(r) = \frac{1}{4\pi \rho r^{2}} \sum_{j}\frac{1}{\sqrt{2\pi\delta^{2}}}e^{-\frac{(r - r_{ij})^{2}}{2\delta^{2}}},
\end{aligned}
\label{rdf1}
\end{equation}
\noindent
where ``i" is the particle index, $\rho$ is the density, $\delta$ is the variance of Gaussian distribution. The variance is used to make the otherwise discontinuous function a continuous one. In this calculation, we assume $\delta=0.09\sigma_{AA}$. The particle level direct correlation function is also calculated from Eq.\ref{direct} using the particle level rdf. Thus we obtain a particle level $\beta\Phi_{r}(\Delta r=0)$.

In calculating $\beta\Phi_{r}(\Delta r=0)$, we have the product of particle level rdf and the direct correlation function. This leads to a term that is the product of rdf and the interaction potential ($U(r)$). As discussed earlier \cite{mohit_pre}, the particle level rdf due to the Gaussian approximation has finite values at distances smaller than the average rdf. Due to this finite value of the rdf at small ``r" its product with the interaction potential, which diverges at small ``r", leads to a large unphysical contribution from this range. This contribution increases at high temperatures as the rdf moves to a smaller `r' and is higher for sharp repulsive potentials like WCA potential. To overcome this challenge, we assume that the interaction potential (-$\beta U_{uv}(r)$) is equal to the potential of mean force, ($ln(g_{uv}(r))$) and we approximate $C^{approx}_{uv}(r) \approx g_{uv}(r) - 1$. In Appendix IV, we show that the predictive power of the theory is marginally better when we use $C^{approx}_{uv}(r)$.

\subsection{Entropy}
The two-body excess entropy for a binary system is written as\cite{s2_charusita},
\begin{equation}
\begin{split}
& S_{2} = \\
& -k_{B}\frac{\rho}{2}\sum_{u,v = 1}^{2}x_{u}x_{v} \int_{0}^{\infty} \{g_{uv}(r)\ln g_{uv}(r) - g_{uv}(r)+1 \}\textbf{dr},
 \end{split}
\label{s2_particle}
\end{equation}
\noindent
The parameters used in this calculation are defined in the previous subsection. For the calculation of the local particle level entropy, the rdf is calculated at the particle level (Eq.\ref{rdf1}).\\

\subsection{Relaxation time}
\label{relaxation}
The relaxation time $\tau_{\alpha}$ is calculated using the overlap function $q(t=\tau_{\alpha})=1/e$. The expression for the overlap function is\cite{overlap_shiladitya}
\begin{equation}
\begin{aligned}
q(t)= \frac{1}{N} \sum_{i=1}^{N} \omega(|r_{i}(t) - r_{i}(0)|),
\end{aligned}
\label{overlap}
\end{equation}
\noindent
 where function $\omega(x) = 1$ when $0 \leq x \leq a$ and $\omega (x) = 0$ otherwise. The time-dependent overlap function is influenced by the selection of the cut-off parameter, denoted as ``a", which we have set to 0.3. This particular value is chosen to ensure that particle positions, which may have slightly drifted apart due to low-amplitude vibrational motion, are still considered the same. In other words, the chosen cut-off parameter allows us to treat particle displacements up to a certain distance as negligible. We set this cut-off parameter comparable to the mean squared displacement (MSD) value observed during the plateau region between the ballistic and diffusive regimes. 

\subsection{Identification of fast particles}
\label{fast_par}

In this study, we intend to connect a structure parameter with the mobility at a local level; there are numerous methods that may be utilized to identify a fast particle or an event\cite{walter_PRL_1997,walter_JCP_2002,harrowell,candelier,smessaert_PRE}, such as performing isoconfigurational runs, detecting irreversible reorganizations\cite{harrowell}, or monitoring mean square displacement over time\cite{walter_JCP_2002}. Here, we apply a technique that Candelier et al.\cite{candelier,smessaert_PRE} first developed, where they calculate a quantity $p_{hop}(i,t)$ that captures for each particle ``i" in a specific time window W = [t1, t2], the cage jumps when the particle's average position varies rapidly. The expression of $p_{hop}(i,t)$ is 

\begin{equation}
  p_{hop}(i,t) = \sqrt{\big<(\vec{r}_i - {\big<\vec{r}_i\big>}_U)^2\big>_V  \big<(\vec{r}_i - {\big<\vec{r}_i\big>}_V)^2\big>_U} \hspace{0.1cm} ,
  \label{phop}
\end{equation}
\noindent
for all t $\in$ W, where averages are performed in surrounding time t, i.e., U = [$t-\Delta t/2$, t] and V = [t, $t+\Delta t/2$] where $\Delta t$ is time in which particle can rearrange. In our study, we take $\Delta t$ as $15\tau$. A small value of $p_{hop}(i,t)$ means the particle is in the same cage, and a large value means the particle is in two separate cages. The threshold value $p_{c}$ for $p_{hop}(i,t)$ above, which particles are said to be rearranging, is the value of mean square displacement where the non-Gaussian parameter\cite{olivier_pre} \scalebox{0.95}{$\alpha_{2}$ = $\frac{3<\Delta r^{4}(t)>}{5<\Delta r^{2}(t)>^{2}} - 1 $} has maximum.

\subsection{Isoconfigurational ensemble}
\label{isoconf}
The iso-configurational ensemble, first introduced by Harrowell et al. \cite{harrowell}, is an ensemble of trajectories that run from an identical configuration of particles with random initial momenta sampled from the Maxwell-Boltzmann distribution for the corresponding temperature. This method removes the uninteresting variation in the particle displacements arising from the choice of initial momenta. If we discover substantial variations in the trajectory-averaged dynamics, we can relate it to a property of the initial configuration. We use this method to quantify the correlation between structure and dynamics, taking 20 different initial configurations apart by at least 75$\tau_{\alpha}$ to make sure they were distinctly different from one another and running 200 different trajectories from each configuration. For the calculation of dynamics, we define mobility as $ \mu^{j}(t) = \frac{1}{N_{IC}}\sum_{i=1}^{N_{IC}} \sqrt{(r_{i}^{j}(t) - r_{i}^{j}(0))^{2}}$. Here $\mu^{j}(t)$ is the mobility of $j^{th}$ particle at time $t$ and $N_{IC}$ is the number of trajectories. We then correlate the structure of the initial configuration with the dynamics at different times using Spearman rank correlation.

\subsection{Spearman rank correlation}
Spearman rank correlation for data with m values is written as,
\begin{equation}
C_{R}(X,Y) = 1 - \frac{6 \sum d_{i}^2}{m(m^2 -1)},
\label{rank_corrl}
\end{equation}
\noindent
where $d_{i}^2$ = $R(X_{i})$ - $R(Y_{i})$
is the difference between the ranks, $R(X_{i})$ and $R(Y_{i})$ of the raw data $X_{i}$ and $Y_{i}$ respectively.\\

\section{CORRELATION BETWEEN STRUCTURE AND DYNAMICS at the MACROSCOPIC Level}
\label{macroscopic_section}

\begin{figure}
\centering
\begin{subfigure}{0.238\textwidth}
\includegraphics[width=1.0\linewidth]{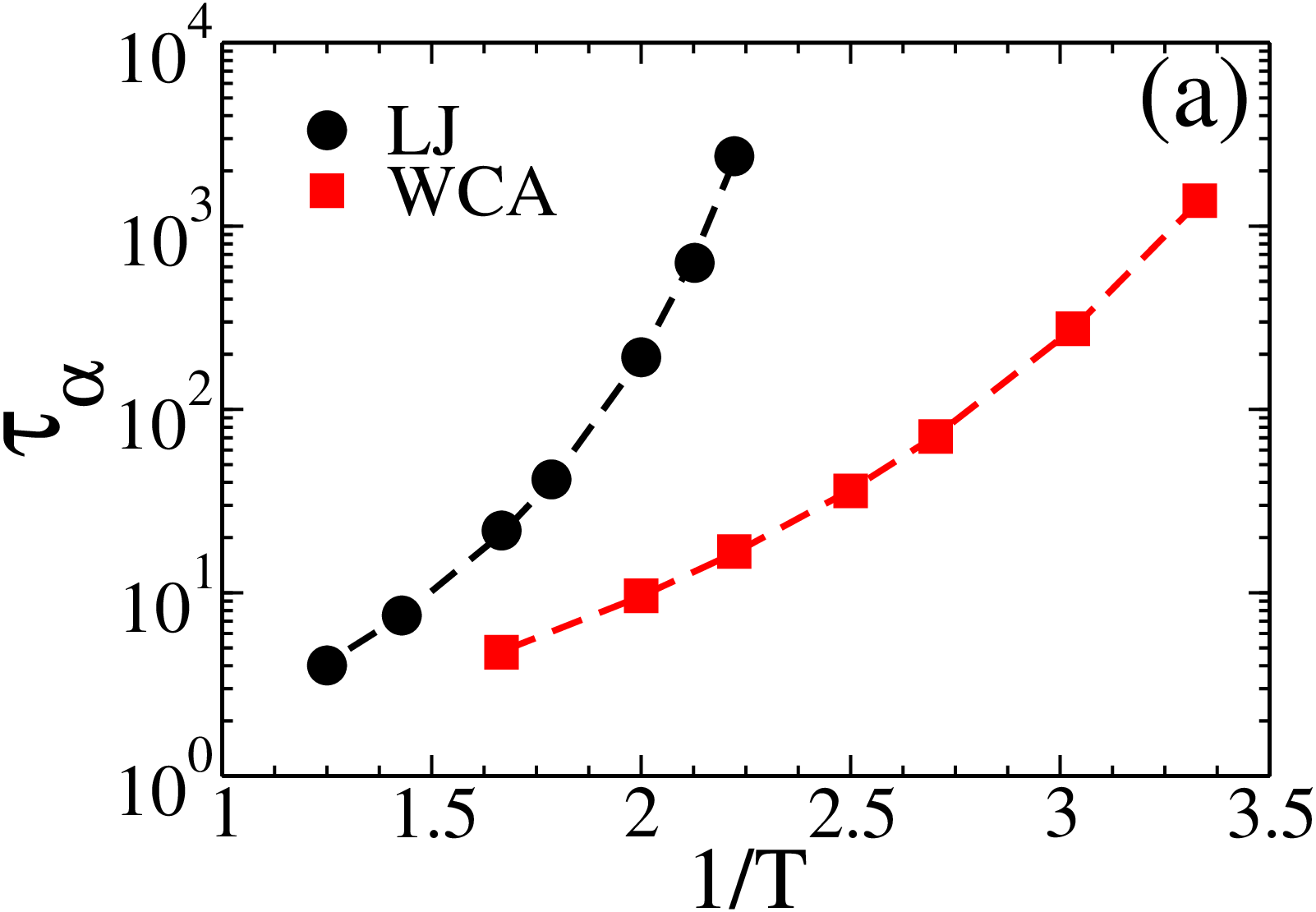}
\end{subfigure}
\begin{subfigure}{0.238\textwidth}
\includegraphics[width=1.0\linewidth]{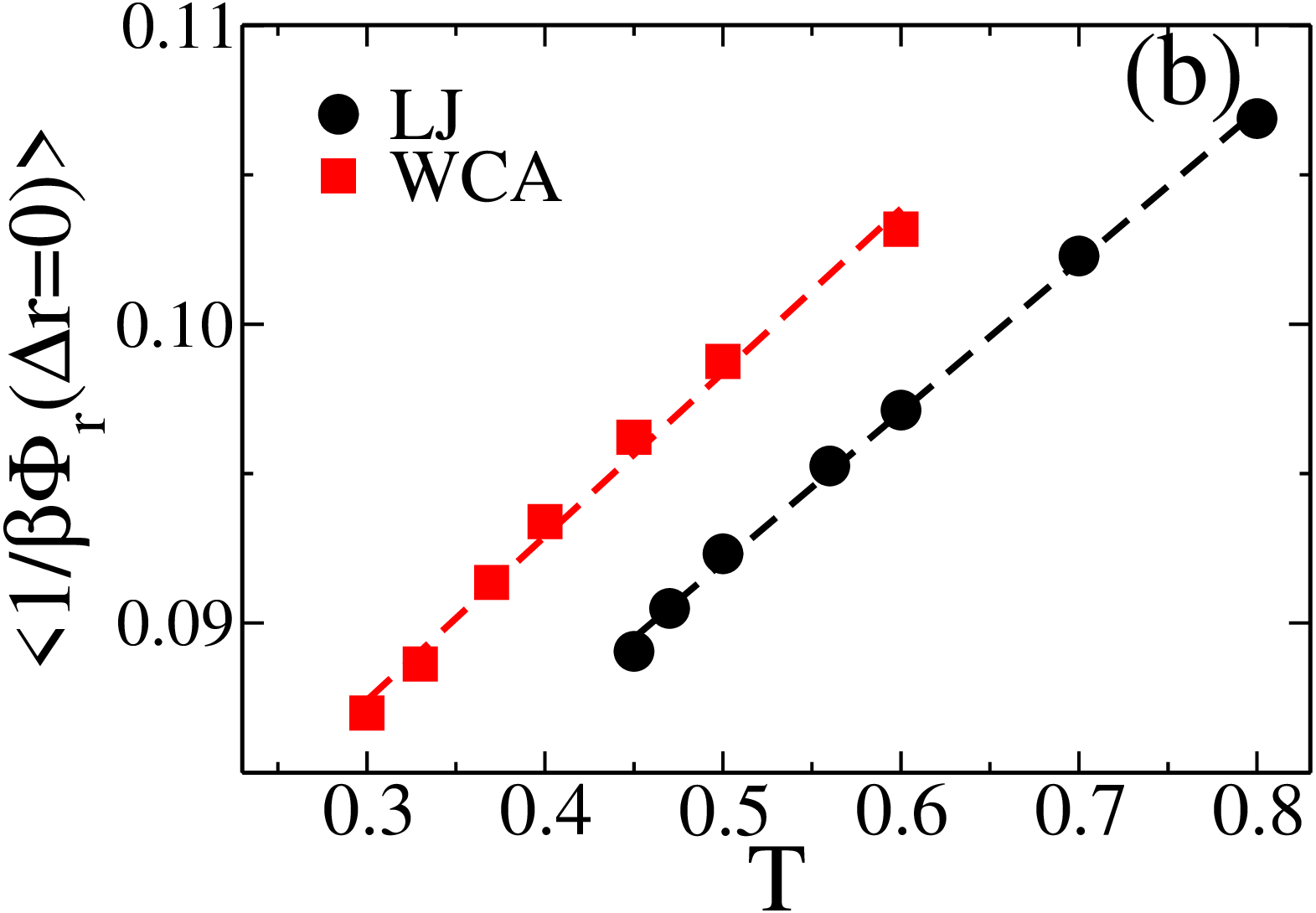}
\end{subfigure}
\vskip\baselineskip
\begin{subfigure}{0.238\textwidth}
\includegraphics[width=1.0\linewidth]{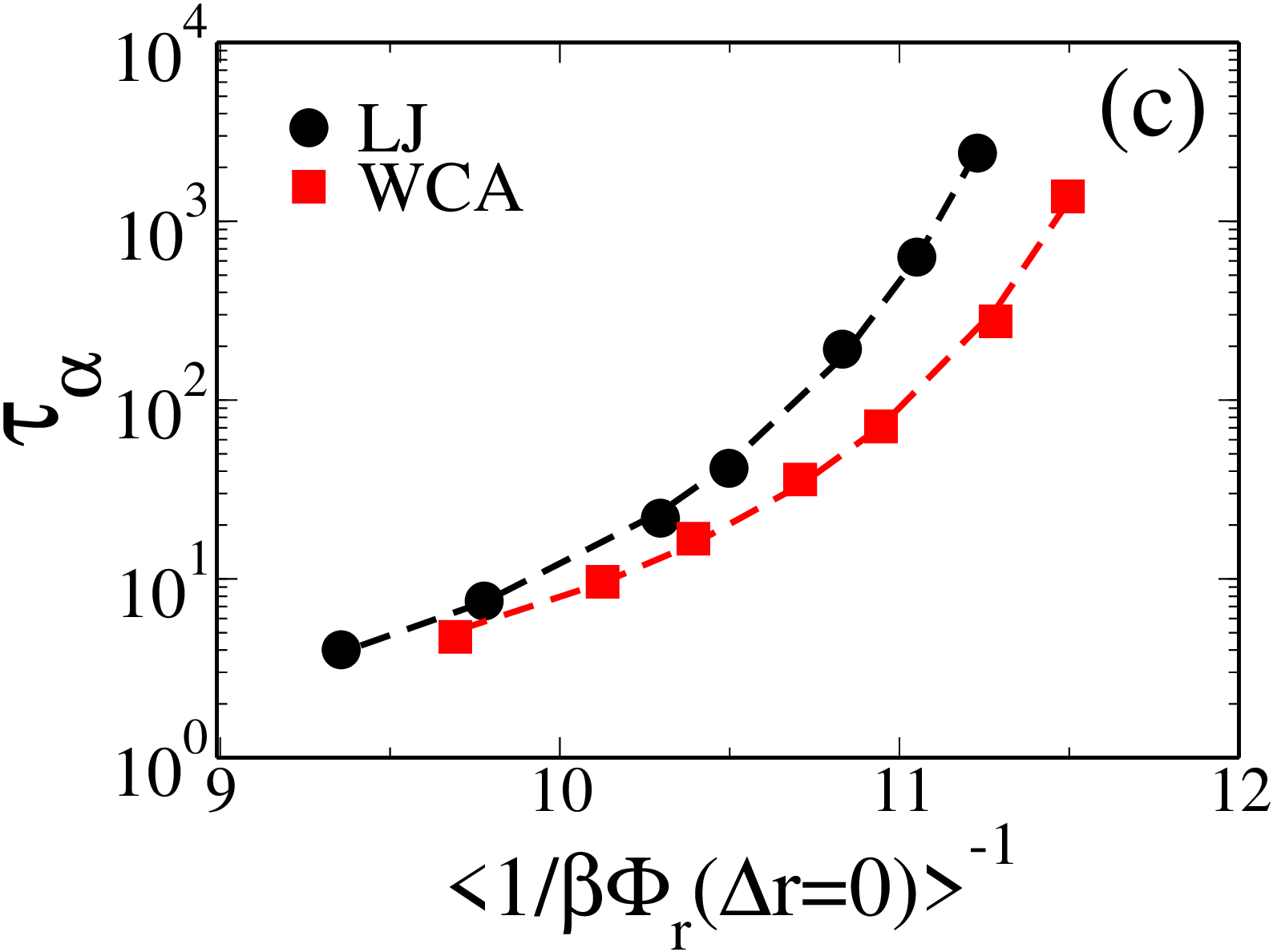}
\end{subfigure}
\begin{subfigure}{0.238\textwidth}
\includegraphics[width=1.0\linewidth]{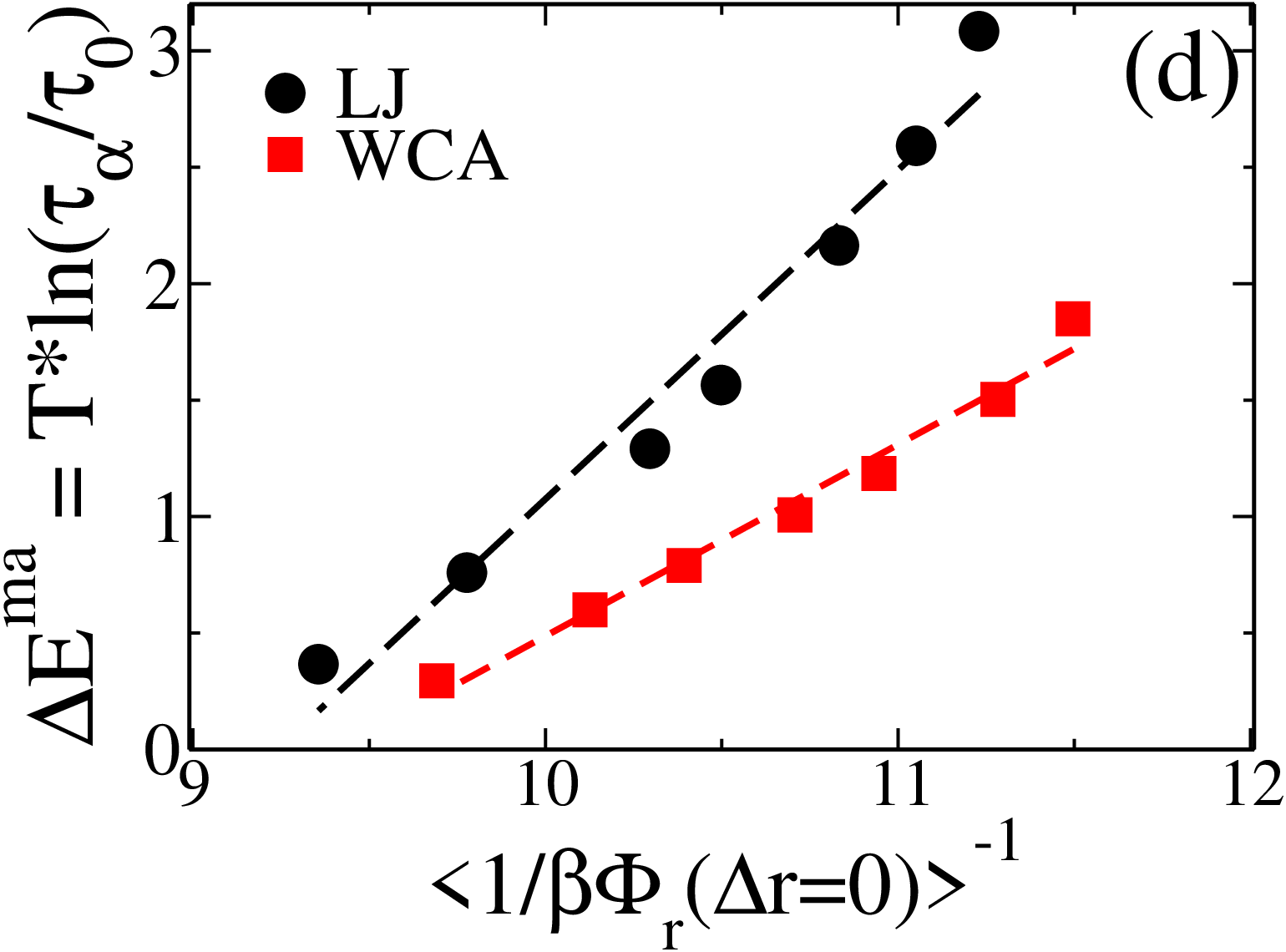}
\end{subfigure}
\caption{\textit{ For $LJ$ and $WCA$ systems \textbf{(a)} $\alpha$ relaxation time ($\tau_{\alpha}$) plotted against the inverse of temperature ($1/T$). The dashed lines are the Vogel-Fulcher-Tammann (VFT) fits. \textbf{(b)} Variation of an average of inverse of depth of caging potential ($\big<1/\beta \Phi_{r}\big>$) with $T$. \textbf{(c)} $\tau_{\alpha}$ plotted against $\big<1/\beta \Phi_{r}\big>^{-1}$. The dashed lines are the VFT fitting against $\big<1/\beta \Phi_{r}\big>$. \textbf{(d)} Activation barrier $\Delta E^{ma}=T*ln(\tau_{\alpha}/\tau_{0})$, as a function of $\big<1/\beta \Phi_{r}\big>^{-1}$ The dashed lines are the linear fit. The slope is system dependent, being higher for the $LJ$ system.}}
\label{macro_study}
\end{figure}

As discussed in the Introduction and also in earlier studies \cite{berthier_tarjus_prl_2009,atreyee_prl,schweizer_prl,olivier_pre,manoj_prl2021,tanaka_prl2020} although the difference in the pair structure of the LJ and WCA systems at the same temperature is marginal the dynamics changes orders of magnitude. As shown in Fig. \ref{macro_study}(a), the dynamics of the LJ system grows much faster with the lowering of temperature. The temperature dependence of the dynamics can be fitted to the well known Vogel-Fulcher-Tammann (VFT) form\cite{VFT}, $\tau_{\alpha} = \tau_{0}\exp(T_{0}/K_{T}(T-T_{0}))$.
The predicted divergence temperature of the dynamics, $T_{0}$, is higher for the LJ system. The rate at which the dynamics diverges is given by $K_{T}$, which is the kinetic fragility of the system. The more fragile LJ system shows a higher rate of divergence \cite{atreyee_prl,manoj_unravel}. It is often asked if there is any structural origin of this higher divergence rate of the dynamics, and if so, is it the pair structure or higher order terms\cite{berthier_tarjus_prl_2009,atreyee_prl,schweizer_prl,tanaka_prl2020}. In this work, we study the correlation of the dynamics with the structure using our recently proposed structural order parameter (SOP), which is the inverse of the depth of the mean-field caging potential\cite{manoj_prl2021}. In Appendix II, we show that the distribution of the SOP at per particle level has a temperature dependence. In Fig. \ref{macro_study}(b), we plot the average SOP($\big<1/\beta\Phi_{r}\big>$). Although the pair structures of the two systems are similar, there is a difference in their average SOP. At the same temperature, the LJ system has a lower value of the SOP, which implies that the caging potential is deeper. This is similar to what has been observed in the two body entropy \cite{atreyee_prl, manoj_unravel}, ML softness parameter \cite{olivier_pre} and many body structural order parameter \cite{tanaka_prl2020}. Since the SOP is linearly proportional to temperature, the relaxation times of the two systems can also be fitted to the VFT form, but now in terms of the SOP (Fig. \ref{macro_study}(c)), $\tau_{\alpha} = \tau_{0}\exp(\big<1/\beta \Phi_{r}\big>^{0}/K_{\Phi}(\big<1/\beta \Phi_{r}\big> - \big<1/\beta \Phi_{r}\big>^{0}))$. $K_{\Phi}$ is the fragility parameter w.r.t $\big<1/\beta \Phi_{r}\big>$. The relaxation time of the LJ and WCA systems, when plotted against $\big<1/\beta \Phi_{r}\big>^{0}/K_{\Phi}(\big<1/\beta \Phi_{r}\big> - \big<1/\beta \Phi_{r}\big>^{0})$, shows a master plot (Fig. \ref{Fig_masterplot} in Appendix I). However, this master plot contains system-specific parameters. We also calculate the effective activation energy from the relaxation time, $\Delta E^{ma}=T*ln(\tau_{\alpha}/\tau_{0}) $, and plot it as a function of $<1/\beta \Phi_{r}>$. As shown in earlier work for the macroscopic SOP \cite{manoj_prl2021}, $\Delta E^{ma}$ is linearly dependent on the average depth of the caging potential, $\Delta E^{ma}=B+C*\big<1/\beta \Phi_{r}\big>^{-1}$ where $B$ and $C$ are system specific fitting parameters. The slope, $C$, has information on the coupling of the dynamics with the SOP. Note that the temperature dependence of the activation energy, which is again related to the kinetic fragility, depends on the temperature dependence of the SOP and how strongly the dynamics is coupled with the SOP. It has been shown earlier that for strong liquids, independent of the temperature dependence of the SOP, $C \rightarrow 0$, this leads to the temperature independence of the activation energy \cite{manoj_prl2021}. Since the temperature dependence of the SOP is similar for the WCA and LJ systems (slopes of the $\big<1/\beta \Phi_{r}\big>$ vs. $T$ plots in Fig. \ref{macro_study}(b)) thus, it is the difference in the slope value between the $\Delta E^{ma}$ and $\big<1/\beta \Phi_{r}\big>^{-1}$ plot, $C$, which is the major contributor to the difference in the kinetic fragility of the two systems. The value of $C$ is lower for the less fragile WCA system. Thus from the macroscopic analysis, it appears that the attractive LJ system has a stronger coupling between the structure and dynamics, which leads to its higher kinetic fragility. Interestingly similar system-specific correlation between dynamics and other structural order parameters has been reported earlier \cite{olivier_pre,tanaka_prl2020,tanaka_nature,coslovich2020}. In this work, we try to unravel the origin of this system specificity. To do that, we realize that the macroscopic parameters and their correlations do not have enough information. Thus in the next section, we investigate the structure dynamics correlation at the microscopic level.

\section{CORRELATION BETWEEN STRUCTURE AND DYNAMICS AT MICROSCOPIC LEVEL}
\label{microscopic_section}

\subsection{Correlation between SOP and fast particles}
\begin{figure}
\centering
\begin{subfigure}{0.238\textwidth}
\includegraphics[width=1.0\linewidth]{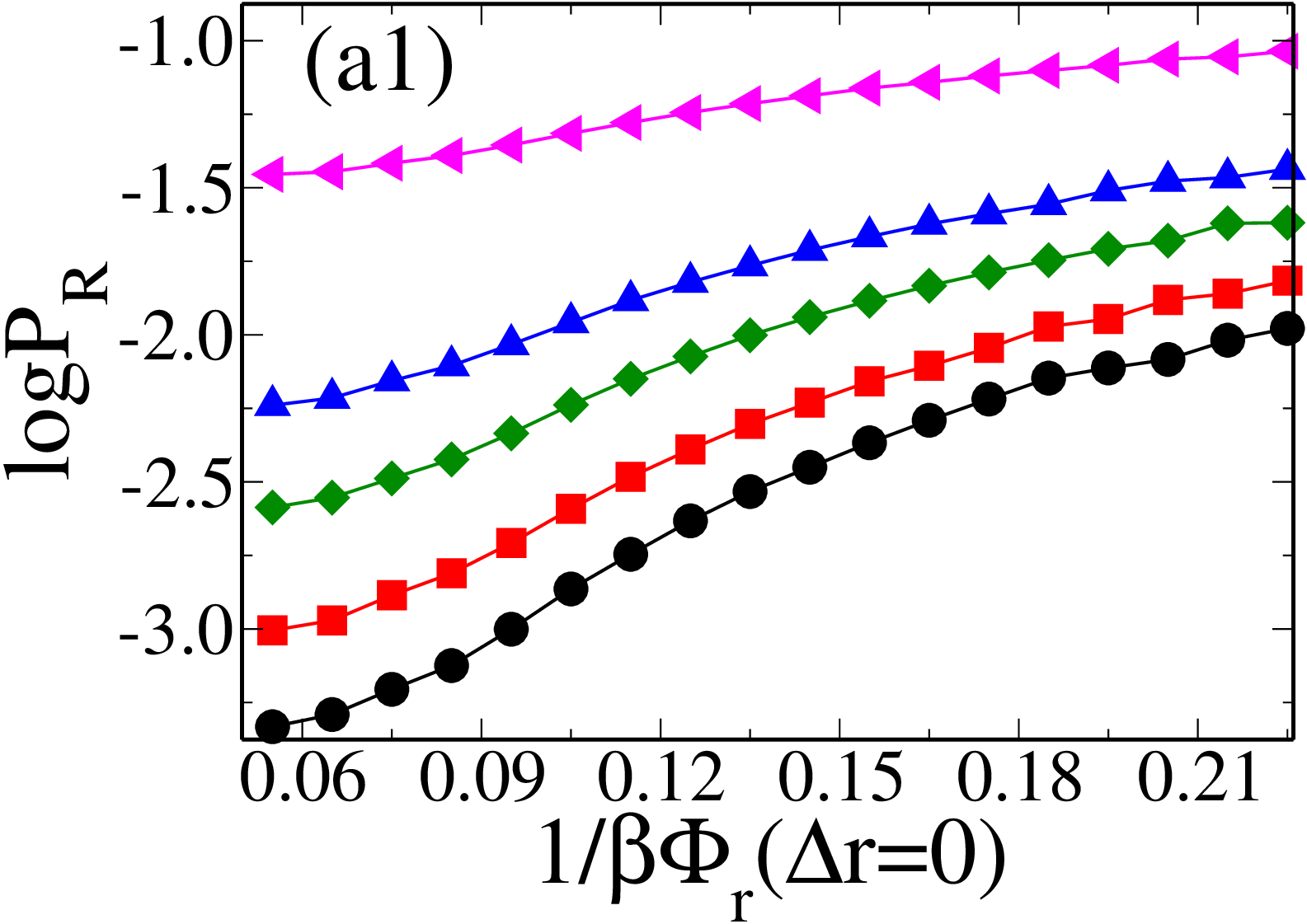}
\end{subfigure}
\begin{subfigure}{0.238\textwidth}
\includegraphics[width=1.0\linewidth]{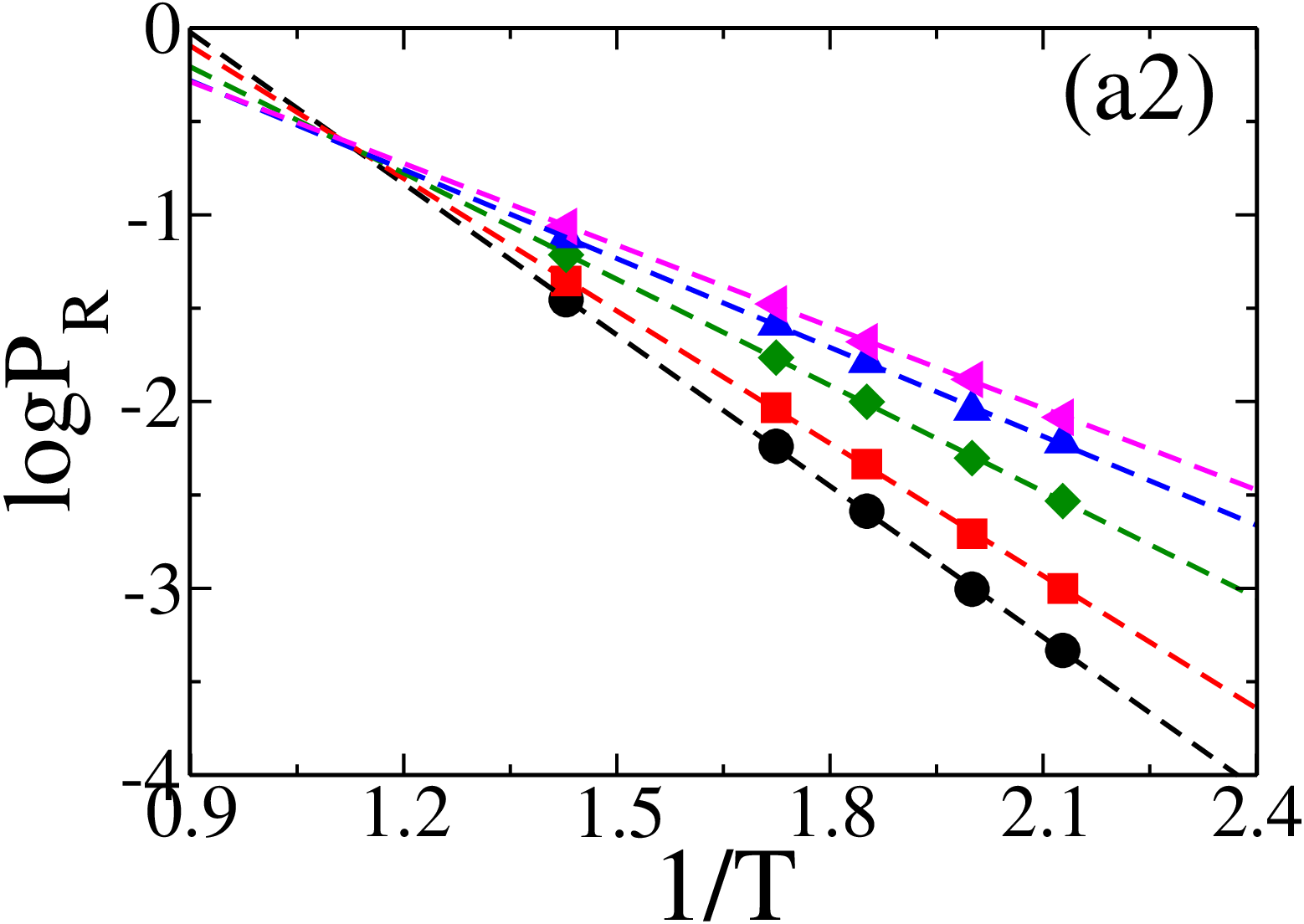}
\end{subfigure}
\vskip\baselineskip
\begin{subfigure}{0.238\textwidth}
\includegraphics[width=1.0\linewidth]{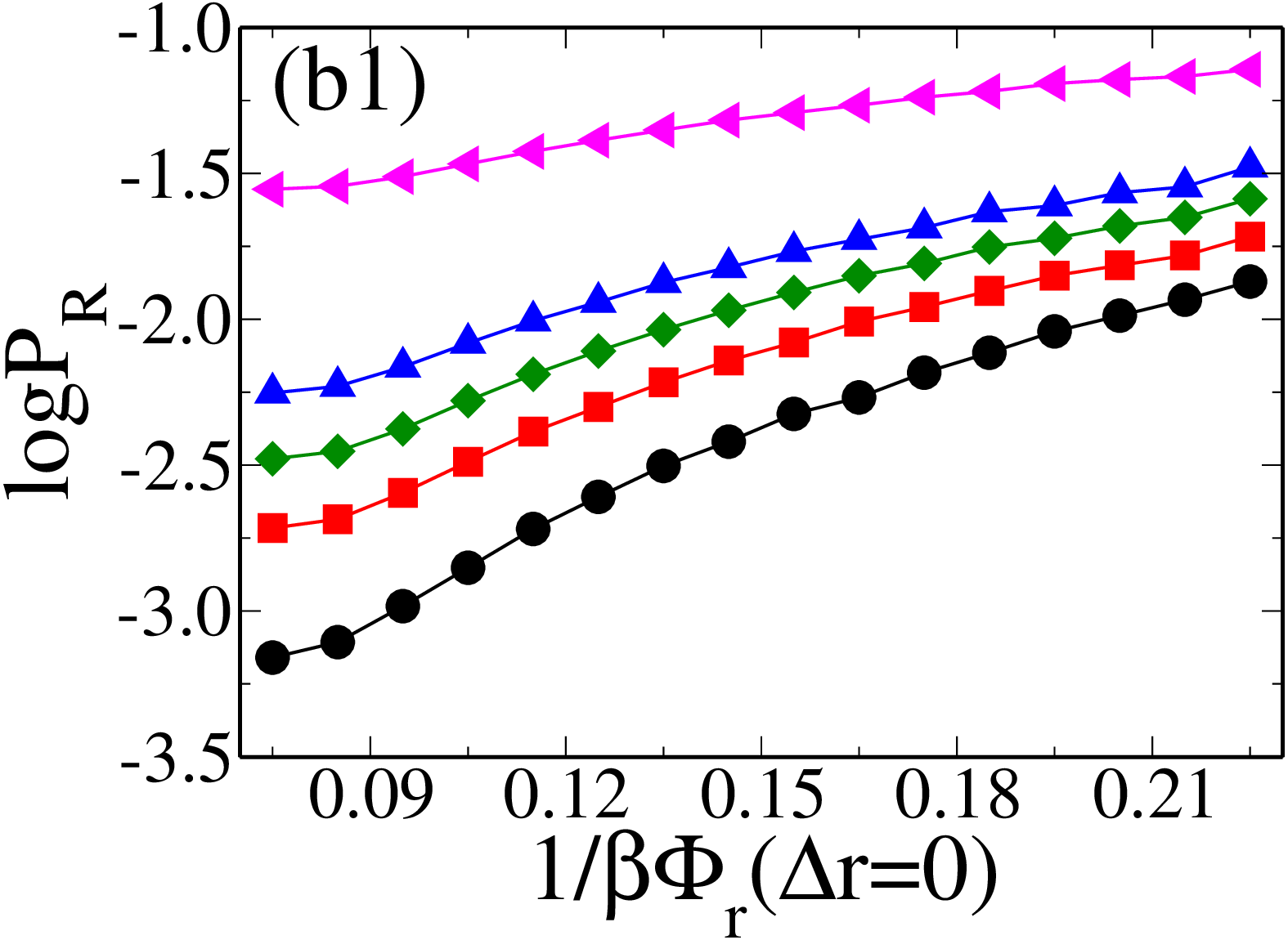}
\end{subfigure}
\begin{subfigure}{0.238\textwidth}
\includegraphics[width=1.0\linewidth]{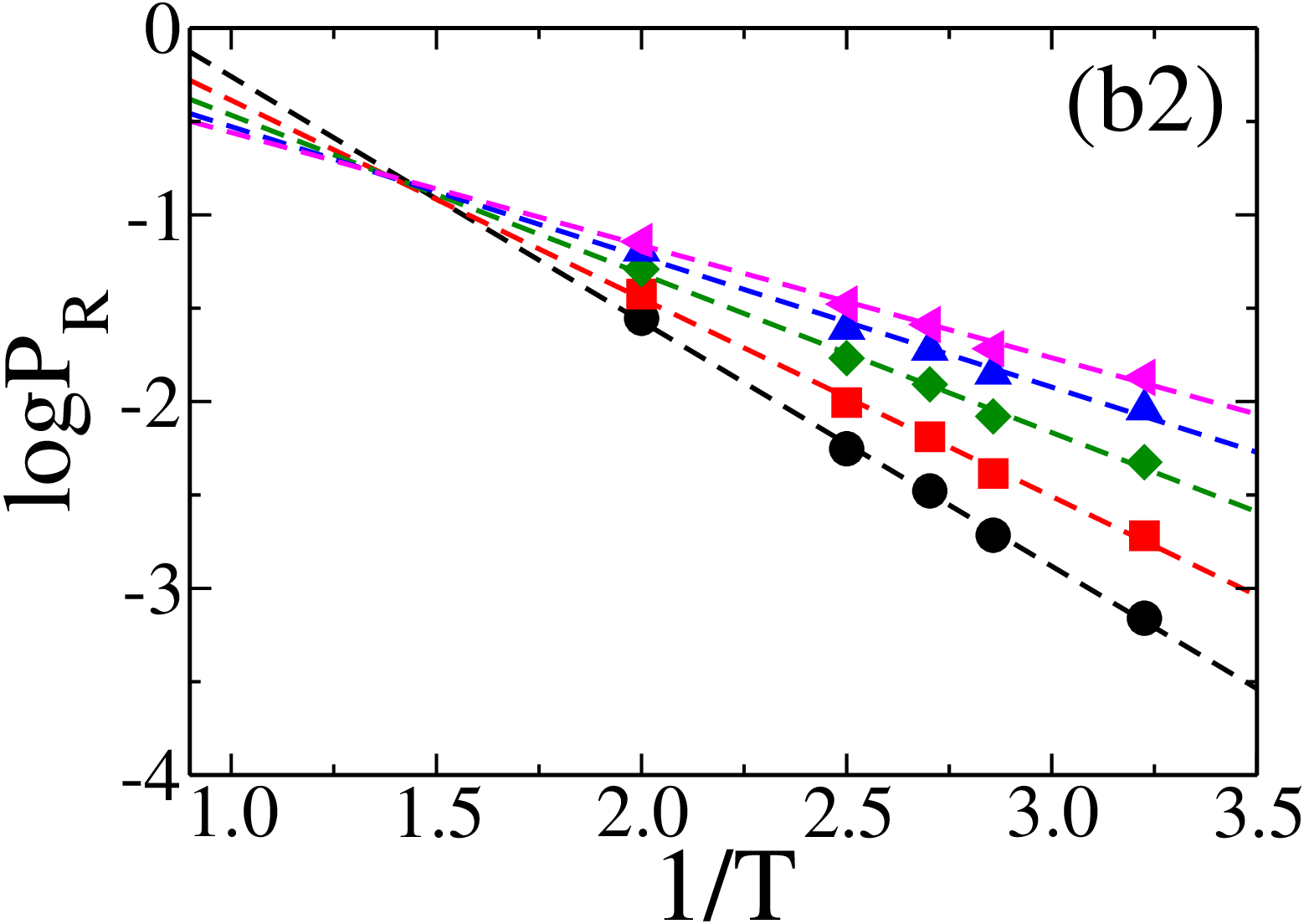}
\end{subfigure}
\caption{\textit{Probability of rearrangement, $logP_{R}$, plotted against SOP, $1/\beta \Phi_{r}$ at different temperatures $T$ for \textbf{(a1)} $LJ$ system(T = 0.47(black,circle), T = 0.50(red, square), T = 0.54(dark green, diamond), T = 0.58(blue, up triangle), T = 0.70(magenta,left triangle)) \textbf{(b1)} $WCA$ system(T = 0.31(black,circle), T = 0.35(red, square), T = 0.37(dark green, diamond), T = 0.40(blue, up triangle), T = 0.50(magenta, left triangle)). $logP_{R}$ plotted against $1/T$ for different $1/\beta \Phi_{r}$ values (lowest(black,circle) to highest(magenta,left triangle) for \textbf{(a2)} $LJ$ system \textbf{(b2)} $WCA$ system. The dashed lines are the linear fit, and the solid lines are the guide to the eye. The base of the log is 10. The temperature where the dashed lines cross, $T_{cross}$ = 0.88 for the LJ and $T_{cross}$ = 0.69 for the WCA system.}}
\label{Fig_prs}
\end{figure}

We next analyze the correlation between structure and dynamics at the microscopic level. The method of identification of the fast rearranging particles is given in Section \ref{fast_par}. In Figs. \ref{Fig_prs}(a1) and \ref{Fig_prs}(b1), we plot the fraction of particles with a specific SOP value that undergoes rearrangement, $P_{R}(1/\beta \Phi_{r})$ as a function of $1/\beta \Phi_{r}$ at different temperatures. At high temperatures, the lines are almost horizontal, suggesting that the displacement probability is independent of the value of the SOP. The $P_{R}(1/\beta \Phi_{r})$ at low temperatures shows a strong $1/\beta \Phi_{r}$ dependence. Thus, as expressed by the slope of the curves, both for the LJ and the WCA systems, the correlation between the dynamics and the SOP increases with a decrease in temperature. We also plot the $P_{R}(1/\beta \Phi_{r})$ against $1/T$ (Figs. \ref{Fig_prs}(a2) and \ref{Fig_prs}(b2)) for different $1/\beta \Phi_{r}$ values and find that it can be expressed in an Arrhenius form, $P_{R}(1/\beta \Phi_{r})=P_{0}(1/\beta \Phi_{r})exp[-
 \Delta E(1/\beta \Phi_{r})/T]$ where the activation energy is a function of the SOP and is higher for smaller $1/\beta \Phi_{r}$ values. The plots cross at a certain temperature ($T_{cross}=0.88$ for LJ and $T_{cross}=0.69$ for WCA) which describes the limiting temperature where the theory is valid \cite{mohit_pre} and has been identified earlier as the onset temperature of the supercooled liquid \cite{liu_nature,mohit_pre}. As explained in our previous work, the breakdown of the theoretical formulation is expected above the onset temperature where we are in the liquid regime, and there is no decoupling between the short and longtime dynamics; thus, both the cage and the caging potential are ill-defined\cite{mohit_pre}. We find that the onset temperature predicted using this method is similar to that of the respective systems obtained using other methods \cite{atreyee_onset}. Until now, the correlation between the SOP and the dynamics appears similar in both systems, suggesting SOP is a good predictor of the dynamics.

\begin{figure}
\centering
\begin{subfigure}{0.32\textwidth}
\includegraphics[width=1.0\linewidth]{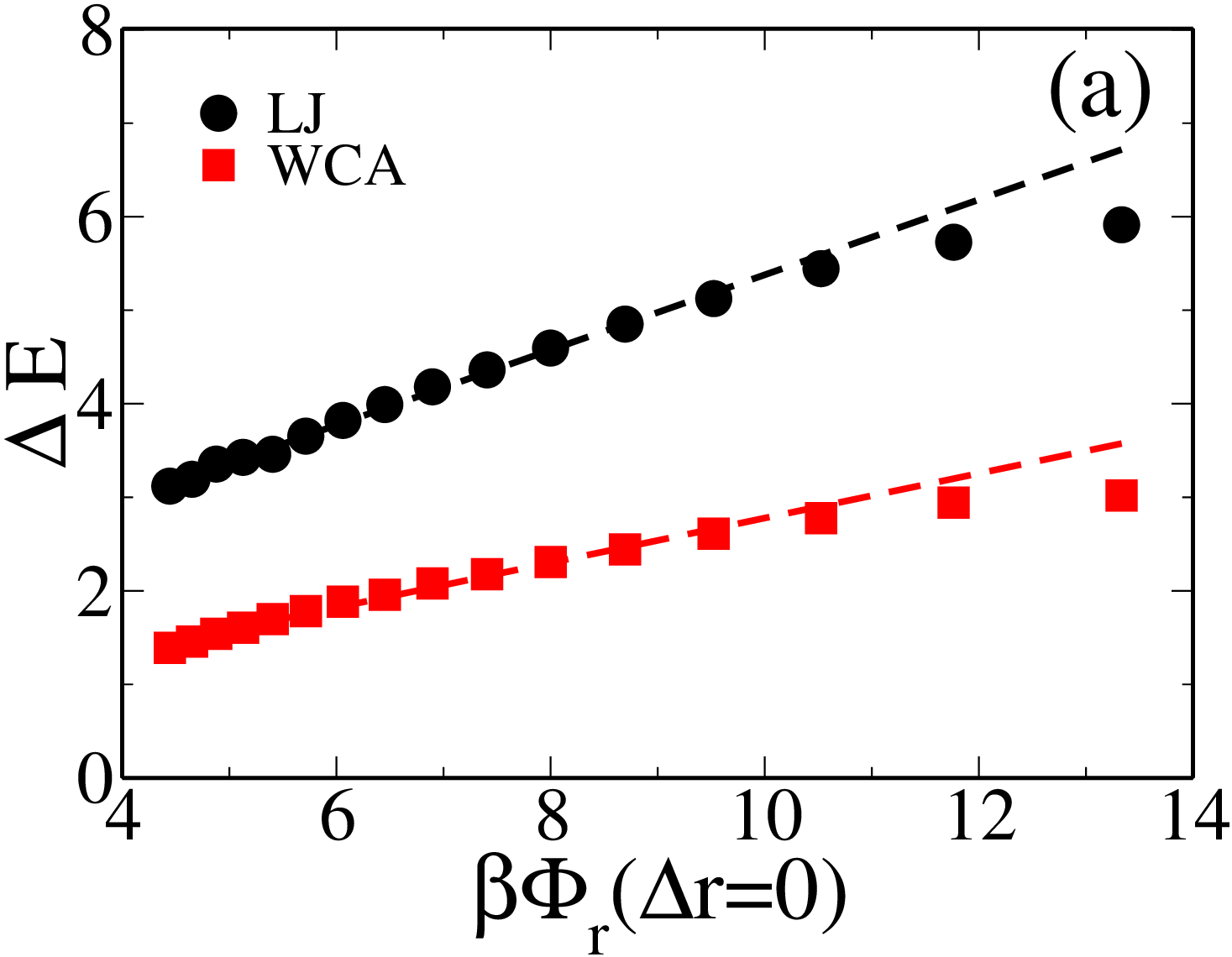}
\end{subfigure}
\vskip\baselineskip
\begin{subfigure}{0.34\textwidth}
\includegraphics[width=1.0\linewidth]{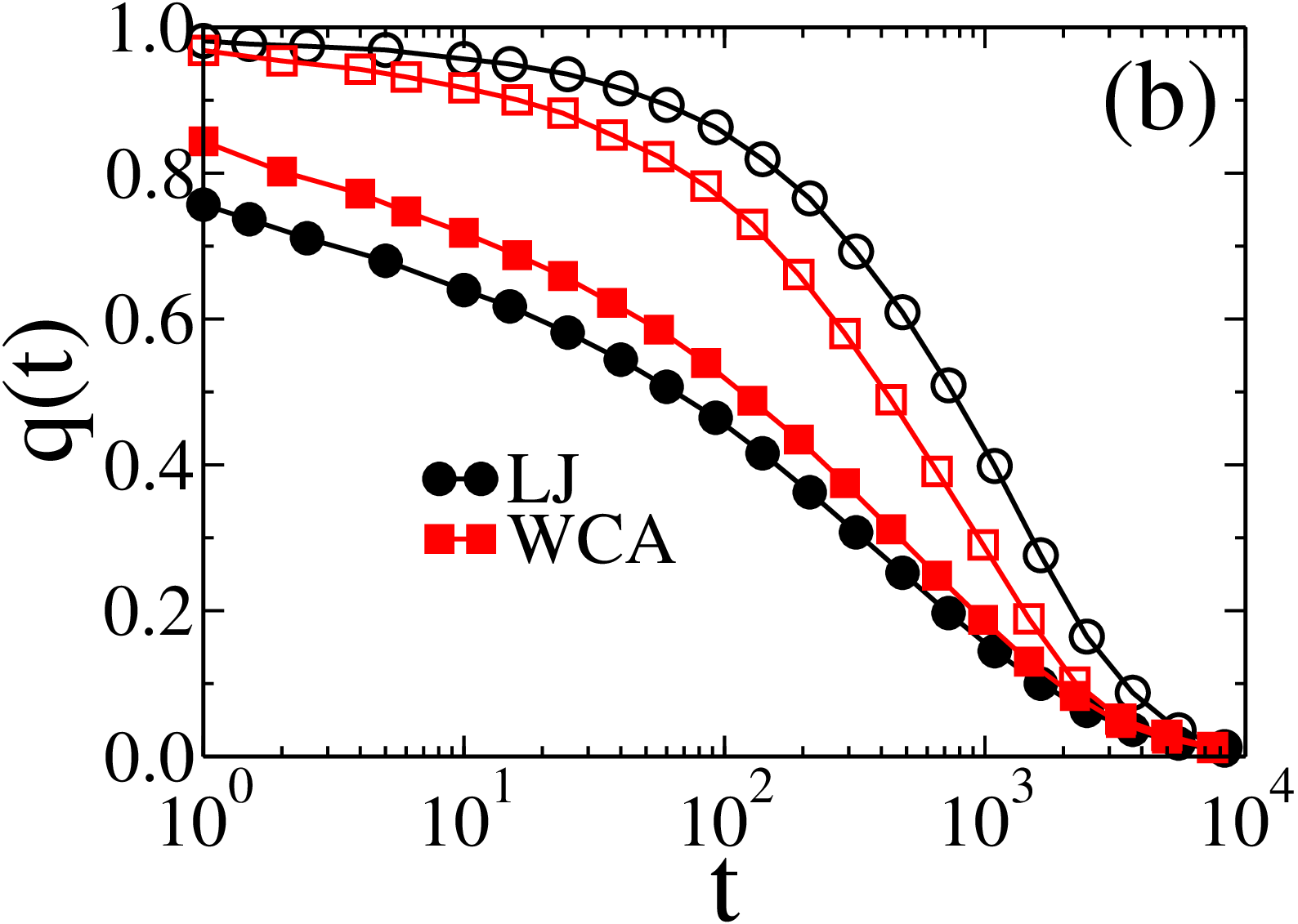}
\end{subfigure}
\caption{\textit{\textbf{(a)} Microscopic activation energy ($\Delta E$) plotted against depth of caging potential ($\beta \Phi_{r}$) for $LJ$ and $WCA$ systems. The high slope for the $LJ$ system reflects the higher correlation between SOP and dynamics, \textbf{(b)} Overlap functions of a few hardest and softest particles for $LJ$ and $WCA$ systems. Open symbols are for the hard and filled symbols for soft particles. The difference is wider in $LJ$ than $WCA$, again reflecting the higher correlation between SOP and dynamics in the $LJ$ system. The dashed lines are the linear fit, and the solid lines are the guide to the eye. }}
\label{Fig_comp}
\end{figure}

In Fig. \ref{Fig_comp}(a) we plot the microscopic activation energy $\Delta E$ against $\beta \Phi_{r}$. We find that similar to the macroscopic activation energy (Fig. \ref{macro_study}(d)), compared to the WCA system, the LJ system has a stronger dependence on the SOP. This observation is similar to that reported in the ML study\cite{olivier_pre}.
 We now compare the dynamics of a few softest and hardest particles at temperatures where the relaxation time of the two systems are similar (T=0.47 for LJ and T=0.31 for WCA). We find that the difference in the dynamics of the hardest and softest particles is wider for the LJ system(Fig. \ref{Fig_comp}(b)). 

 The microscopic analysis done so far does throw light on the origin of the difference in the SOP of the two systems at the same temperature. It also corroborates the observations made in our macroscopic study that the dynamics of the more attractive LJ system have a stronger coupling to the SOP. However, the origin of this stronger coupling remains elusive. 
 
\subsection{Correlation between SOP and mobility}
\begin{figure*}
\centering
\begin{subfigure}{0.27\textwidth}
\includegraphics[width=0.98\linewidth]{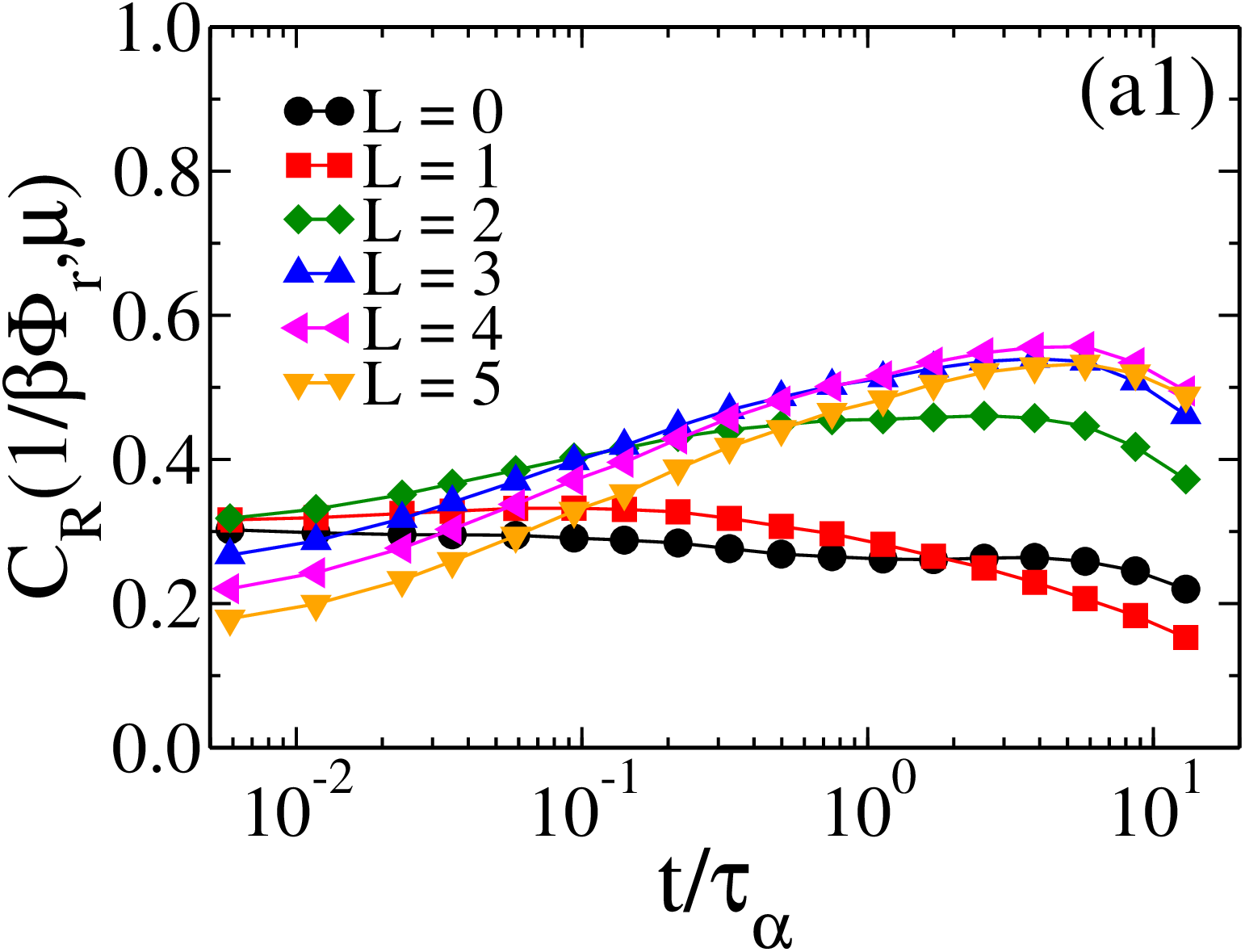}
\end{subfigure}
\begin{subfigure}{0.29\textwidth}
\includegraphics[width=1.0\linewidth]{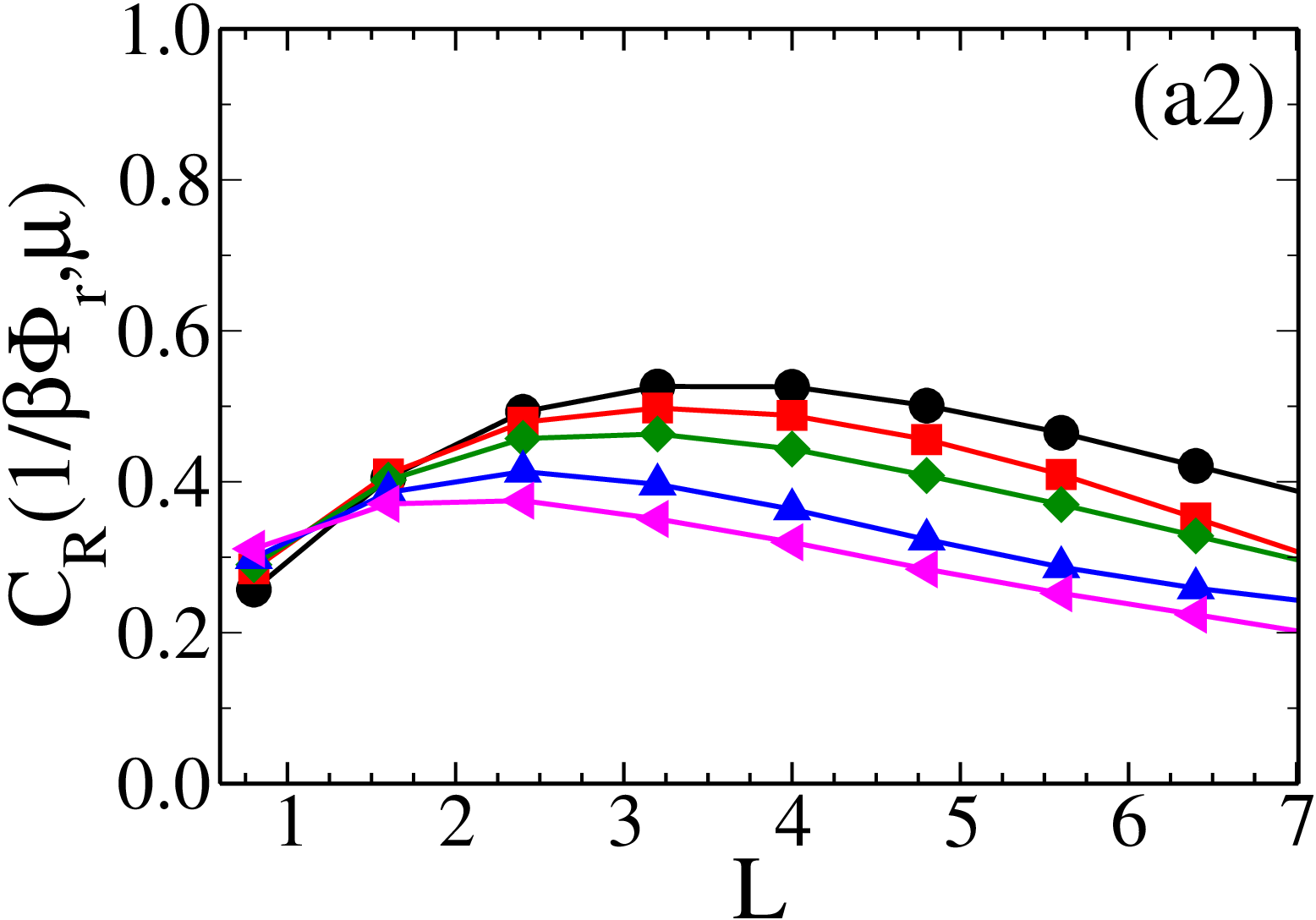}
\end{subfigure}
\begin{subfigure}{0.29\textwidth}
\includegraphics[width=1.0\linewidth]{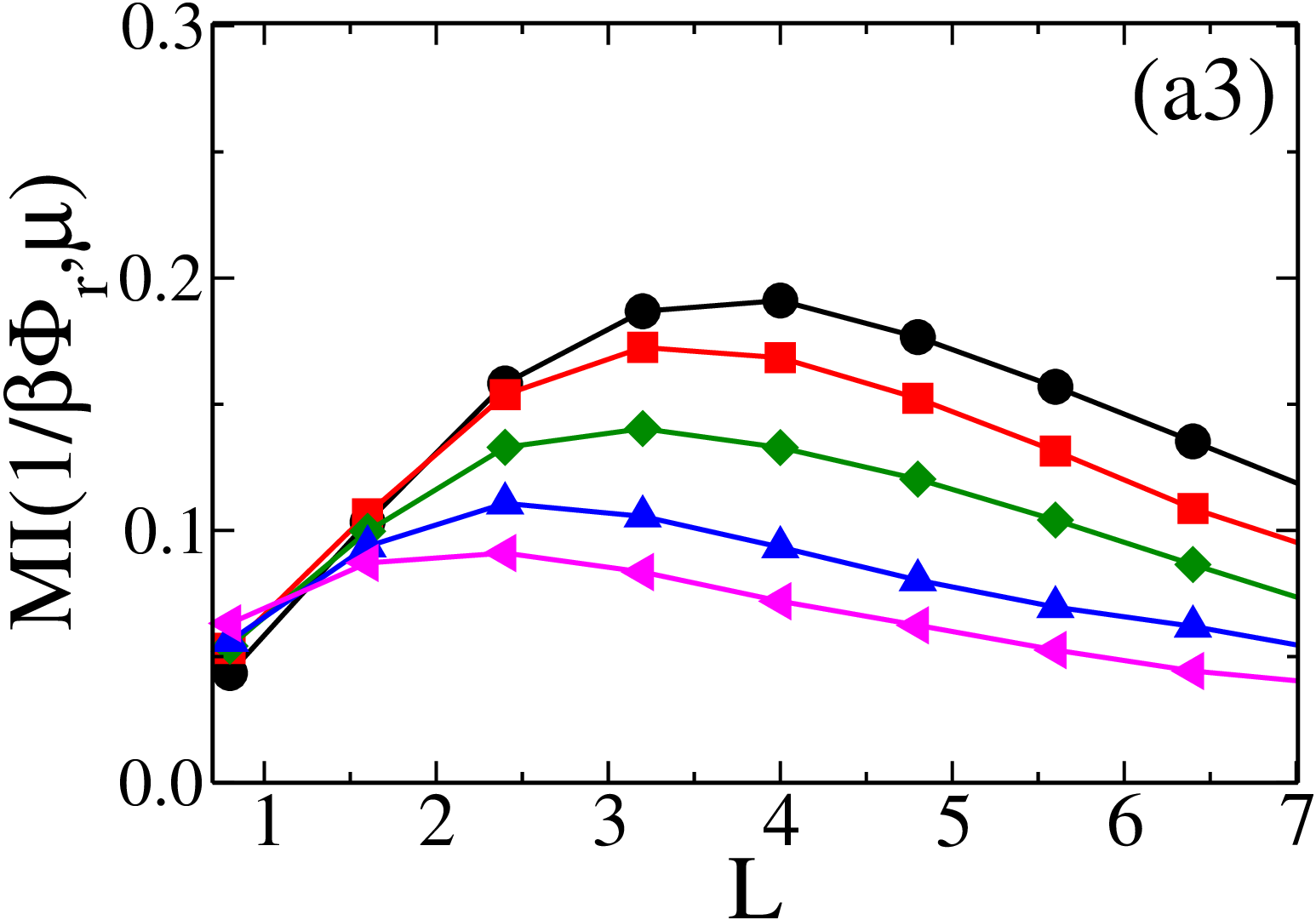}
\end{subfigure}
\vskip\baselineskip
\begin{subfigure}{0.27\textwidth}
\includegraphics[width=0.98\linewidth]{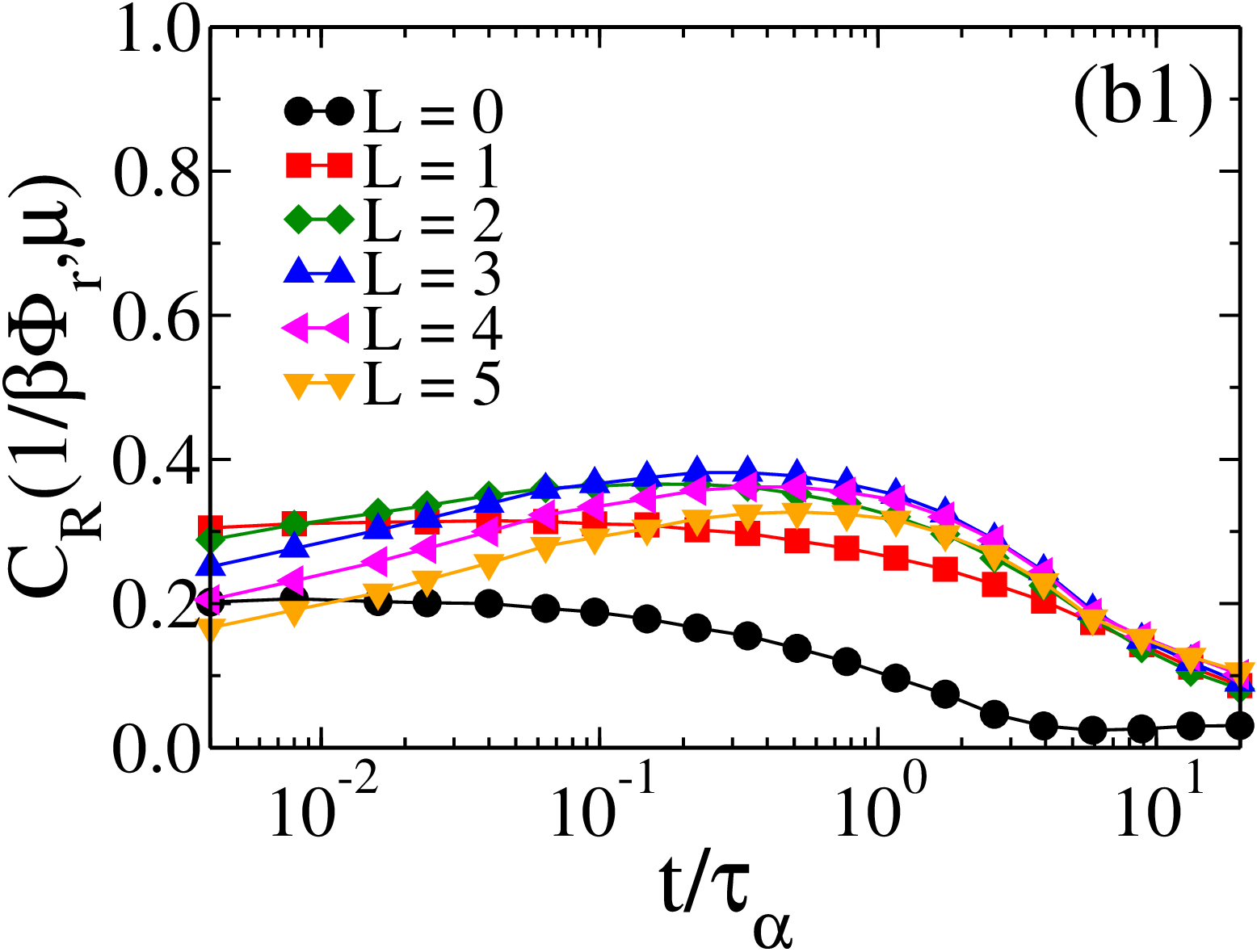}
\end{subfigure}
\begin{subfigure}{0.29\textwidth}
\includegraphics[width=1.0\linewidth]{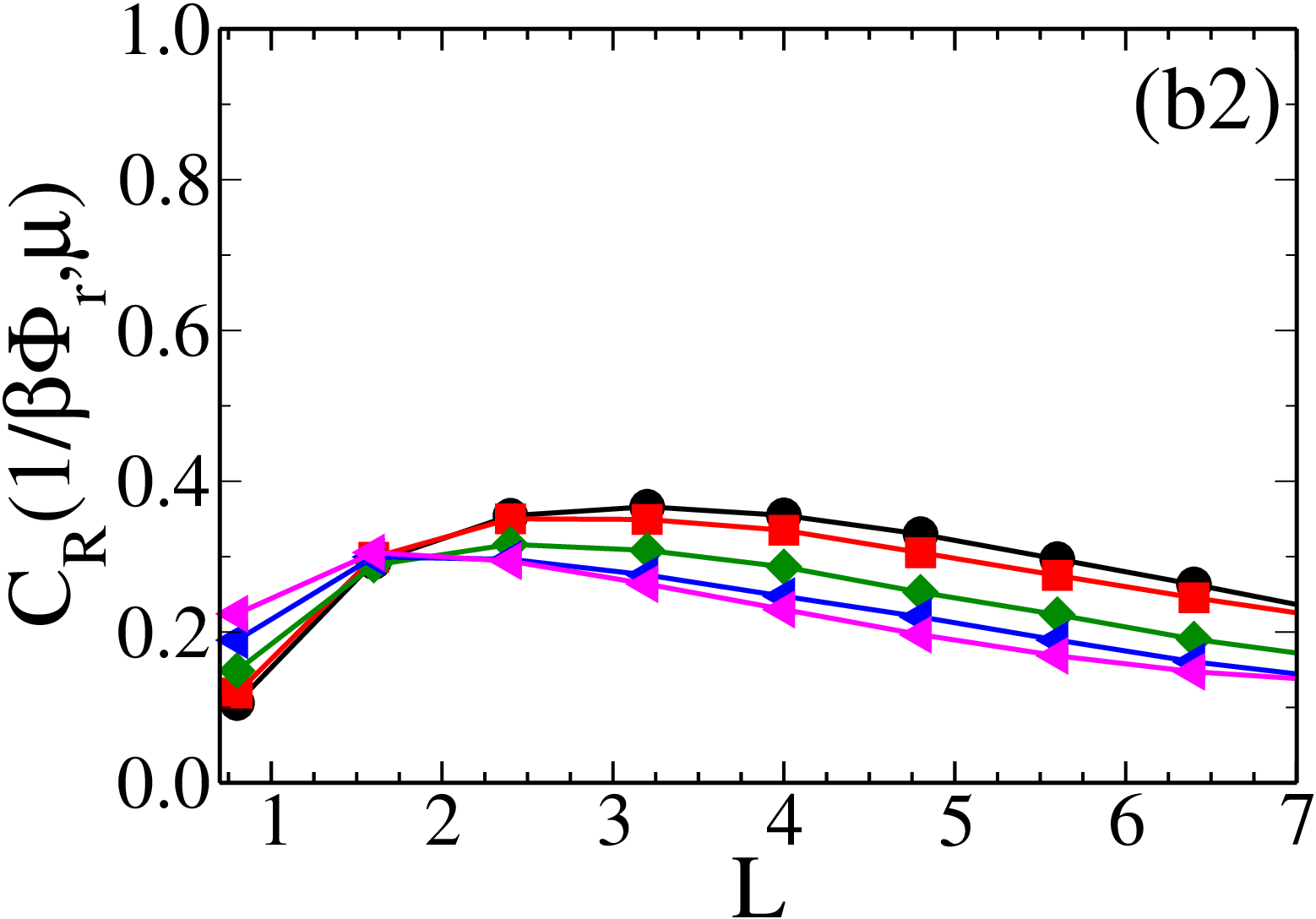}
\end{subfigure}
\begin{subfigure}{0.29\textwidth}
\includegraphics[width=1.0\linewidth]{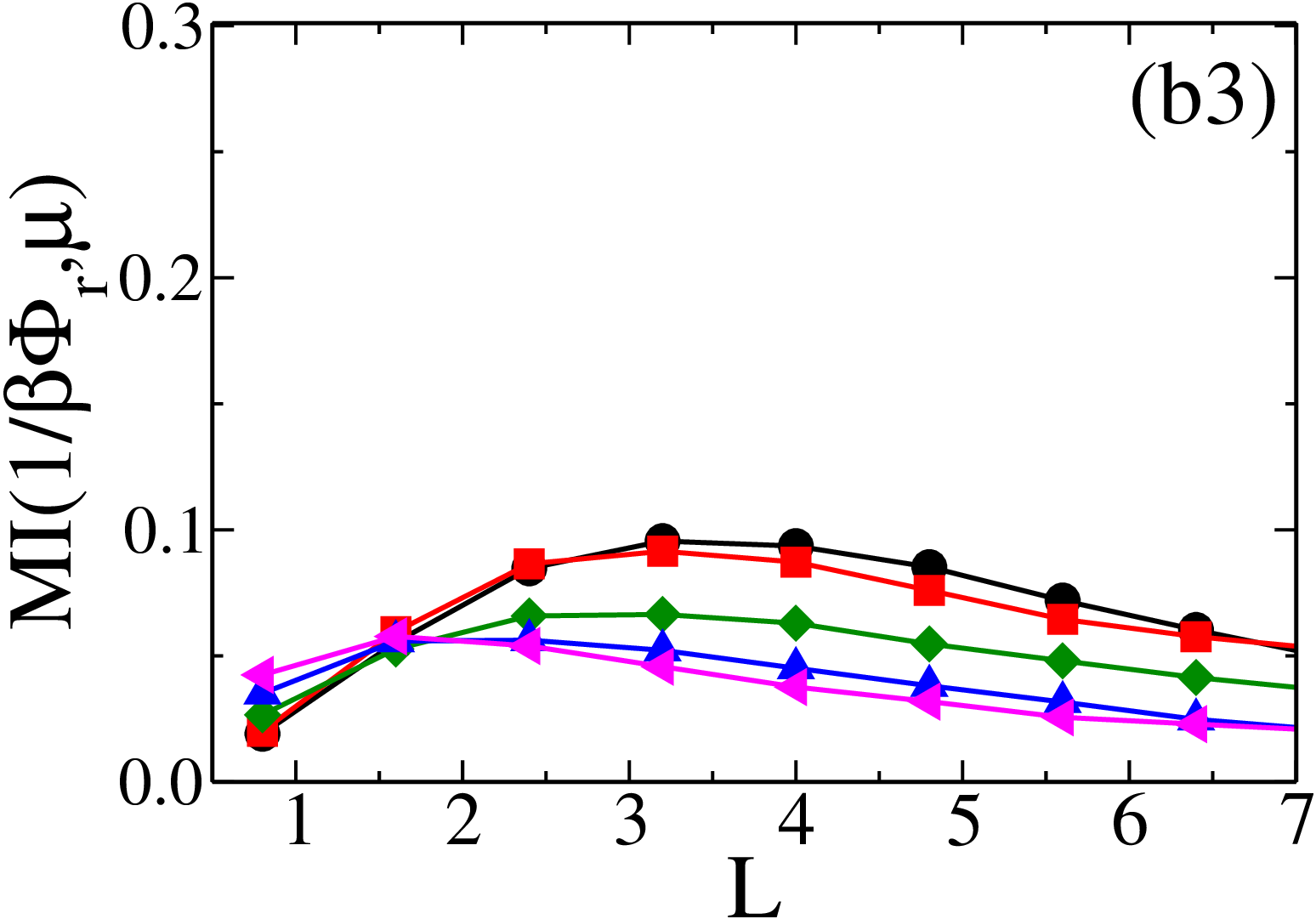}
\end{subfigure}
\caption{\textit{For \textbf{LJ}(top panel) and \textbf{WCA}(bottom panel) \textbf{(a1) and (b1)} Spearman rank correlation($C_{R}(1/\beta\Phi_{r},\mu)$) between mobility($\mu$) and SOP($1/\beta \Phi_{r}$) at different coarse graining lengths ($L$) plotted against time scaled by $\alpha$ relaxation time ($t/\tau_{\alpha}$). \textbf{(a2) and (b2)} $C_{R}(1/\beta\Phi_{r},\mu)$ between $\mu$ (calculated at $\tau_{\alpha}$) and $1/\beta \Phi_{r}$ as a function of $L$ for different $T$ (T = 0.47(black, circle), T = 0.50(red, square), T = 0.54(dark green, diamond), T = 0.58(blue, up triangle), T = 0.70(magenta,left triangle) in top panel and T = 0.31(black,circle), T = 0.33(red, square), T = 0.37(dark green, diamond), T = 0.50(blue, up triangle), T = 0.60(magenta,left triangle) in bottom panel). \textbf{(a3) and (b3)} Mutual information(MI) between $\mu$ (calculated at $\tau_{\alpha}$) and $1/\beta \Phi_{r}$ as a function of $L$ for the same $T$ as in (a2) and (b2). The solid lines are a guide to the eye. }}
\label{Fig_isoconf}
\end{figure*}

For further analysis, we next perform isoconfigurational runs (Section \ref{isoconf}). In Figs. \ref{Fig_isoconf}(a1) and \ref{Fig_isoconf}(b1) we plot the Spearman rank correlation, $C_{R}(1/\beta\Phi_{r},\mu)$ between mobility $\mu$ and the SOP as a function of $t/\tau_{\alpha}$, where $\tau_{\alpha}$ is the $\alpha$ relaxation time. This correlation against time is plotted for different coarse graining lengths, $L$, over which the SOP is coarse grained. A parameter $X$ for the $i^{th}$ particle when coarse grained over length L is defined as $\Bar{X}_{i}(L) = \sum_{j}X_{j}P(r_{j}-r_{i})/\sum_{j}P(r_{j}-r_{i})$, where $P(x)$ = $exp(-x/L)$, under the hypothesis that the influence of the neighbourhood decays exponentially with distance.
When the SOP is not coarse grained or for small values of $L$, we find the correlation between structure and dynamics is maximum at short times, after which the correlation drops. However, for higher values of $L$, the $C_{R}(1/\beta\Phi_{r},\mu)$ starts from a smaller value and grows non-monotonically with time. For the LJ system, the peak is obtained beyond $\tau_{\alpha}$, whereas for the WCA system, the peak appears below $\tau_{\alpha}$. We find that the position of the peak, $t/\tau^{\alpha}=x^{*}$ is independent of temperature (not shown here) and is dependent on the system and also on the order parameter. Similar system specific peak positions of the correlation between mobility and other order parameters have been observed before \cite{coslovich2020,coslovich_reichman, paddyMI_nature}. It needs further systematic investigation to understand what decides the peak position.

To further investigate the dependence on the coarse graining length, we plot the $C_{R}(1/\beta\Phi_{r},\mu)$ between SOP and the mobility at $t=\tau_{\alpha}$(Figs. \ref{Fig_isoconf}(a2) and \ref{Fig_isoconf}(b2)) and also at $t=x^{*}\tau_{\alpha}$ (in Appendix III ), as a function of coarse graining length $L$. In both cases, the $C_{R}(1/\beta\Phi_{r},\mu)$ shows a non-monotonic $L$ dependence. The correlation is maximum at $L=L_{max}$ and $L_{max}$ grows with the lowering of temperature. This implies that the dynamics at long times is not described by a single particle SOP. The coarse grained SOP, which has contributions from neighbouring particles over a lengthscale $L$, contributes to the long time dynamics of the central particle. This increase in the correlation between the structure and dynamics after coarse graining is quite well known \cite{harrowell_prl2006,berthier_jack_PRE,coslovich2020,tanaka_nature,tanaka_prl2020,tanaka_prx}. 
At the same temperature, the $L_{max}$ for a WCA system is smaller than the LJ system. Tong and Tanaka \cite{tanaka_nature,tanaka_prl2020,tanaka_prx} suggested that the $L_{max}$ is the predicted static length scale. Note that the static lengthscale of a less fragile WCA system is expected to be smaller than the more fragile LJ system. The study of the correlation between $L_{max}$ and static lengthscale is beyond the scope of the present work and will be presented in a future work \cite{sanket_mohit}.

We also calculate the Mutual information(MI) between the SOP and the dynamics. MI between two variables x and y is calculated by using the expression $MI(x,y) = \int \int p_{j}(x,y)\ln_{2}[p_{j}(x,y)/p(x)p(y)]dxdy$, where $p_{j}(x,y)$ is the joint probability distribution for both variables and $p(x)$ and $p(y)$ are the marginal probability distributions\cite{paddyMI_nature,coslovich2020} . The MI results are similar to that obtained from the Spearman rank correlation. For both systems, the MI as a function of $t/ \tau_{\alpha}$ also peaks at the same position as the Spearman rank correlation (shown in Appendix III). In Figs. \ref{Fig_isoconf}(a3) and \ref{Fig_isoconf}(b3), we plot the MI between $\mu (t= \tau_{\alpha})$ and $1/\beta \Phi_{r}$ as a function of $L$ at different temperatures for the LJ and WCA systems, respectively. We find that the MI also shows a non-monotonic $L$ dependence. The value of $L$ where MI is maximum is temperature dependent and increases with a decrease in the temperature. The values are similar to the $L_{max}$ values obtained from the Spearman rank correlation. Finally, in terms of Spearman rank correlation and also MI, we find the correlation between structure and mobility is always higher for the LJ system.

\begin{figure}
\centering
\begin{subfigure}{0.238\textwidth}
\includegraphics[width=1\linewidth]{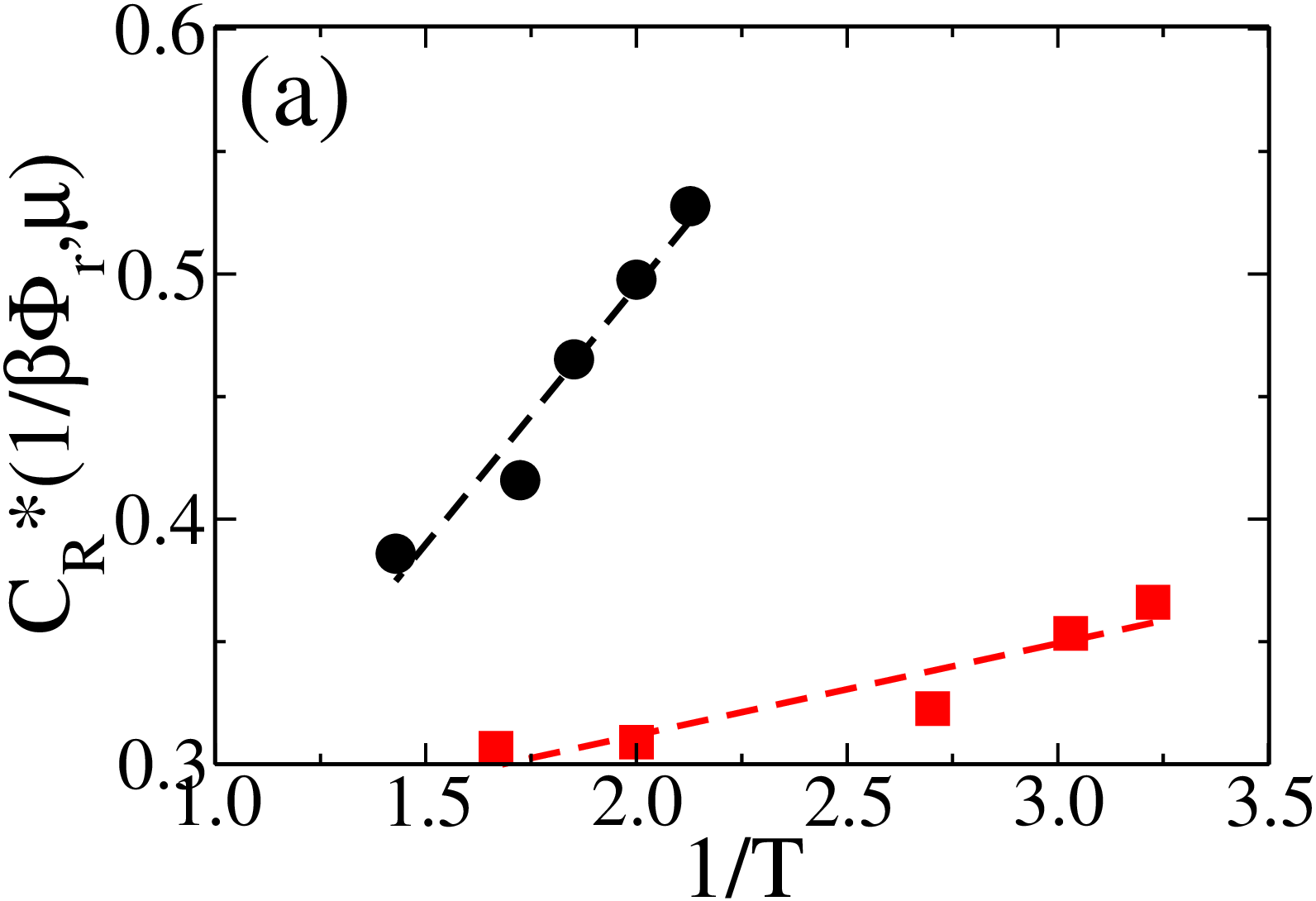}
\end{subfigure}
\begin{subfigure}{0.238\textwidth}
\includegraphics[width=0.9\linewidth]{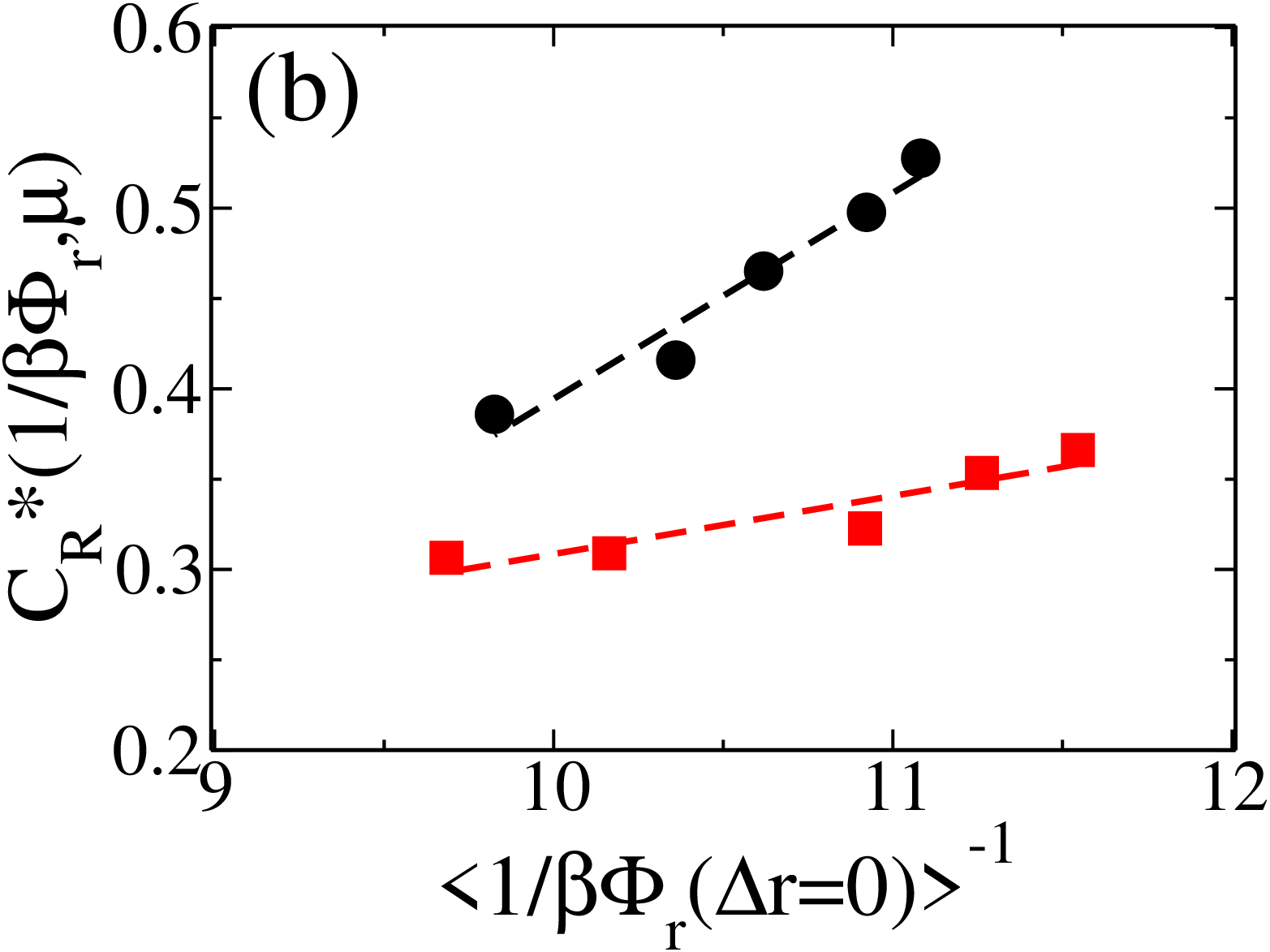}
\end{subfigure}
\vskip\baselineskip
\begin{subfigure}{0.238\textwidth}
\includegraphics[width=1\linewidth]{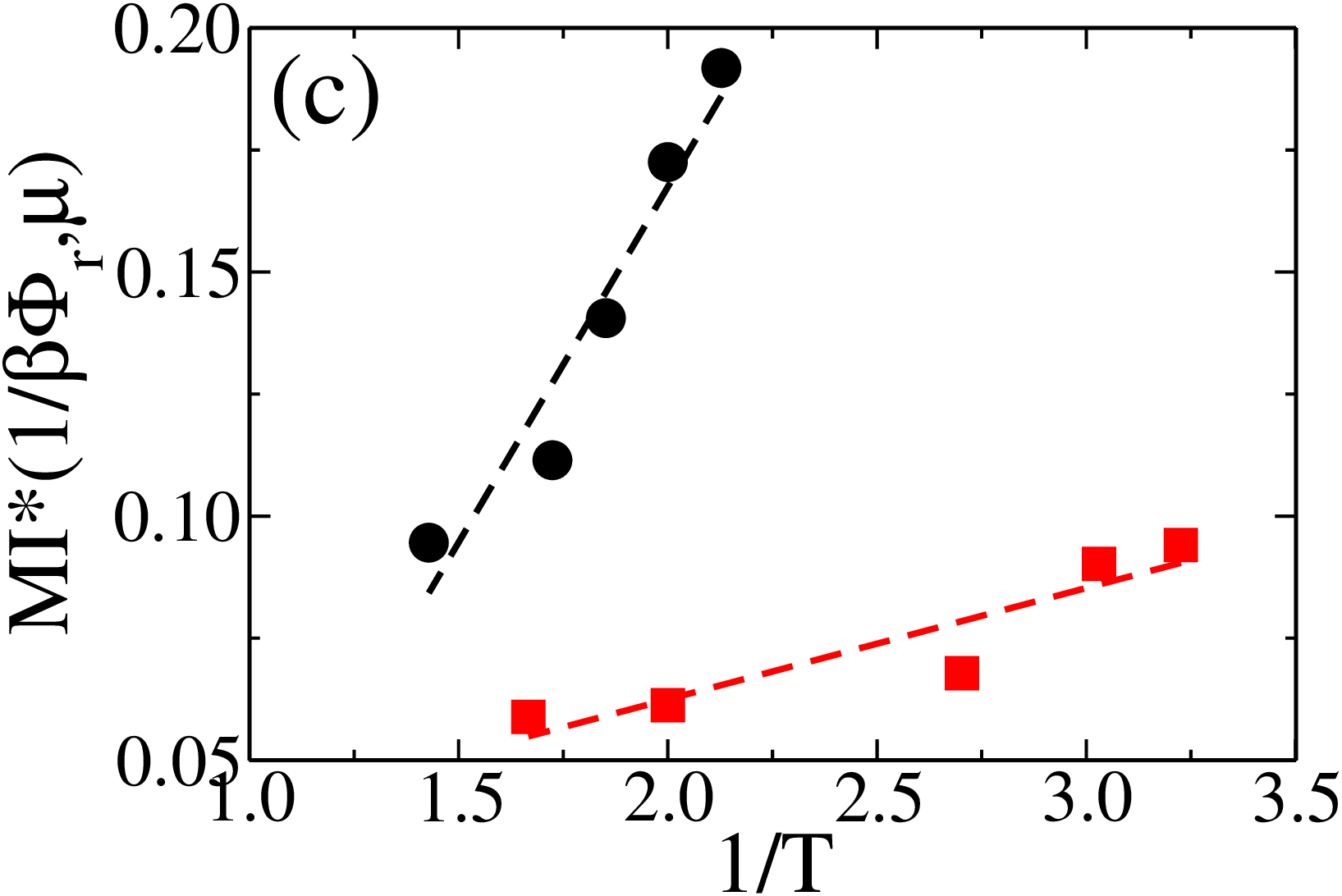}
\end{subfigure}
\begin{subfigure}{0.238\textwidth}
\includegraphics[width=0.9\linewidth]{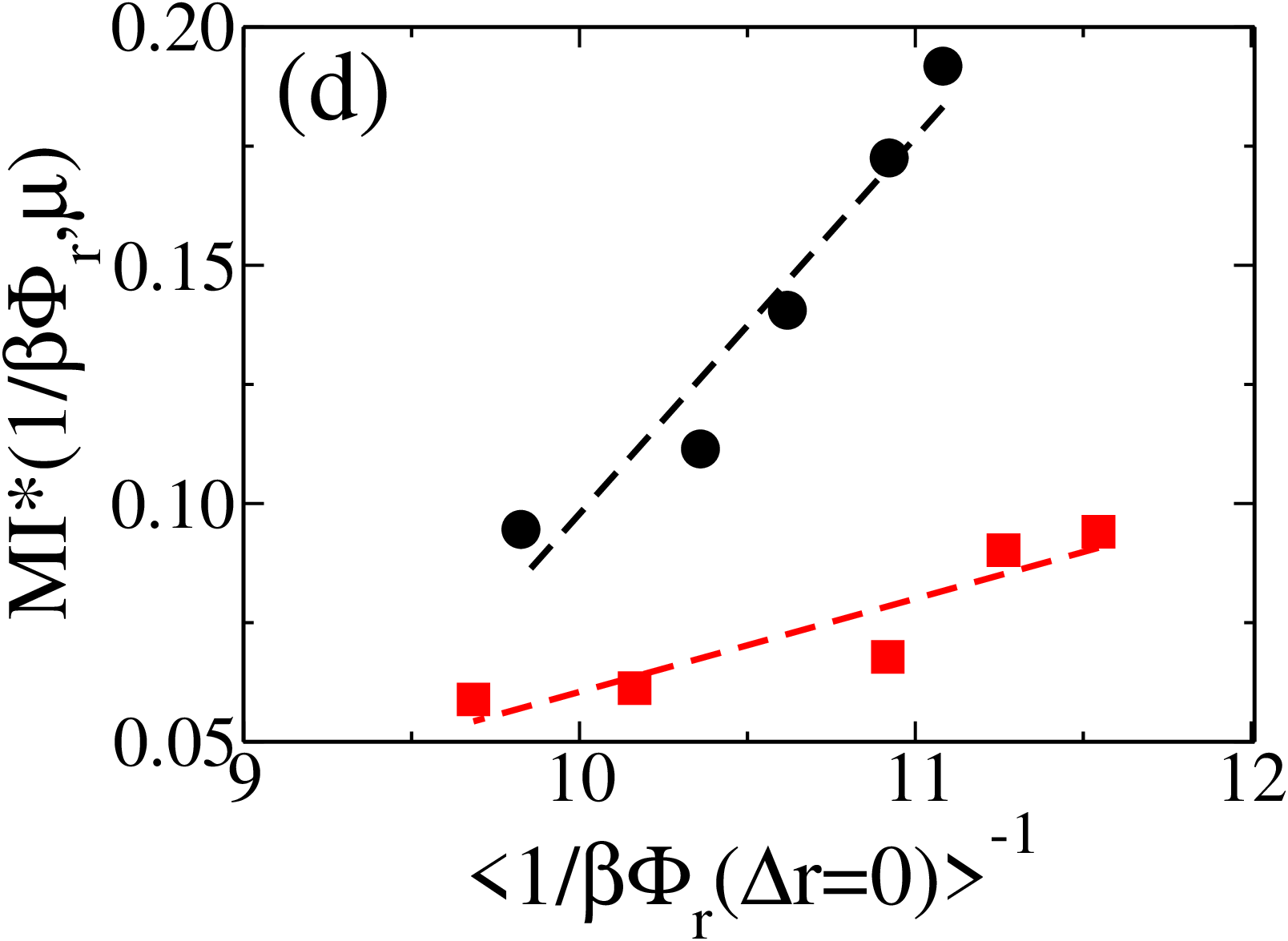}
\end{subfigure}
\vskip\baselineskip
\begin{subfigure}{0.238\textwidth}
\includegraphics[width=1\linewidth]{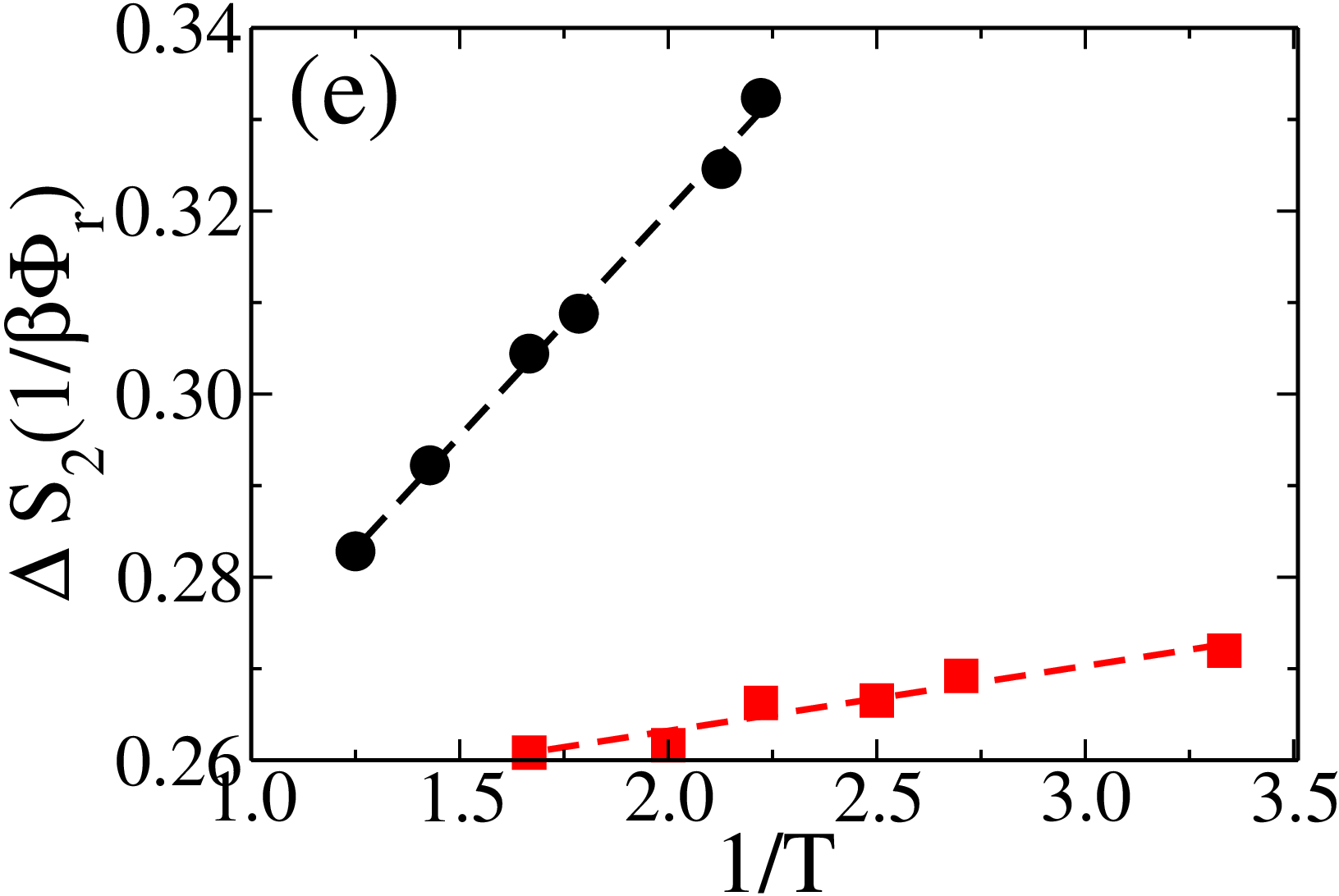}
\end{subfigure}
\begin{subfigure}{0.238\textwidth}
\includegraphics[width=0.9\linewidth]{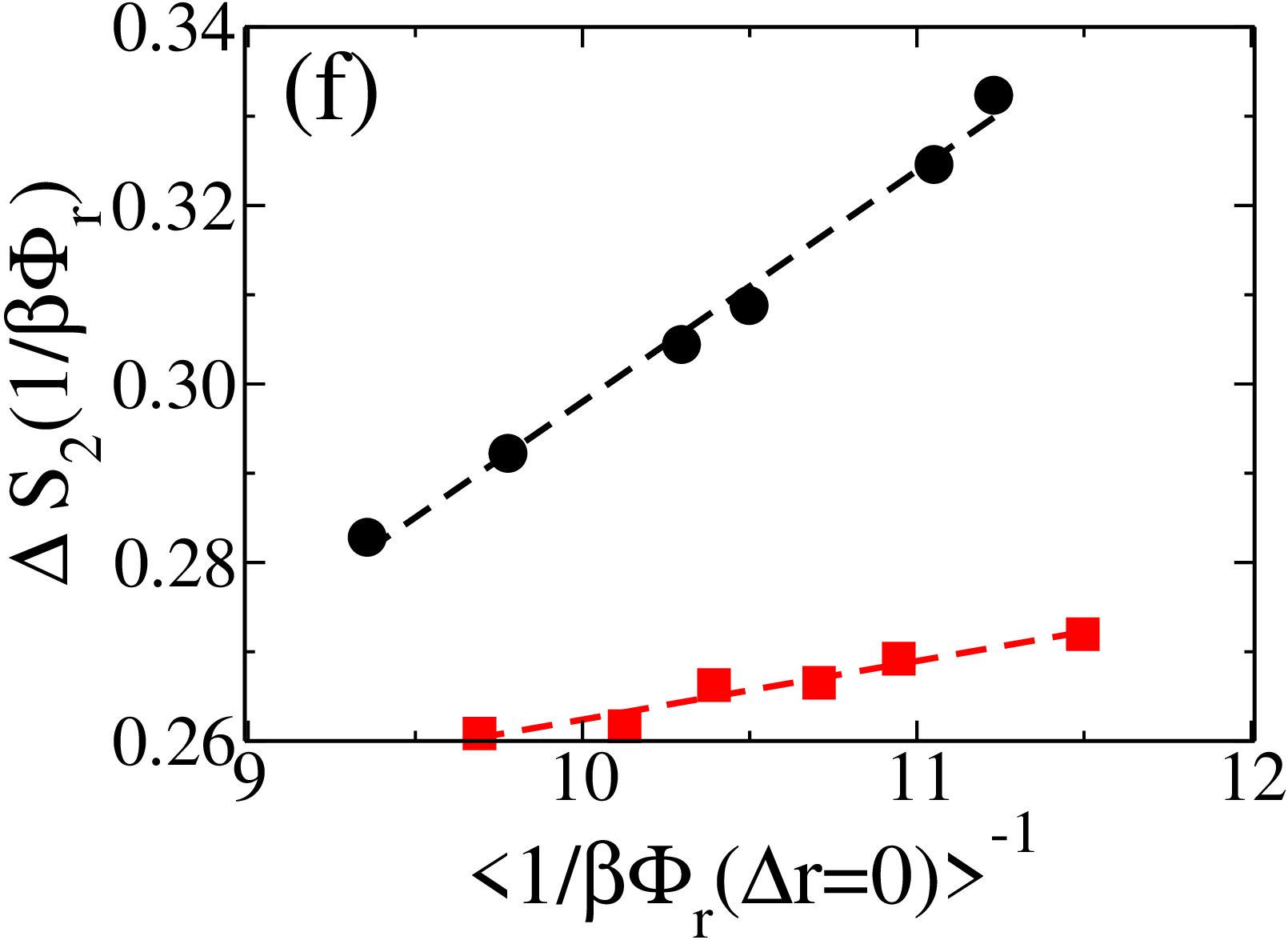}
\end{subfigure}

\caption{\textit{For \textbf{LJ}(black, circle) and \textbf{WCA}(red, square) system, the position of the peak in Figs. \ref{Fig_isoconf}(a2) and \ref{Fig_isoconf}(b2) ($C_{R}^{*}$) when plotted against \textbf{(a)} inverse of temperature ($1/T$) \textbf{(b)} average of depth of caging potential ($\big<1/\beta \Phi_{r}\big>^{-1}$). The position of the peak in Figs. \ref{Fig_isoconf}(a3) and \ref{Fig_isoconf}(b3) ($MI^{*}$) when plotted against \textbf{(c)} $1/T$ and \textbf{(d)} $\big<1/\beta \Phi_{r}\big>^{-1}$. $\Delta S_2$ (calculated using Eq.\ref{entropy_diff}) when plotted against \textbf{(e)} $1/T$ and \textbf{(f)} $\big<1/\beta \Phi_{r}\big>^{-1}$. The dashed lines are linear fits.}}
\label{Fig_Lmax}
\end{figure}

\subsection{Community information}
We next plot the peak values of $C_{R}(1/\beta\Phi_{r},\mu)$ vs. $L$ plot (Figs. \ref{Fig_isoconf}(a2) and \ref{Fig_isoconf}(b2)), $C_{R}^{*}(1/\beta\Phi_{r},\mu)$ and MI($1/\beta\Phi_{r},\mu$) vs. $L$ plot (Figs. \ref{Fig_isoconf}(a3) and \ref{Fig_isoconf}(b3)) , MI$^{*}(1/\beta\Phi_{r},\mu)$ against the temperature (Figs. \ref{Fig_Lmax}(a) and \ref{Fig_Lmax}(c)). Compared to the WCA system, both $C_{R}^{*}(1/\beta\Phi_{r},\mu)$ and MI$^{*}(1/\beta\Phi_{r},\mu)$ for the LJ system have a stronger temperature dependence. To understand this better, we next perform an analysis inspired by the recent work of Coslovich and co-workers \cite{coslovich2020}. In their work, depending on the local structure given by the radial distribution function, they have divided the system into two types of particles belonging to the locally ordered and locally disordered communities. The authors have shown that compared to the KA system with LJ interaction potential (our LJ system) in the more fragile Wahnström (WAHN) system, the MI between the dynamics and the structural community is higher. The authors have also shown that for the WAHN system, the community information which differentiates the local structure of the ordered and disordered communities has a stronger temperature dependence. Thus they suggested that the community information, which is completely structural in nature, can be considered a proxy of structural fragility. Their study also suggests that higher community information is correlated with higher structure dynamics correlation.

Following their work \cite{coslovich2020}, the community information for a binary system can be expressed as,

\begin{equation}
 \Delta S_{2} = \sum_{k=1}^{2} \int_{0}^{R} 4\pi r^2 \sum_{\beta}\rho_{\beta}f_{k}^{\alpha}g_{k}^{\alpha\beta}(r)\ln \Big(\frac{g_{k}^{\alpha\beta}(r)}{g^{\alpha\beta}(r)} \Big) dr,
\label{entropy_diff}
\end{equation}
\noindent 
where $\alpha$ is the tagged particle, $\beta$ is the neighbouring particle, $\rho_{\beta}$ is the density of these neighbours, $f_{k}^{\alpha}$ is the fraction of $\alpha$ particle in the community $k$, $g_{k}^{\alpha\beta}(r)$ is the partial rdf for particles in the community $k$ and $g^{\alpha\beta}(r)$ is partial rdf for all particles irrespective of their community type. Note that this community information is similar but not identical to Mutual information. Instead of the probability distribution of a single particle property used in the MI, here, the rdf is used, which is not normalized and has information about the interparticle distances. If the rdf of the different communities are different from each other and different from the average rdf, then the community information is high. 

In our study, there is no fixed number of communities since our SOP distribution for all the particles is continuous. However, in the simplest case, we can divide the system into two communities. We assume that particles whose SOP are higher than the average SOP value, $\big<1/\beta \Phi_{r}\big>$, are soft particles, and whose SOP are lower than $\big<1/\beta \Phi_{r}\big>$ are hard particles. Thus, we have a fraction of hard and soft particles for both LJ and WCA systems. We find that this fraction is temperature dependent, and as we go to lower temperatures, the fraction of hard particles increases. We now calculate the community information of the hard and soft particles in terms of their local rdf as given by Eq.\ref{entropy_diff}. We find that the community information of the more fragile LJ system shows a stronger temperature dependence (Fig. \ref{Fig_Lmax}(e)). This implies that as the system goes to lower temperatures, the difference in the local structure of the hard and soft particles increases sharply for the LJ system. The results also clearly show that our SOP, the inverse of depth of the local caging potential, is a good predictor of the local structure. Similar to the earlier study,\cite{coslovich2020}, the increase in community information is accompanied by the increase in $C_{R}^{*}(1/\beta\Phi_{r},\mu)$ and MI$^{*}(1/\beta\Phi_{r},\mu)$. Note that the former is only structural in nature, whereas the latter two correlate the structure with the dynamics. Thus we can say that when the local structure of the different communities is different, the dynamics is better correlated with the structure.

Note that this study aims to understand the system specific role the SOP plays in the dynamics. Thus we also plot $C_{R}^{*}(1/\beta\Phi_{r},\mu)$, $MI^{*}(1/\beta\Phi_{r},\mu)$ and $\Delta S_{2}(1/\beta\Phi_{r})$ as a function of inverse of average SOP $\big<1/\beta \Phi_{r}\big>^{-1}$ (Figs. \ref{Fig_Lmax}(b), \ref{Fig_Lmax}(d) and \ref{Fig_Lmax}(f)). $C_{R}^{*}$ and $MI^{*}$ show system specific dependence on the average SOP value. At the same value of average SOP, the correlation between the structure and dynamics is higher for the LJ system, and the correlation also grows faster with the decrease of the average SOP value. For the same value of average SOP, the difference in the local structure of the hard and soft particles is also wider for the LJ system leading to the higher values of $\Delta S_{2}(1/\beta\Phi_{r})$. Thus not only in terms of temperature but also in terms of average SOP, the community information of these two systems is different.

Since $\Delta S_{2}(1/\beta\Phi_{r})$ vs. $\big<1/\beta \Phi_{r}\big>^{-1}$ shows a linear behaviour, we can write $\Delta S_{2}(1/\beta\Phi_{r}) = a_{0} + a_{1}*\big<1/\beta \Phi_{r}\big>^{-1}$. As mentioned before, Coslovich and coworkers suggested that the temperature dependence of the community information can be used as a proxy for structural fragility \cite{coslovich2020}. Here we find that the temperature dependence of $\Delta S_{2}(1/\beta\Phi_{r})$ is dependent on the temperature dependence of $\big<1/\beta \Phi_{r}\big>$ and also the parameter $a_{1}$. Thus the study shows that there may be two contributions to structural fragility. The first is the temperature dependence of the average structure, which is captured by the average quantities like $\big<1/\beta \Phi_{r}\big>$ and other such structural order parameters. However, a second contribution comes from the variation of the local structures of different communities (hard and soft) for the same value of the average quantity. The latter contribution can be measured in terms of the slope of the $\Delta S_{2}(1/\beta\Phi_{r})$ vs. $\big<1/\beta \Phi_{r}\big>^{-1}$ plot, $a_{1}$. The slope is higher for the LJ system. 
This is quite similar to what we observe for the kinetic fragility from the $\Delta E^{ma}$ vs. $\big<1/\beta \Phi_{r}\big>^{-1}$ relationship (Fig. \ref{macro_study}(d)). Thus it appears that the system specificity of the correlation between the dynamics and the SOP is connected to how different the local structures are for the hard and soft particles and how well we can describe them in terms of different communities. 

\section{Comparing Different Order Parameters
\label{section_diffpara}}
In this section, we compare the performance of the mean field caging potential with other local order parameters. Since our SOP represents a form of the potential, the first obvious choice is the local potential energy $e_{loc}$. Also, since we use the information of the radial distribution function to describe our caging potential we compare it with another order parameter that also uses the radial distribution function, namely the local two body excess entropy $S_{2}$. These comparisons provide some interesting insight.

\begin{figure}[h!]
\centering
\begin{subfigure}{0.23\textwidth}
\includegraphics[width=1.0\linewidth]{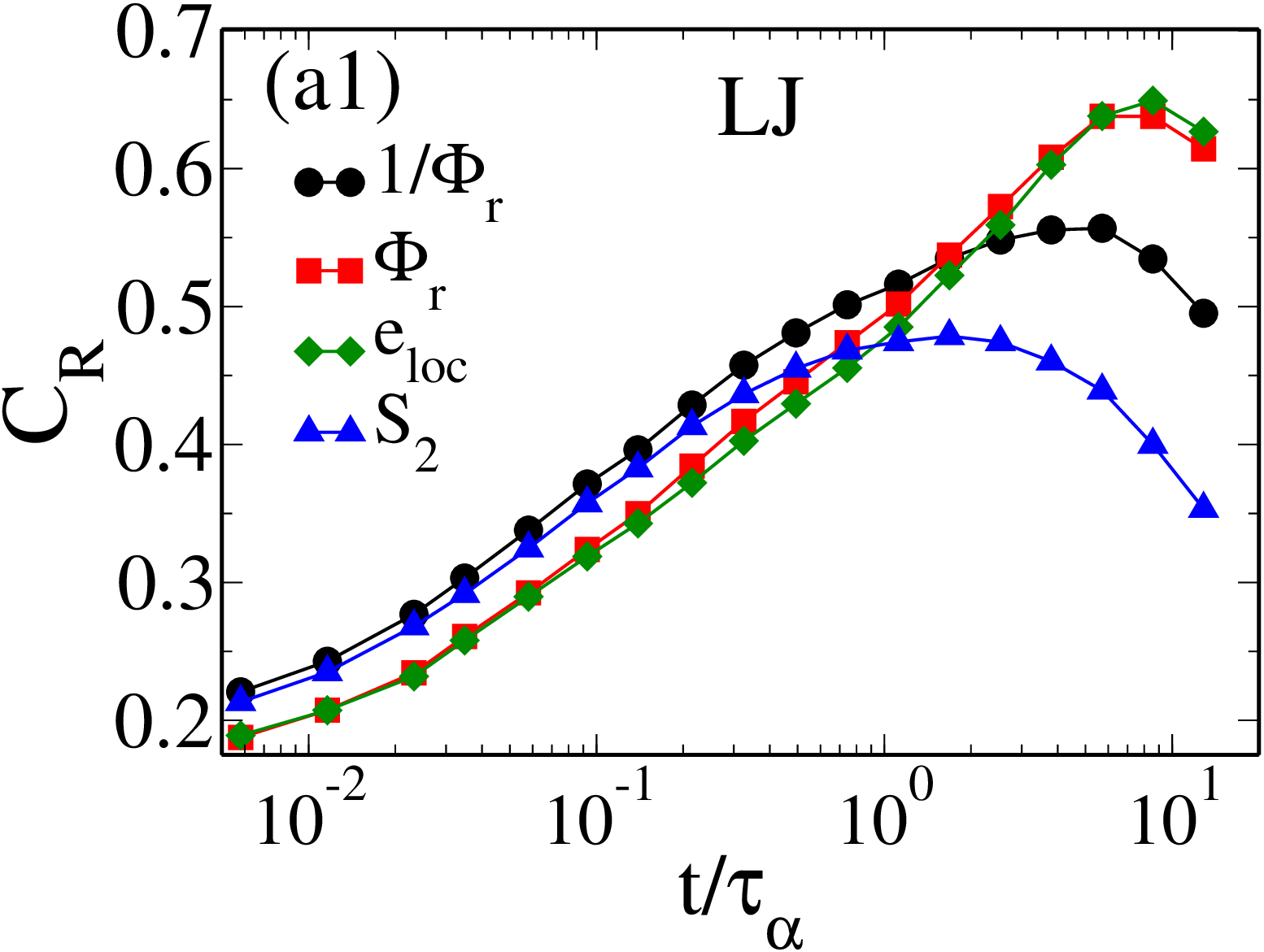}
\end{subfigure}
\begin{subfigure}{0.238\textwidth}
\includegraphics[width=1.0\linewidth]{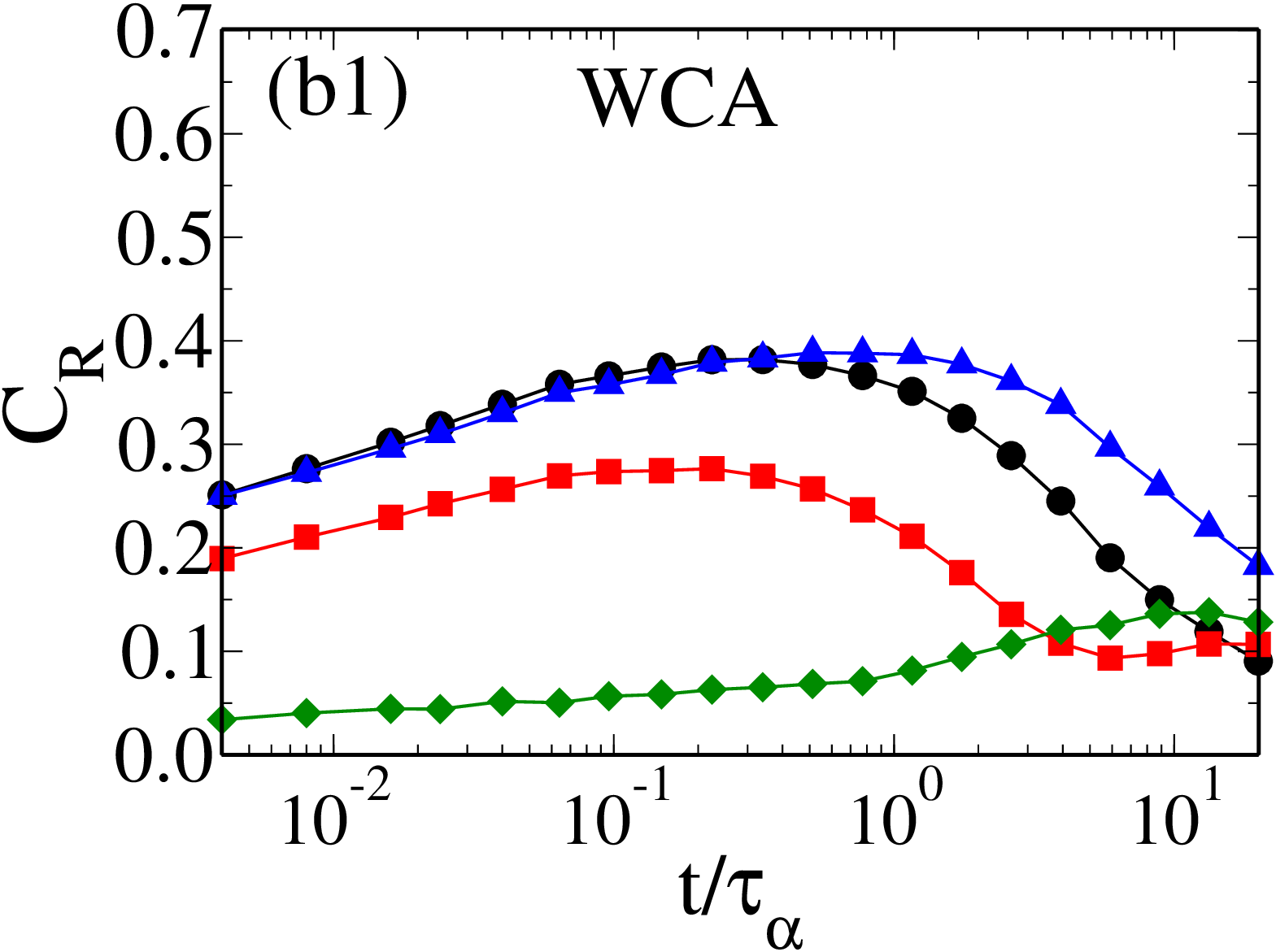}
\end{subfigure}
\vskip\baselineskip
\begin{subfigure}{0.23\textwidth}
\includegraphics[width=1.0\linewidth]{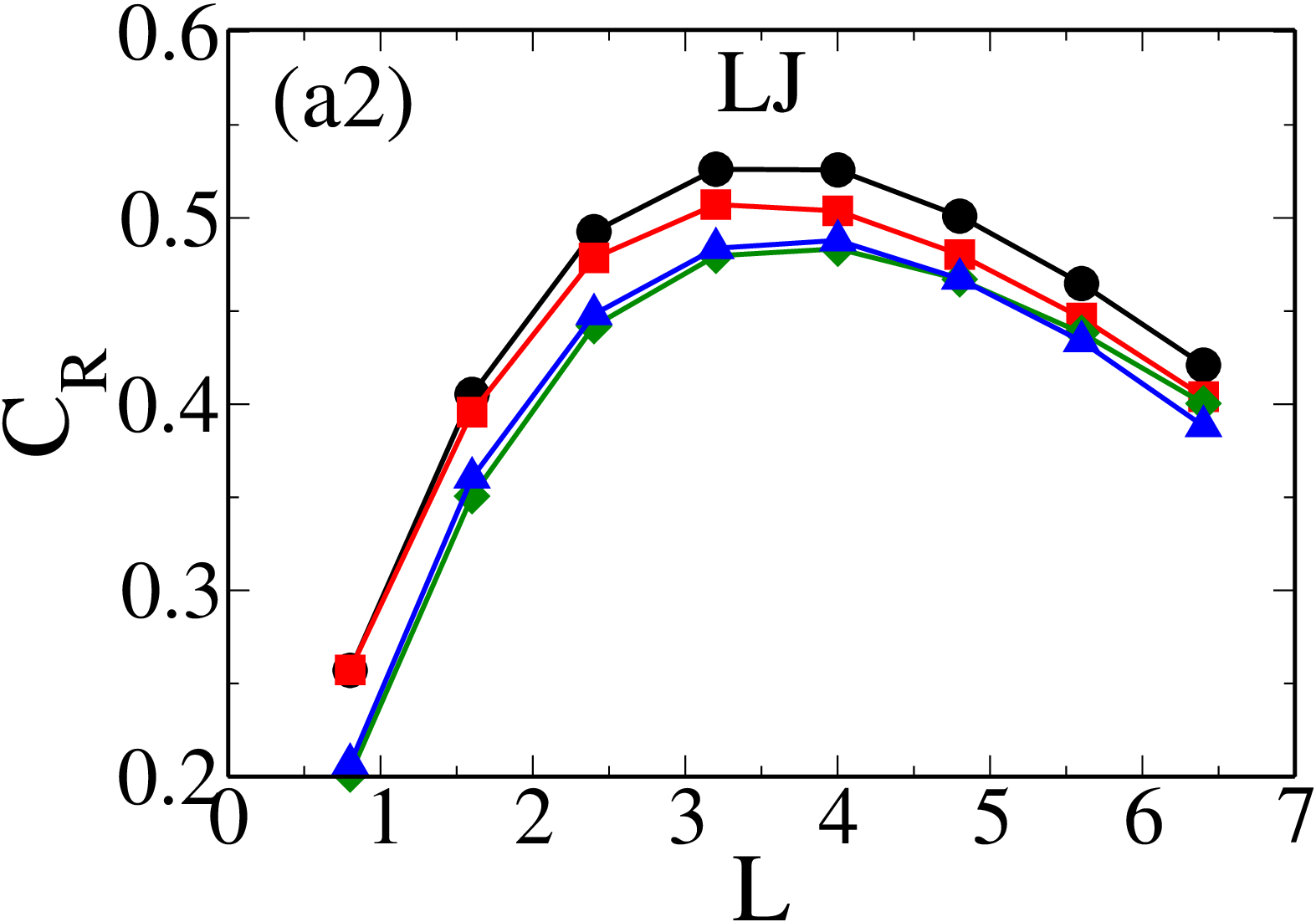}
\end{subfigure}
\begin{subfigure}{0.238\textwidth}
\includegraphics[width=1.0\linewidth]{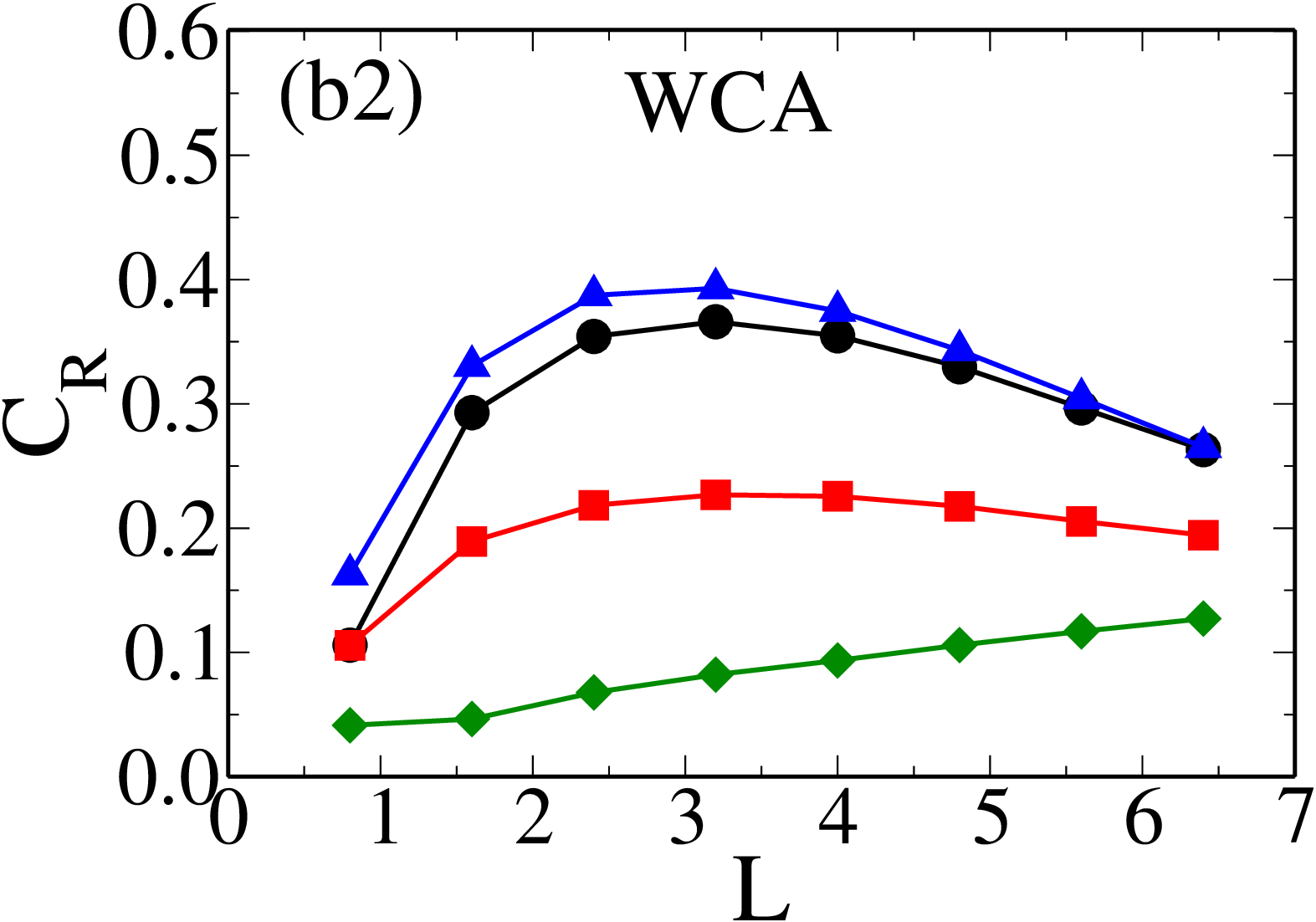}
\end{subfigure}
\caption{\textit{ \textbf{(a1) and (b1)} Spearman rank correlation ($C_{R}$) between mobility($\mu$) and coarse grained order parameters for LJ and WCA systems, plotted against $t/\tau_{\alpha}$, respectively. The coarse graining is done at $L=L_{max}$, which for the LJ and WCA systems are 4 and 3, respectively. \textbf{(a2) and (b2)} $C_{R}$ between $\mu$ (calculated at $\tau_{\alpha}$) and different order parameters( color coding is similar to Fig. \ref{Fig_diffparameter}(a1) ) as a function of $L$ for the LJ and WCA systems, respectively. The plots are at temperatures where the dynamics of the two systems are similar, T = 0.47 for LJ and T = 0.31 for WCA).}}
\label{Fig_diffparameter}
\end{figure}

In Fig. \ref{Fig_diffparameter}(a1) and \ref{Fig_diffparameter}(b1), we plot the Spearman correlation of the different order parameters with the mobility as a function of scaled time. The temperatures are chosen such that the relaxation time of the two systems are similar (T=0.47 for LJ and T=0.31 for WCA). The order parameters are coarse grained until $L=L_{max}$, where $L_{max}=4$ for the LJ system and $L_{max}=3$ for the WCA system. We find that for the LJ system below $\tau_{\alpha}$, the $C_{R}(1/\beta \Phi_r,\mu)$ has the highest value. However beyond $\tau_{\alpha}$ the correlation between $\mu$ and $e_{loc}$ increases. A similar observation was made earlier \cite{paddy_prl}. In this figure, we also plot $C_{R}(\beta \Phi_r,\mu)$, which we find traces $C_{R}(e_{loc},\mu)$. Note that without coarse graining the order parameters, $C_{R}(\beta \Phi_r,\mu)$ is identical to $C_{R}(1/\beta \Phi_r,\mu)$. However, the coarse grained $\beta \Phi_{r}$ gets dominated by the neighbours in a more structured environment and thus in a deeper potential well (hard particles), whereas coarse grained $1/\beta \Phi_{r}$ gets dominated by the neighbours in a loosely packed environment and thus in a shallow potential well (soft particles). Our study shows that for the LJ system at shorter times ($t<\tau_{\alpha}$), the correlation between dynamics and SOP is dominated by soft particles, whereas at longer times ($t>\tau_{\alpha}$) it is dominated by hard particles. We also plot $C_{R}(S_{2},\mu)$. We find that at shorter times, this correlation is similar to $C_{R}(1/\beta \Phi_r,\mu)$. However, at longer times, it is weaker than $C_{R}(1/\beta \Phi_r,\mu)$. The observation for the WCA system is quite different from that for the LJ system. First of all, the large growth in correlation with time observed in the LJ system is not present in the WCA system. We find that in the WCA system, $C_{R}(S_{2},\mu)$ is highest at all times, and $C_{R}(e_{loc},\mu)$ has a low value suggesting a poor correlation between local energy and dynamics. A similar observation between the local inherent potential energy and dynamics was made earlier for the repulsive inverse power law system \cite{harrowell_prl2006}. Interestingly we also find that $C_{R}(1/\beta \Phi_r,\mu)$ is larger than $C_{R}(\beta \Phi_r,\mu)$ at all times. This means that the soft particles in the WCA system always dominate the correlation between dynamics and SOP. 

In Fig. \ref{Fig_diffparameter}(a2) and \ref{Fig_diffparameter}(b2), we plot the correlations between $\mu(t=\tau_{\alpha})$ and the different order parameters as a function of L at $T=0.47$ for the LJ system and at $T=0.31$ for the WCA system. We find that in the LJ system, all the correlations peak at a similar value of $L$, also the correlation between our SOP and mobility (calculated at $t=\tau_{\alpha}$) is maximum. For the WCA system, although the correlations between $\mu$ and other order parameters peak at similar $L$ values, $C_{R}(e_{loc},\mu)$ keeps growing and does not show any non-monotonic $L$ dependence. The present analysis suggests, for the attractive LJ system, the mean field caging potential is closely associated with the local potential energy, which is not the case for the repulsive WCA system. On the other hand, for both systems, the inverse of the mean field caging potential is closely associated with the local pair excess entropy. It also shows that in the attractive system, dynamics is best correlated with the enthalpy, whereas in the repulsive systems, the dynamics is best correlated with the entropy, and the mean field caging potential can capture both these contributions. 

\begin{figure}[h!]
\centering
\begin{subfigure}{0.238\textwidth}
\includegraphics[width=0.98\linewidth]{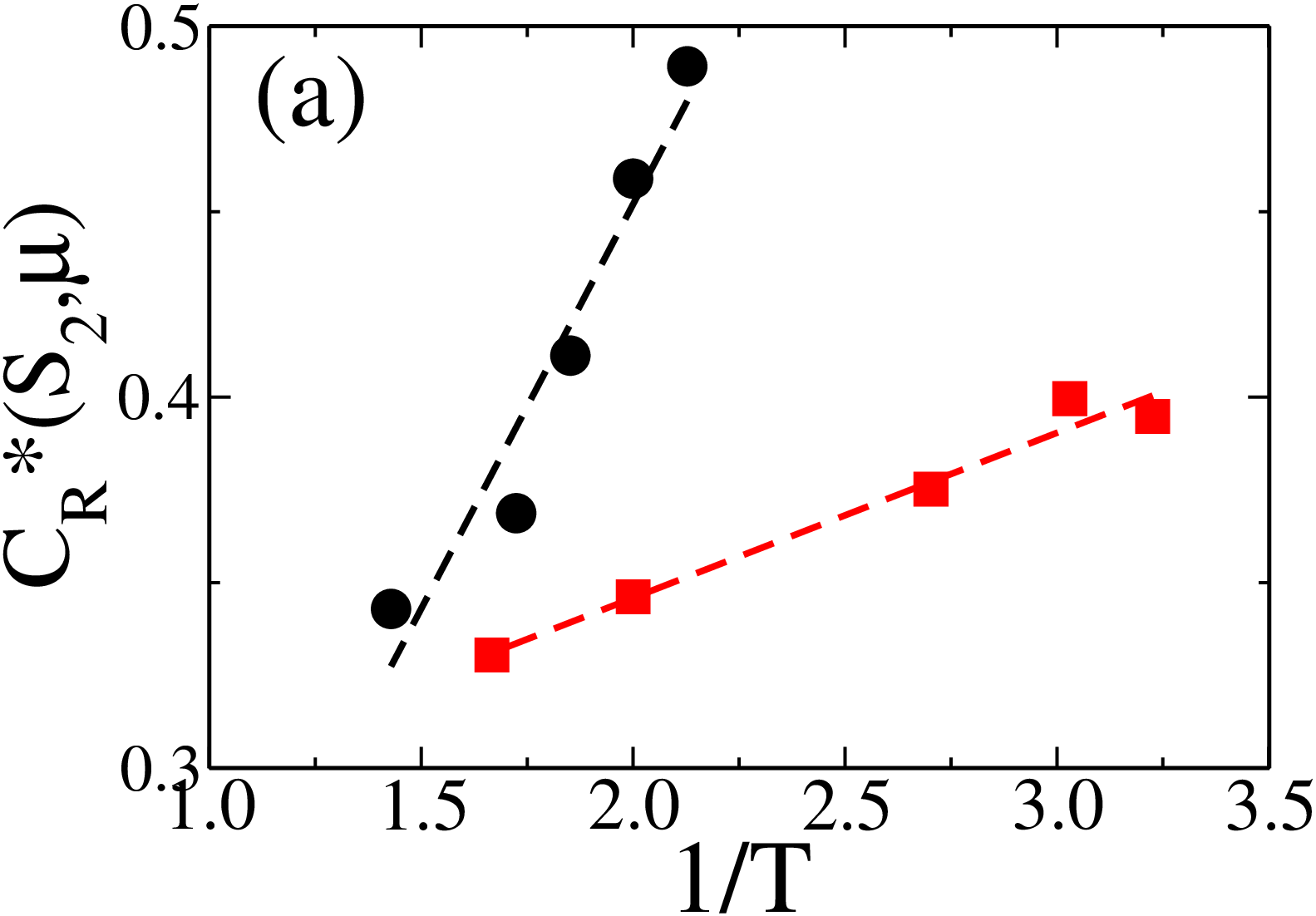}
\end{subfigure}
\begin{subfigure}{0.238\textwidth}
\includegraphics[width=1.0\linewidth]{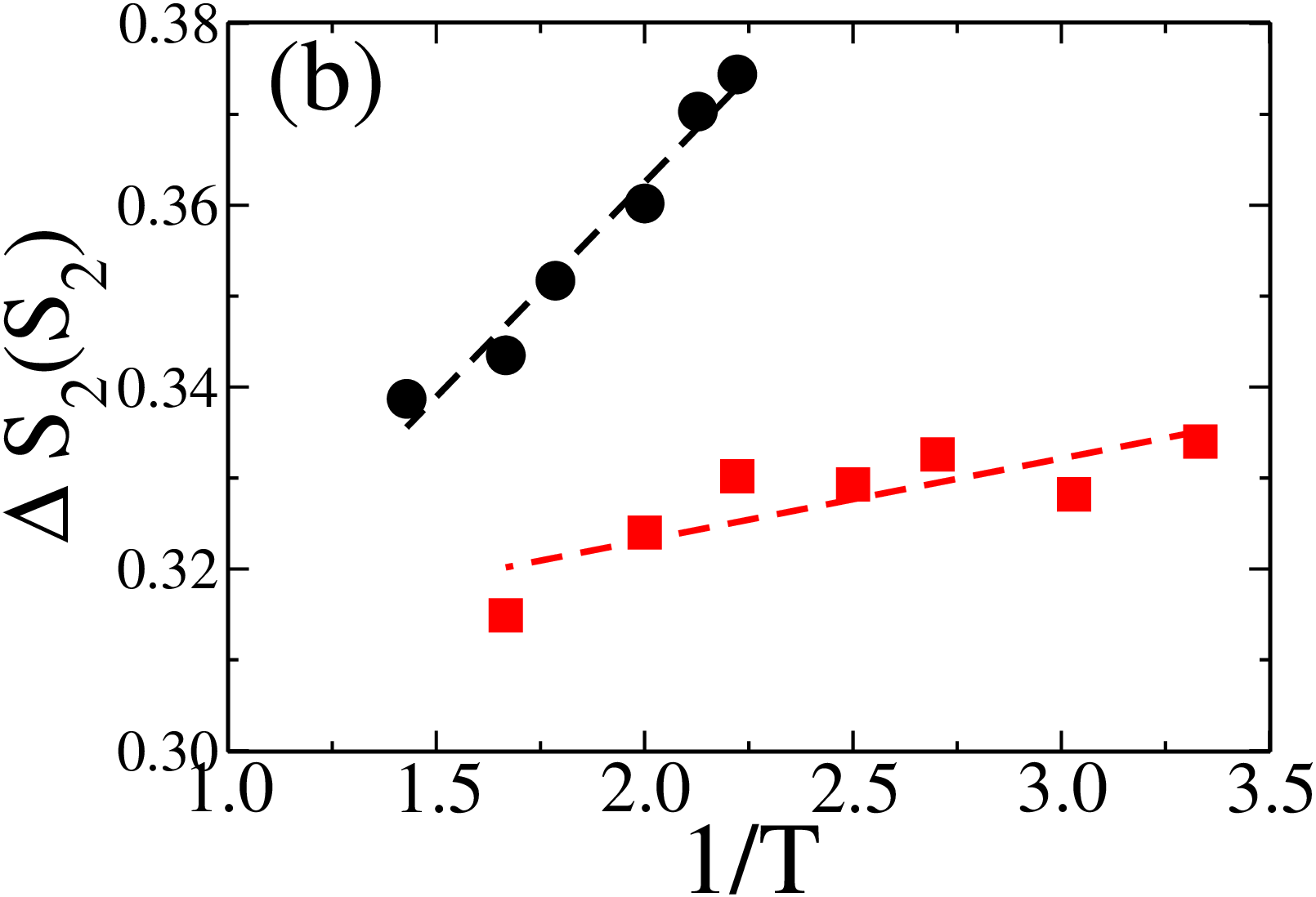}
\end{subfigure}
\caption{\textit{For \textbf{LJ}(black, circle) and \textbf{WCA}(red, square) systems, \textbf{(a)} the position of the peak $C_{R}^{*}$ (as obtained from Figs. \ref{Fig_diff}(a) and \ref{Fig_diff}(b)) plotted against the inverse of temperature ($1/T$).\textbf{(b)} $\Delta S_2$ (calculated using Eq.\ref{entropy_diff}) vs. $1/T$ where the communities are identified in terms of $S_{2}$.}}
\label{Fig_s2}
\end{figure}

We next compare the rate of increase in correlation between $S_{2}$ and $\mu$ as a function of temperature by plotting the peak of $C_{R}(S_{2},\mu)$ vs. $L$ plot (Figs. \ref{Fig_diff}(a) and \ref{Fig_diff}(b) in Appendix), $C^{*}_{R}(S_{2},\mu)$ against $T$ for the WCA and LJ systems (Fig. \ref{Fig_s2}(a)). We find that the correlation for the LJ system has a stronger $T$ dependence compared to the WCA system. We next calculate the community information $\Delta S_{2}(S_{2})$ where the communities are identified according to the value of local $S_{2}$. From the distribution of local $S_{2}$, we create two communities, one which has entropy higher than the average value (high entropy group) and the other which has entropy lower than the average value (low entropy group). We then calculate the $\Delta S_{2}(S_{2})$ for these two communities. The temperature dependence of the $\Delta S_{2}(S_{2})$ for the LJ and the WCA systems are plotted in Fig.\ref{Fig_s2}(b). We find that compared to the WCA system, the temperature dependence of $\Delta S_{2}(S_{2})$ is higher for the LJ system, a trend also observed in the case of $C^{*}_{R}$. These results appear similar to what we observe when we use the mean field caging potential to describe the communities. When communities are created in terms $e_{loc}$, it does not show any meaningful result. Also, we cannot compare the temperature dependence of the peak of the correlation $C^{*}_{R}(e_{loc},\mu)$ as for the WCA system, we cannot identify the peak value (refer to Fig. \ref{Fig_diff} (d) in Appendix III).\\

\section{Conclusion}
\label{conclusion_section}
In this work, we compare the correlation between structure and dynamics for two binary systems, one interacting via attractive LJ interaction and the other via its repulsive counterpart, the WCA interaction. It is well known that these two systems have similar pair structures, but the difference in dynamics is large, with the LJ system showing a much faster divergence of the relaxation time. The analysis is performed at both macroscopic and microscopic levels. The macroscopic study predicts that although the recently defined macroscopic structural order parameters of the LJ and the WCA systems are different, that difference is not enough to predict the difference in the dynamics. The dynamics as a function of the structural order parameter predicts a stronger coupling between the structure and dynamics in the LJ system. It is further corroborated by studying the correlation between structure and dynamics at the microscopic level. This study aims to identify the origin of this system specific coupling in the LJ and WCA systems. 

In order to understand the origin of this higher correlation, we perform an analysis which is inspired by the recent work of Coslovich and coworkers \cite{coslovich2020}, where they defined a quantity called community information and showed that the larger the difference in the structure between different communities, the higher the community information. 
They suggested this community information can be used as a proxy for structural fragility.

We study the difference between the local structure of hard and soft particles as described by our SOP and quantify the community information for the LJ and WCA systems. We find that for both systems, our SOP is a good parameter to describe the different communities. However, the community information of the LJ system shows a much stronger temperature dependence. Thus, the attractive interaction may have a higher contribution towards creating communities with distinct local structures. Also, we show that for the same value of the average SOP, the community information is higher for the LJ system. This shows that similar system specificity, as observed in the structure dynamics correlation, also exists in the correlation between the community information and structure, and they may be correlated. Thus our study suggests that for systems where the structural communities are better defined, there is a higher correlation between the SOP and the dynamics and the system specific structure dynamics correlation may have a structural origin. 

We also compare the correlation between dynamics and our SOP with the correlation between dynamics and two other order parameters, namely the local energy and the local two body excess entropy. The analysis suggests that the dynamics in the attractive LJ system is primarily controlled by the enthalpy, whereas the entropy controls that in the repulsive WCA system, and our SOP is capable of capturing both enthalpic and entropic contributions. 

In one of our earlier studies on a large class of systems at the macroscopic level, we discussed this system specific structure dynamics correlation \cite{manoj_prl2021}. 
We have shown that this system specific structure dynamics correlation, along with the temperature dependence of the average SOP, is connected to kinetic fragility \cite{manoj_prl2021}. 
If temperature dependence of the community information can be related to structural fragility, then structural fragility may have two contributions. One is the variation of the average structure and structural order parameter with temperature, which is probably the dominant contributor in differentiating the WAHN model from the LJ model \cite{manoj_prl2021,coslovich2020} and not so dominant in differentiating the LJ model from the WCA model (as structurally they are similar). The other is the difference in the local structure of the different communities for the same average value of the SOP, like the LJ system having much higher community information than the WCA system. It will be interesting to extend the present study to a wider class of systems and connect the community information of the local structure based on the SOP to the system specificity of the structure dynamics correlation and the kinetic fragility, a work to be addressed in the future. Also, note that the community information will depend on the order parameter used in defining the community. Understanding the order parameter dependence of the community information will also be interesting. \\ 

\textbf{Appendix I: Masterplot between relaxation time and SOP }\\

\begin{figure}[h!]
\begin{subfigure}{0.32\textwidth}
\includegraphics[width=1.0\linewidth]{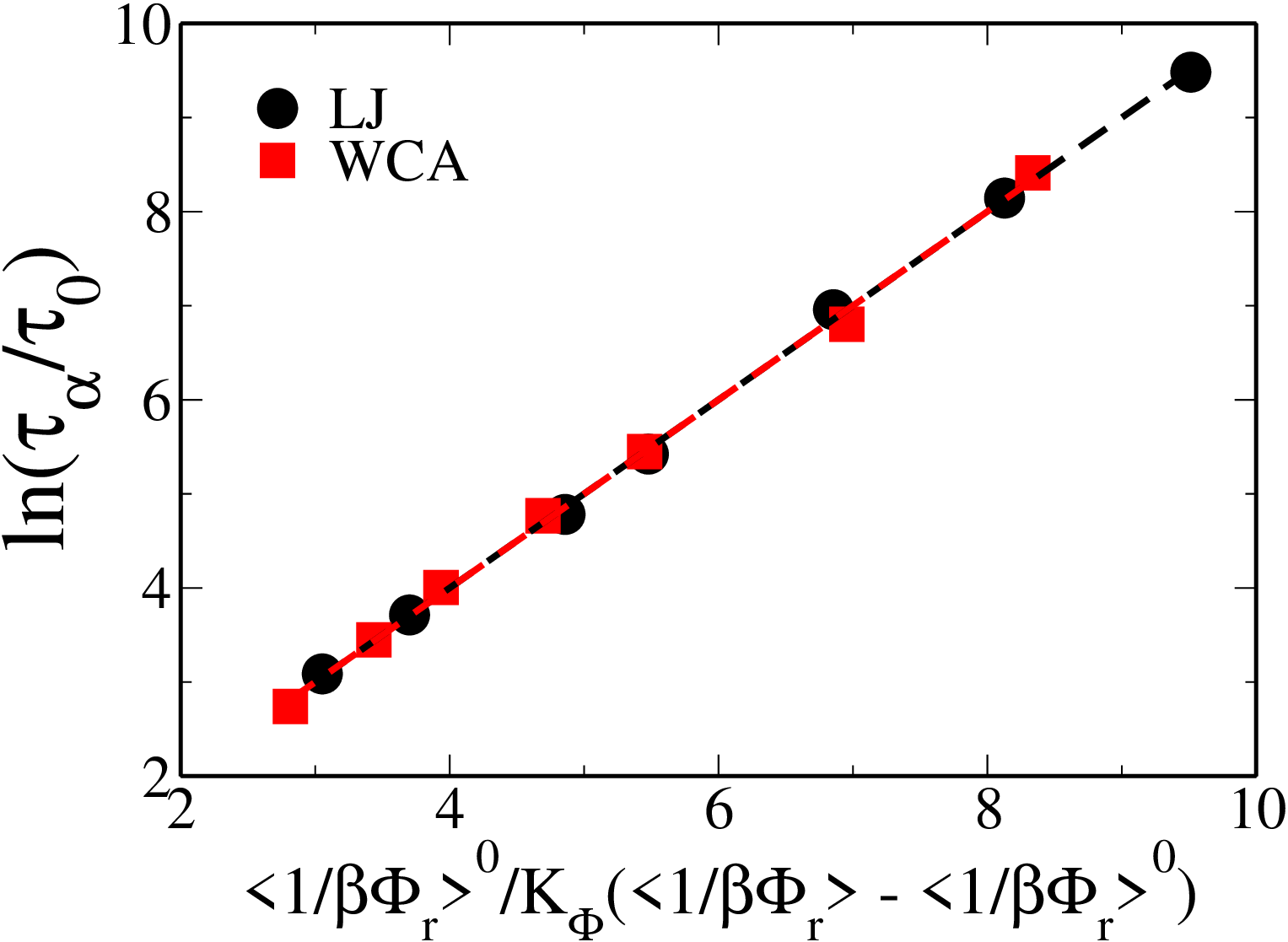}
\end{subfigure}
\caption{\textit{Master curve for both LJ and WCA systems, when scaled $\alpha$ relaxation time is plotted against $\big<1/\beta \Phi_{r}\big>^{0}/K_{\Phi}(\big<1/\beta \Phi_{r}\big> - \big<1/\beta \Phi_{r}\big>^{0})$. }}
\label{Fig_masterplot}
\end{figure}

In this study, we investigated the SOP dependence of the relaxation times of two systems by fitting them to the Vogel-Fulcher-Tammann (VFT) form. By analyzing the fitting results, we obtained the values of $\big<1/\beta \Phi_{r}\big>$ and $K_{\Phi}$, which represent the SOP value where the relaxation time will diverge and the associated fragility, respectively. To gain further insights, we constructed a master plot in Fig. \ref{Fig_masterplot}. This plot involves scaling the relaxation times of both LJ and WCA systems by the relaxation time at the onset temperature. The scaled relaxation time is plotted against $\big<1/\beta \Phi_{r}\big>^{0}/K_{\Phi}(\big<1/\beta \Phi_{r}\big> - \big<1/\beta \Phi_{r}\big>^{0})$. Remarkably, the resulting master plot reveals a universal behaviour, indicating that the relaxation times of both systems follow a similar trend.\\ 

\textbf{Appendix II: Distribution of SOP}\\

\begin{figure}
\centering
\begin{subfigure}{0.238\textwidth}
\includegraphics[width=1\linewidth]{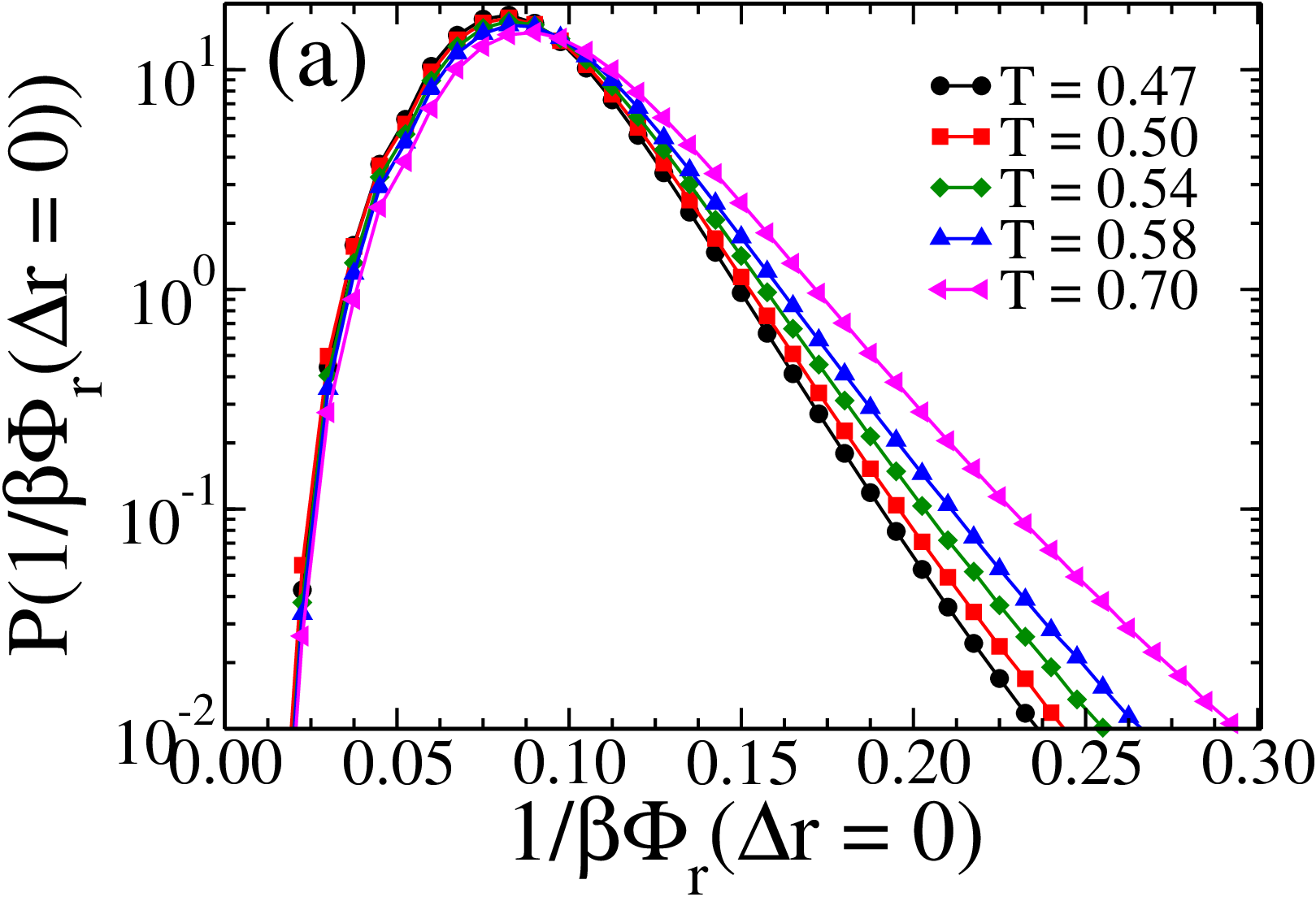}
\end{subfigure}
\begin{subfigure}{0.238\textwidth}
\includegraphics[width=1\linewidth]{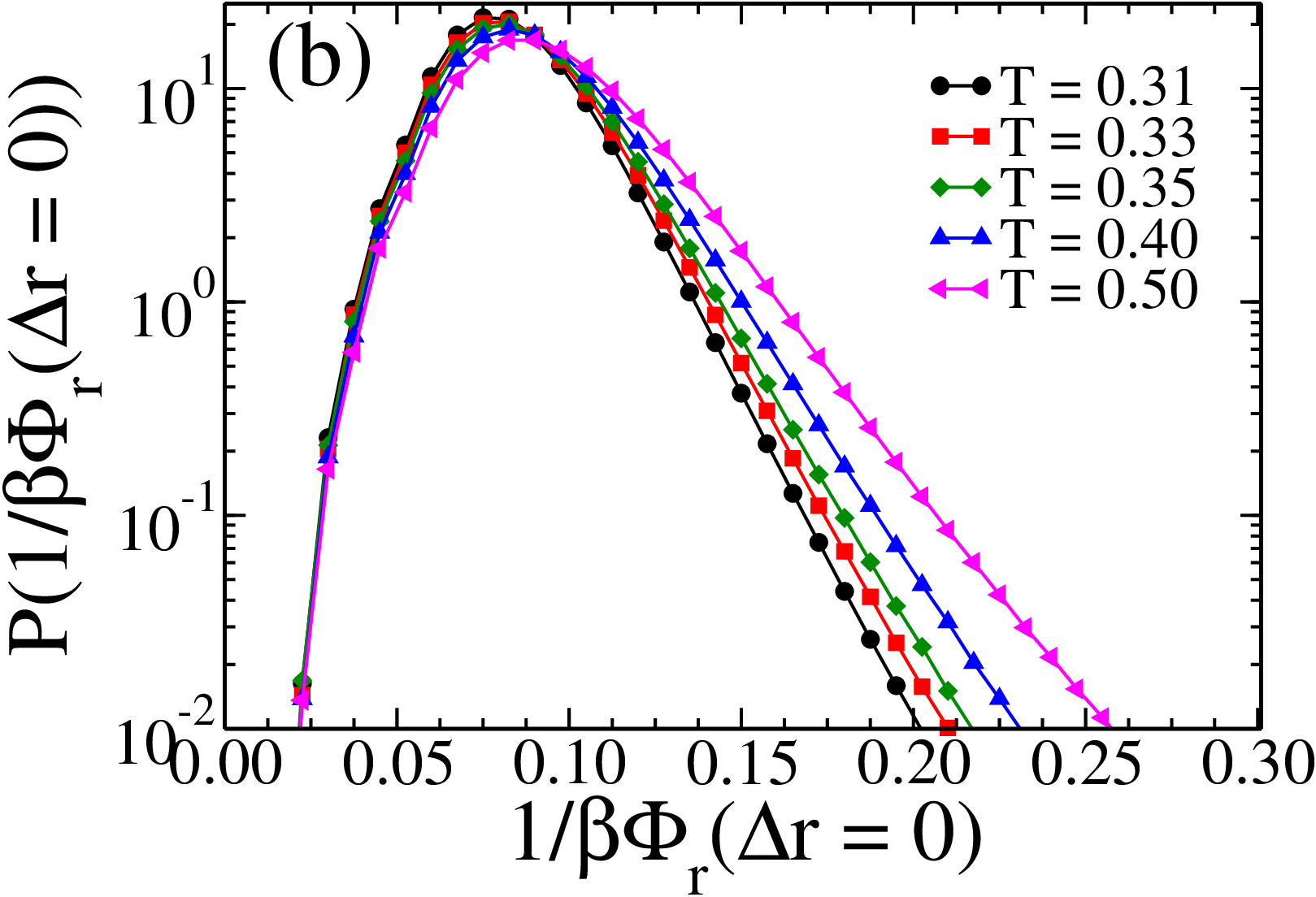}
\end{subfigure}
\vskip\baselineskip
\begin{subfigure}{0.23\textwidth}
\includegraphics[width=1.0\linewidth]{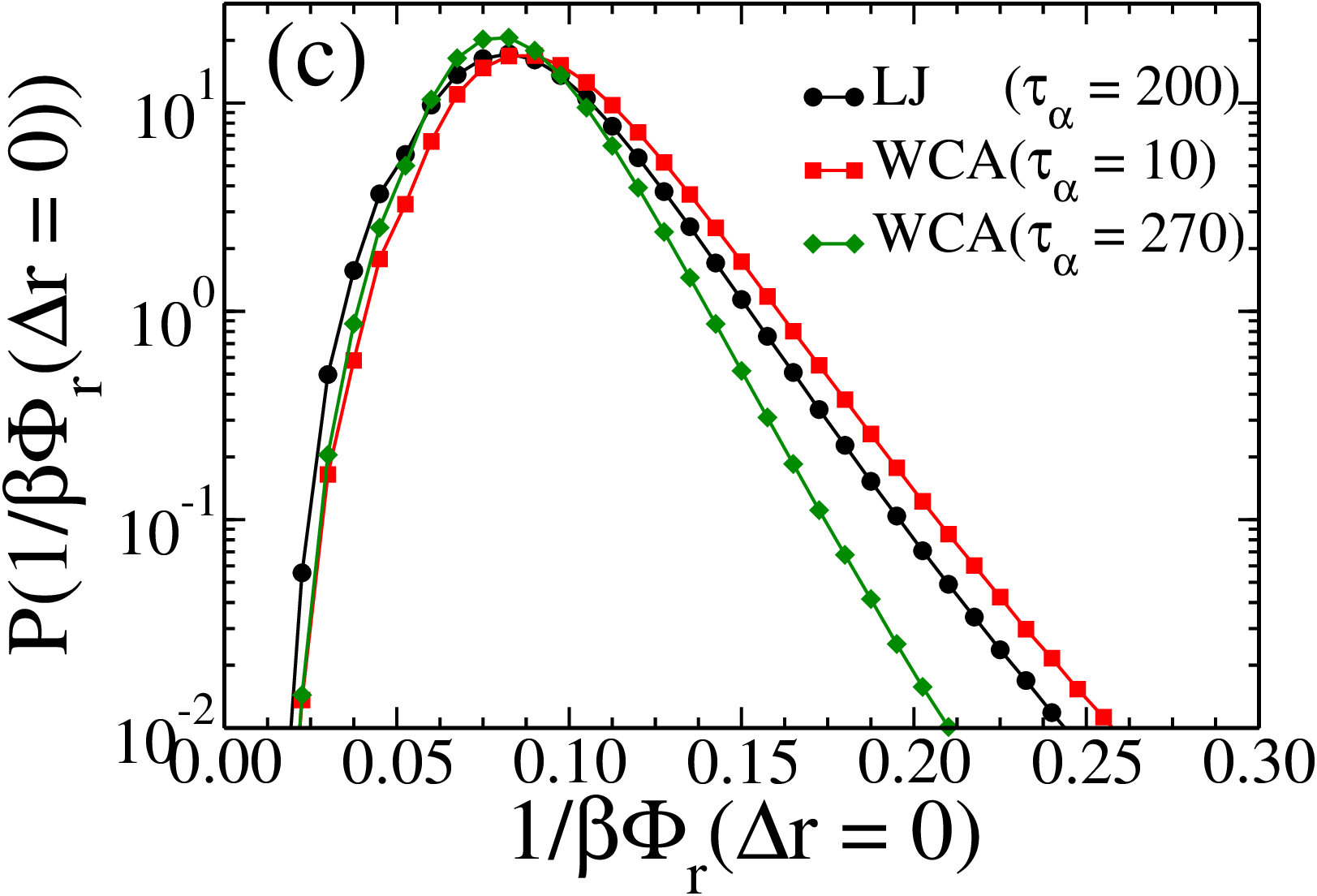}
\end{subfigure}
\begin{subfigure}{0.23\textwidth}
\includegraphics[width=1.0\linewidth]{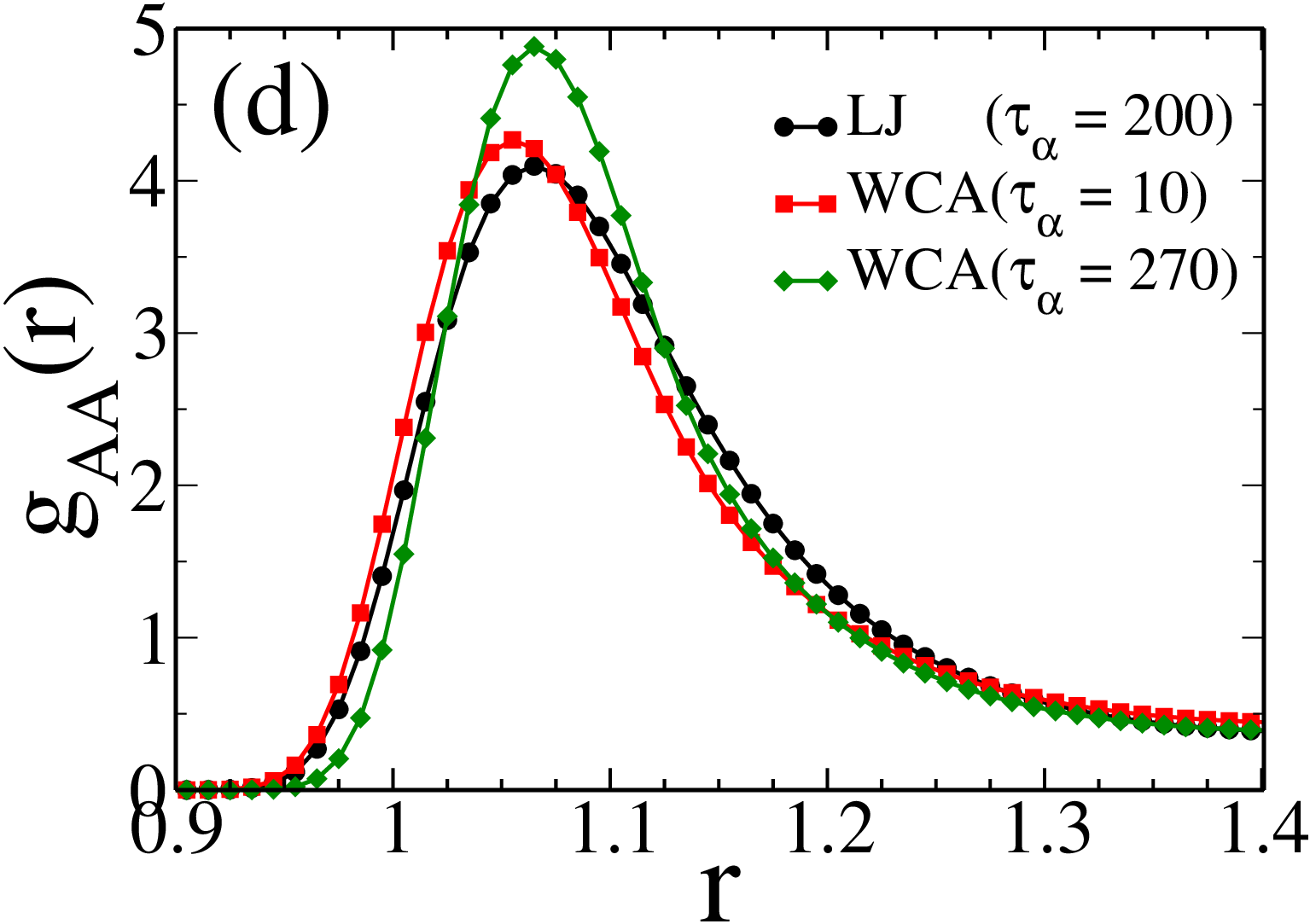}
\end{subfigure}

\caption{\textit{ Distribution of microscopic SOP ($1/\beta \Phi_{r}$) for \textbf{(a)} $LJ$ system, \textbf{(b)} $WCA$ system at different temperatures ($T$). The temperatures are chosen in a way that we can cover the range from below onset to above Mode-Coupling Theory transition temperature, keeping some temperatures common for the two systems and somewhere the relaxation times of the two systems are similar.} \textbf{(c)} Distribution of inverse of depth of caging potential ($1/\beta \Phi_{r}$) for $LJ$ and $WCA$ system at same temperature ($T$ = 0.50) and similar $\alpha$ relaxation time ($\tau_{\alpha}$) ($T$ = 0.50 for $LJ$ and T = 0.33 for $WCA$). \textbf{(d)} Partial radial distribution function ($g_{AA}(r)$) for $LJ$ and $WCA$ system at same $T$ and similar $\tau_{\alpha}$. The solid lines are the guide to the eye.} 
\label{Fig_dist}
\end{figure}

 The calculation of the microscopic SOP is given in Section \ref{sop}. All the microscopic calculations are done for the bigger ``A" particles, which are also larger in number. This is to ensure that there is no size inhomogeneity which we know can also play a role in the dynamics \cite{palak_soft}, and also, the statistics are better for the same set of runs. In Figs. \ref{Fig_dist}(a) and \ref{Fig_dist}(b), we plot the SOP distribution of the LJ and WCA systems, respectively. For both systems, the low SOP regime appears to be temperature independent, and with a decrease in temperature, the tail of the distribution towards the higher value of SOP shifts inwards. A similar observation was reported for another order parameter obtained using ML\cite{olivier_pre}. 
 In Fig. \ref{Fig_dist}(c), we compare the $P(1/\beta\Phi_{r})$ of the LJ and WCA systems at the same temperature and also at a similar relaxation time. At the same temperature, for the WCA system, despite the average rdf being sharper(Fig. \ref{Fig_dist}(d)), we find a longer tail of the $P(1/\beta\Phi_{r})$ distribution towards higher SOP values. Also, compared to the WCA system, the LJ system's $P(1/\beta\Phi_{r})$ has higher contribution from the lowest SOP regime. Thus at the same temperature, the LJ system's $P(1/\beta\Phi_{r})$ is shifted towards smaller SOP values leading to a lower value of $\big<1/\beta \Phi_{r}\big>$ (Fig. \ref{macro_study}(b)). When we compare the distribution of the SOP for the two systems at similar relaxation times, we find that the tail of the distribution for the WCA system at the higher SOP value is much shorter compared to the LJ system. Also, the rdf of the WCA system at T = 0.33 is much sharper. Thus, compared to the LJ system, for the WCA system, the cage structure needs to be better defined to have similar activation energy/dynamics. \\

\textbf{Appendix III: Correlation functions}\\

\begin{figure}[h!]
\centering
\begin{subfigure}{0.23\textwidth}
\includegraphics[width=1.0\linewidth]{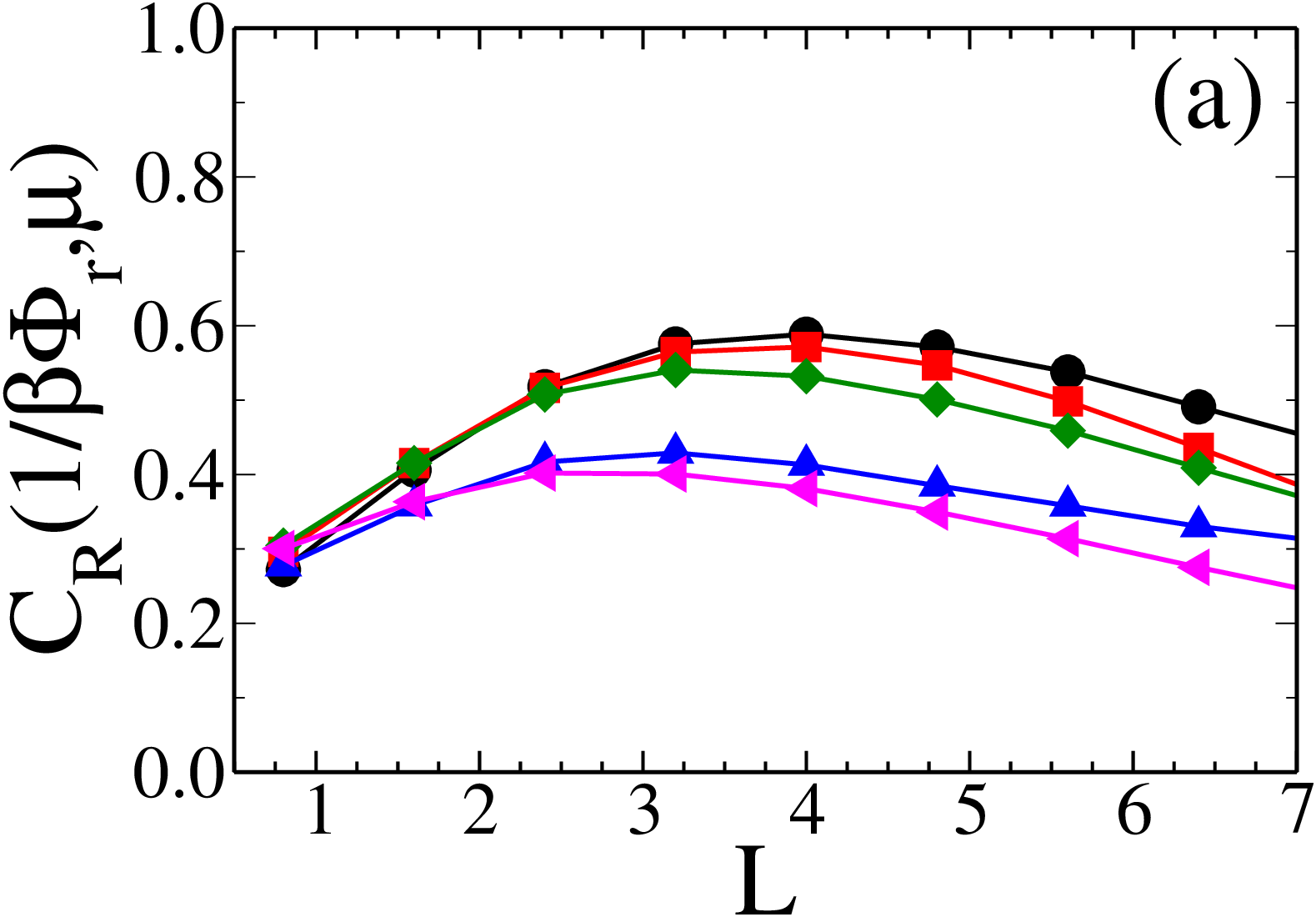}
\end{subfigure}
\begin{subfigure}{0.23\textwidth}
\includegraphics[width=1.0\linewidth]{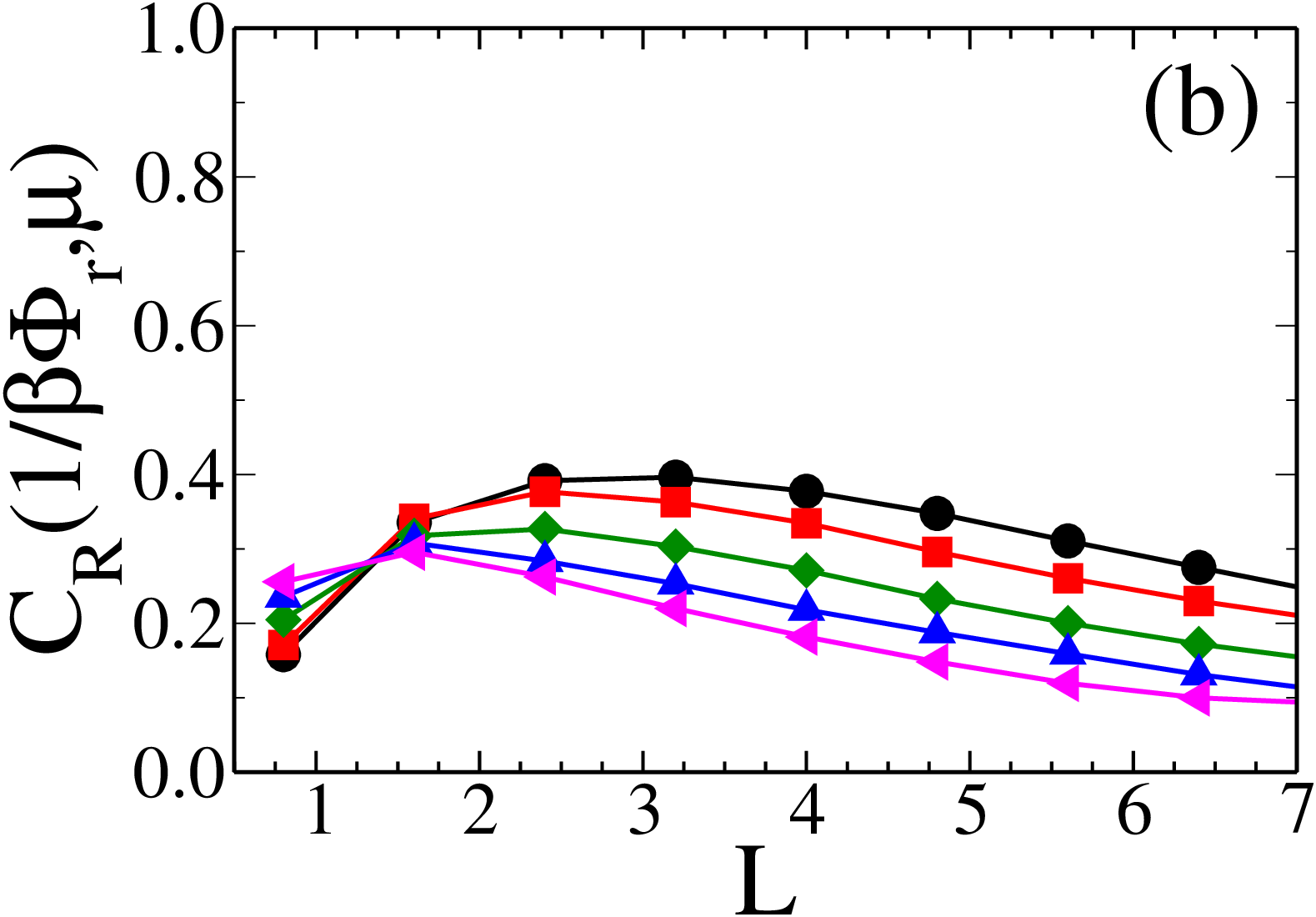}
\end{subfigure}
\vskip\baselineskip
\begin{subfigure}{0.23\textwidth}
\includegraphics[width=1.0\linewidth]{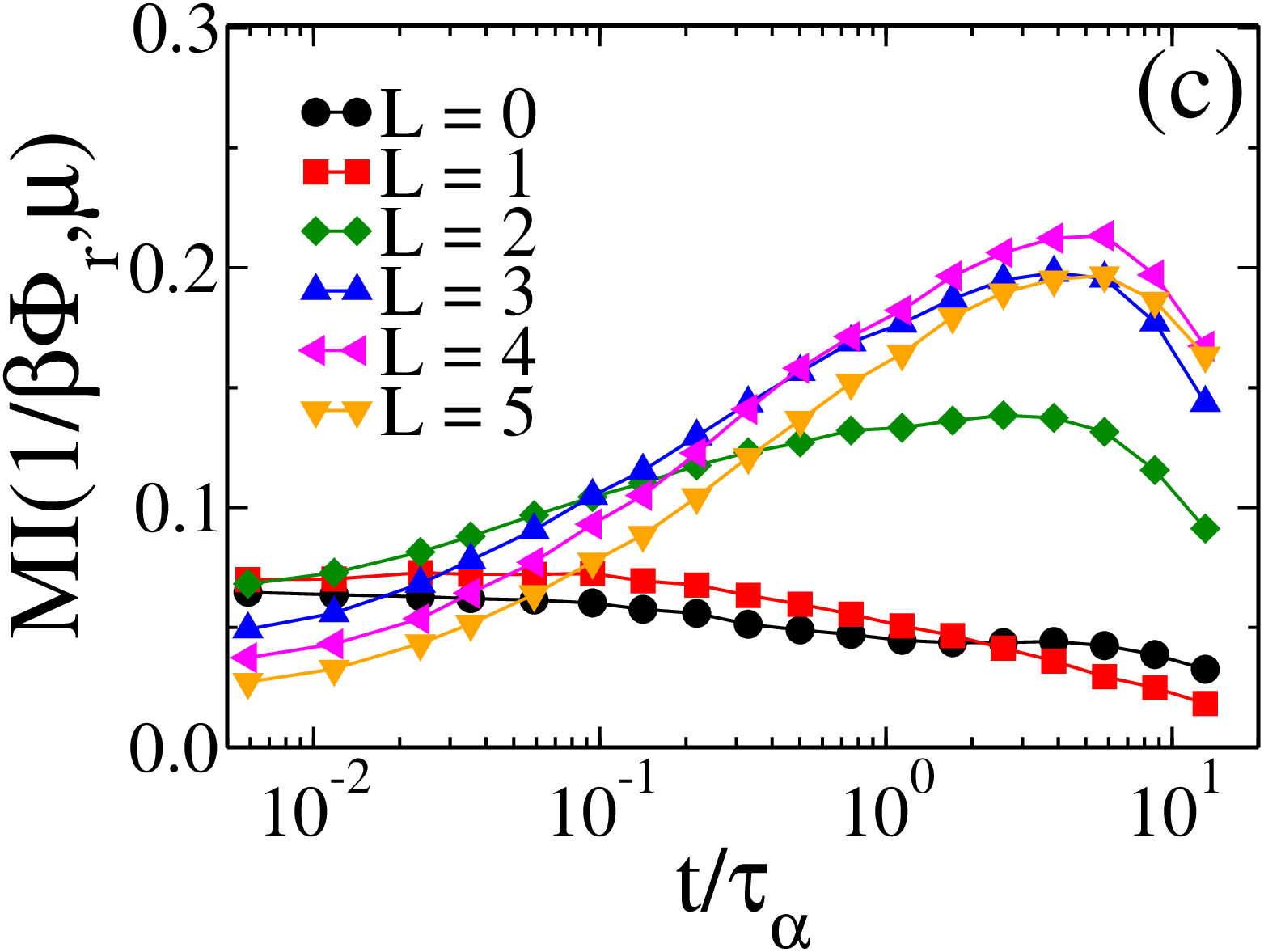}
\end{subfigure}
\begin{subfigure}{0.23\textwidth}
\includegraphics[width=1.0\linewidth]{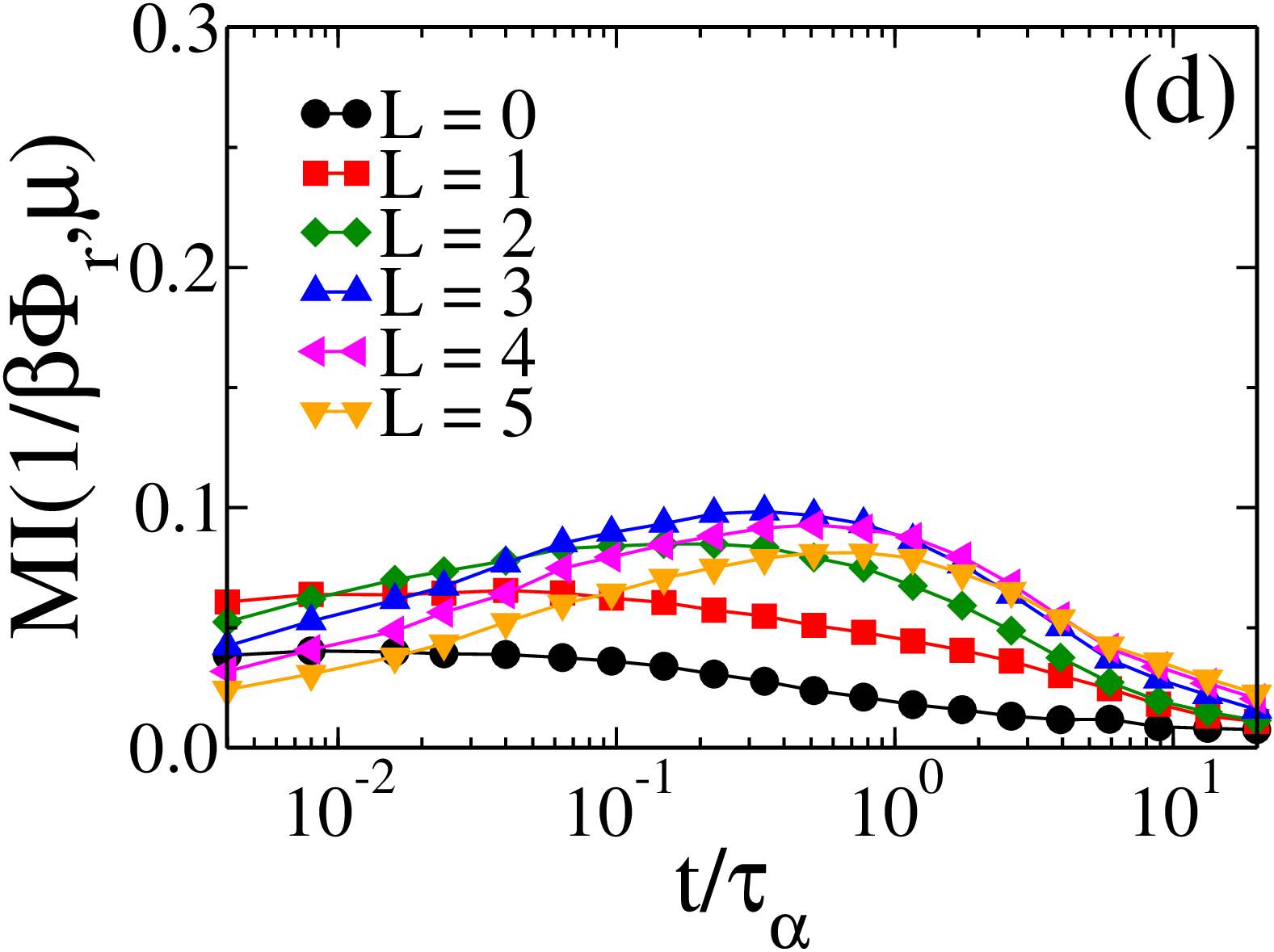}
\end{subfigure}
\caption{\textit{Spearman correlation($C_{R}$) between mobility ($\mu$) and SOP ($1/\beta \Phi_{r}$) plotted against coarse graining length $L$, where $\mu$ is calculated at the time where $C_{R}$ peaks in Figs. \ref{Fig_isoconf}(a1) and \ref{Fig_isoconf}(b1) ($t=x^{*}\tau_{\alpha}$) for \textbf{(a)} $LJ$ system at temperature T = 0.47(black, circle), T = 0.50(red, square), T = 0.54(dark green, diamond), T = 0.58(blue, up triangle), T = 0.70(magenta,left triangle) and \textbf{(b)} $WCA$ system at temperature T = 0.31(black,circle), T = 0.33(red, square), T = 0.37(dark green, diamond), T = 0.50(blue, up triangle), T = 0.60(magenta,left triangle). Mutual information between $\mu$ and $1/\beta \Phi_{r}$ at different $L$ plotted against time scaled by $\alpha$ relaxation time ($t/\tau_{\alpha}$) for \textbf{(c)} $LJ$ system \textbf{(d)} $WCA$ system. }}
\label{Fig_peak}
\end{figure}

In Fig. \ref{Fig_peak}, we present the calculated values of $C_{R}(1/\beta\Phi_{r},\mu)$, representing the correlation between the SOP and the mobility at $t=x^{*}\tau_{\alpha}$ (the peak position in the $C_{R}$ vs. $t/\tau_{\alpha}$ plot, Figs. \ref{Fig_isoconf}(a1) and \ref{Fig_isoconf}(b1)), as a function of the coarse-graining length $L$. As expected from the results shown in Figs. \ref{Fig_isoconf}(a1) and \ref{Fig_isoconf}(b1), the $C_{R}(1/\beta\Phi_{r},\mu)$ values obtained at $t=x^{*}\tau_{\alpha}$ are higher compared to those obtained at $\tau_{\alpha}$. Additionally, we observe a consistent non-monotonic behaviour in the correlation with varying coarse graining length $L$ for both the LJ and WCA systems. Notably, the maximum value of $L$, denoted as $L_{max}$, is higher for the LJ system compared to the WCA system.

The $L_{max}$ value appears to be similar to that when the mobility is calculated at $t=\tau_{\alpha}$.
We also plot the mutual information between the structure order parameter (SOP) of the initial frame and the mobility ($\mu$) as a function of $t/\tau_{\alpha}$. We find that similar to the $C_{R}(1/\beta\Phi_{r},\mu)$ the $MI(1/\beta\Phi_{r},\mu)$ peaks at the same time $t=x^{*}\tau_{\alpha}$ (Figs. \ref{Fig_isoconf}(a1) and \ref{Fig_isoconf}(b1)) for both systems and, the LJ system peaks at a time higher than $\tau_{\alpha}$ and that of the WCA system peaks at a time lower than $\tau_{\alpha}$.\\
\begin{figure}[h!]
\begin{subfigure}{0.23\textwidth}
\includegraphics[width=1.0\linewidth]{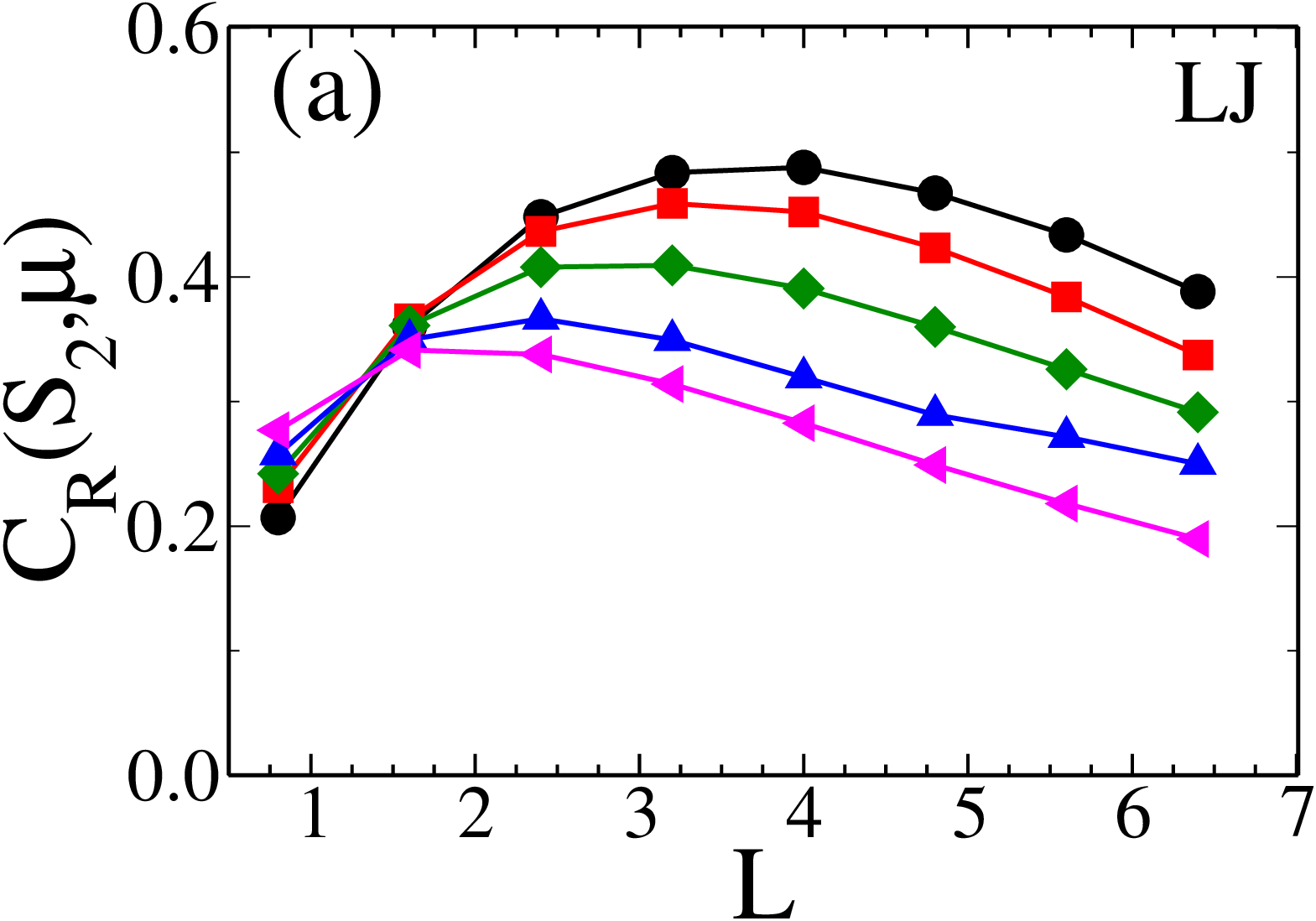}
\end{subfigure}
\begin{subfigure}{0.23\textwidth}
\includegraphics[width=1.0\linewidth]{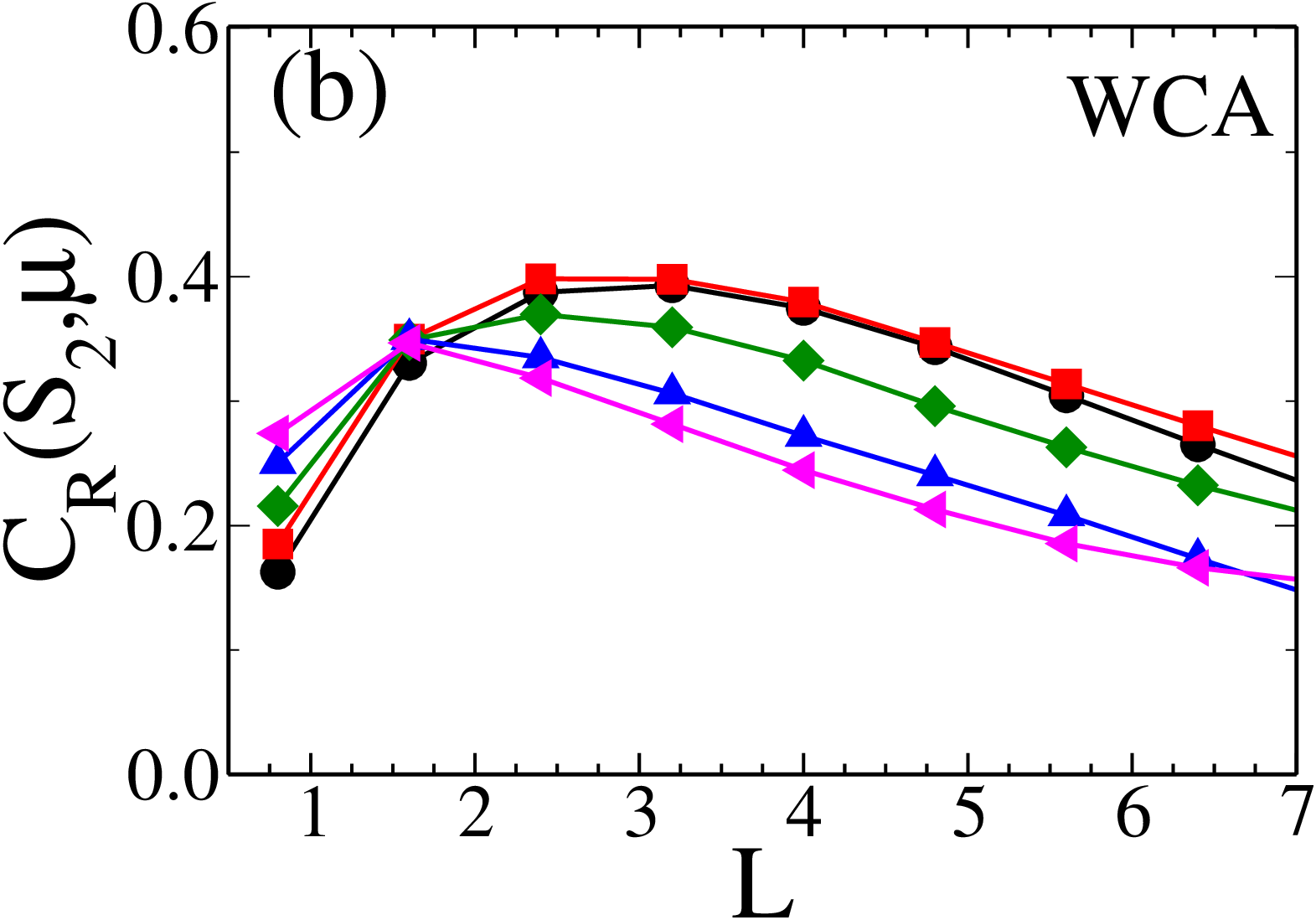}
\end{subfigure}
\vskip\baselineskip
\begin{subfigure}{0.23\textwidth}
\includegraphics[width=1.0\linewidth]{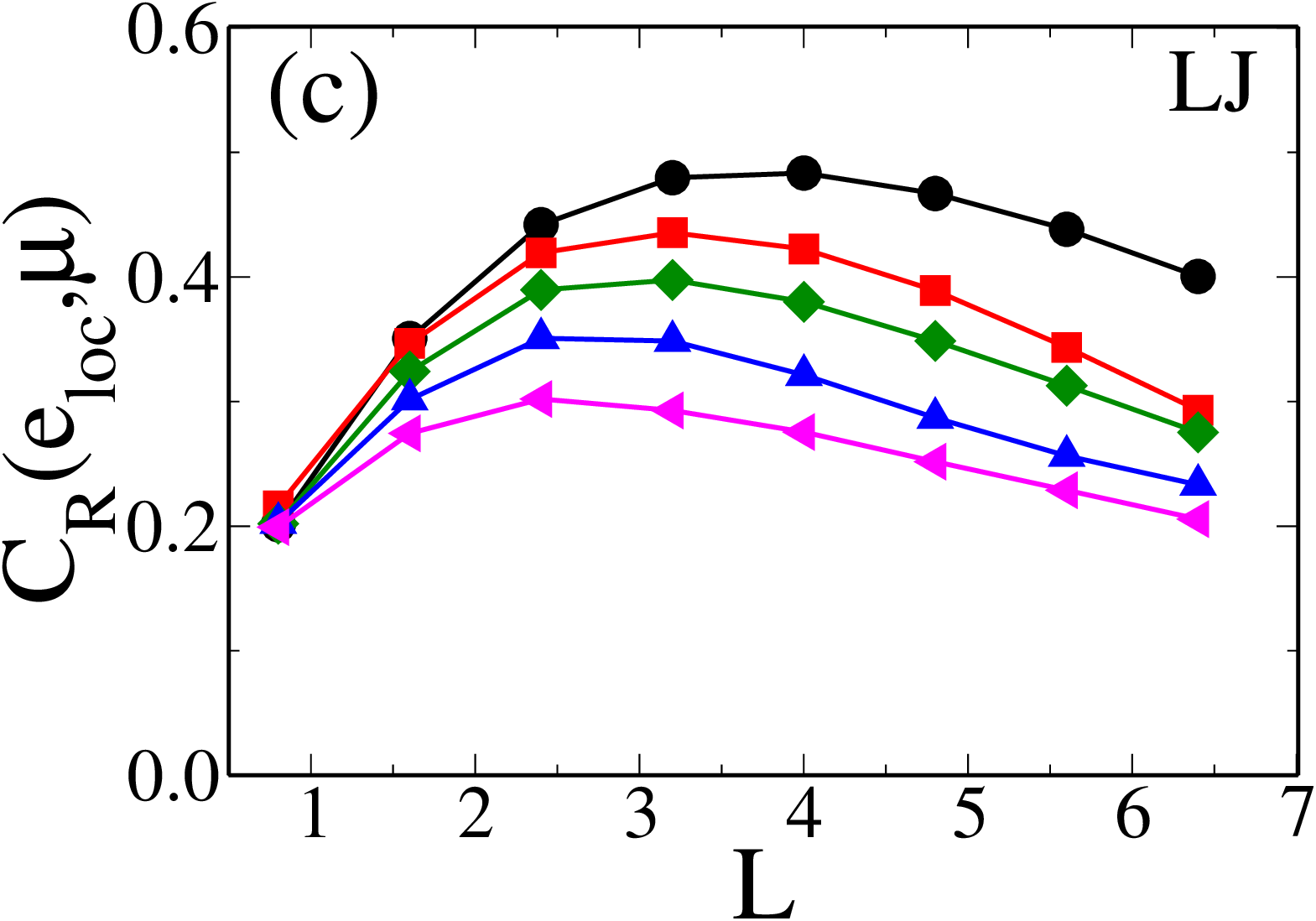}
\end{subfigure}
\begin{subfigure}{0.23\textwidth}
\includegraphics[width=1.0\linewidth]{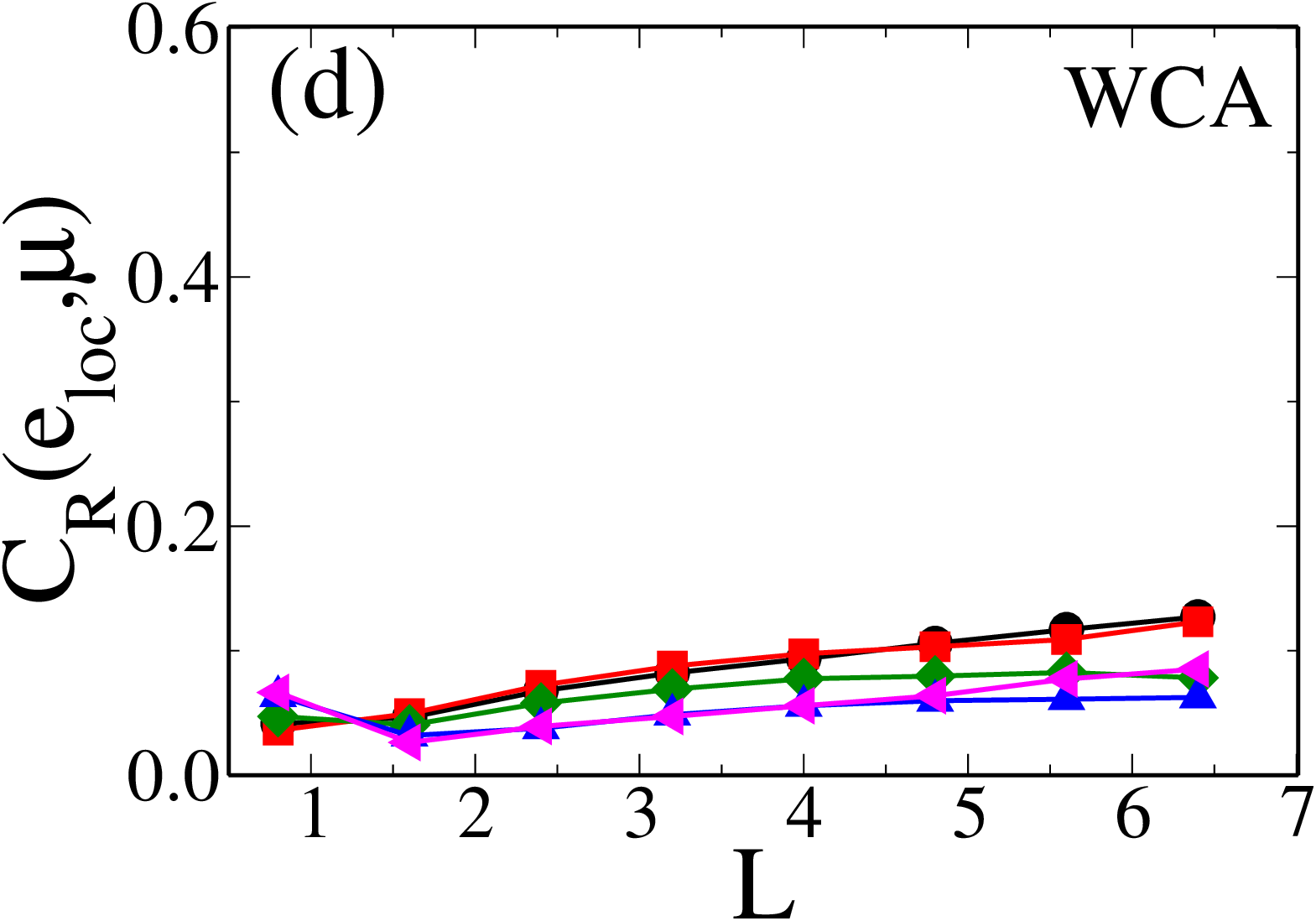}
\end{subfigure}

\caption{\textit{For \textbf{LJ}(right panel) and \textbf{WCA}(left panel): Spearman correlation between mobility,$\mu$ (calculated at $\tau_{\alpha}$) and two body excess entropy,$S_{2}$ (top panel), local potential energy,$e_{loc}$(bottom panel), as a function of $L$ for different $T$ (T = 0.47(black, circle), T = 0.50(red, square), T = 0.54(dark green, diamond), T = 0.58(blue, up triangle), T = 0.70(magenta,left triangle) in left panel and T = 0.31(black,circle), T = 0.33(red, square), T = 0.37(dark green, diamond), T = 0.50(blue, up triangle), T = 0.60(magenta,left triangle) in right panel).}} 
\label{Fig_diff}
\end{figure}
We also study the correlation between two body excess entropy ($S_2$) and mobility ($\mu$) for LJ and WCA systems, respectively (Figs. \ref{Fig_diff}(a) and \ref{Fig_diff}(b)). This correlation is analyzed as a function of varying coarse graining length (L), and we find that it is similar to $C_{R}(1/\beta\Phi_{r},\mu)$, also showing a non-monotonic dependence on L. The correlation reaches its maximum at $L=L_{max}$, and $L_{max}$ increases with decreasing temperature. We find that $C_{R}(S_{2},\mu)$ values are lower for the LJ system but higher for the WCA system when compared to $C_{R}(1/\beta\Phi_{r},\mu)$ (Figs. \ref{Fig_isoconf}(a2) and \ref{Fig_isoconf}(b2)). Subsequently, we have also studied the correlation between local potential energy ($e_{loc}$) and mobility ($\mu$) for both LJ and WCA systems(Figs. \ref{Fig_diff}(c) and \ref{Fig_diff}(d)), varying the coarse graining length (L). For the LJ system, we find similar L dependence as observed for $1/\beta\Phi_{r}$ and $S_{2}$. However, for the WCA system, the correlation is significantly lower and does not exhibit any non-monotonic behaviour with varying L.\\

\textbf{Appendix IV: Use of $C_{uv}^{approx}(r)$ and its impact on the predictive power of SOP}\\

\begin{figure}[h!]
\begin{subfigure}{0.23\textwidth}
\includegraphics[width=1.0\linewidth]{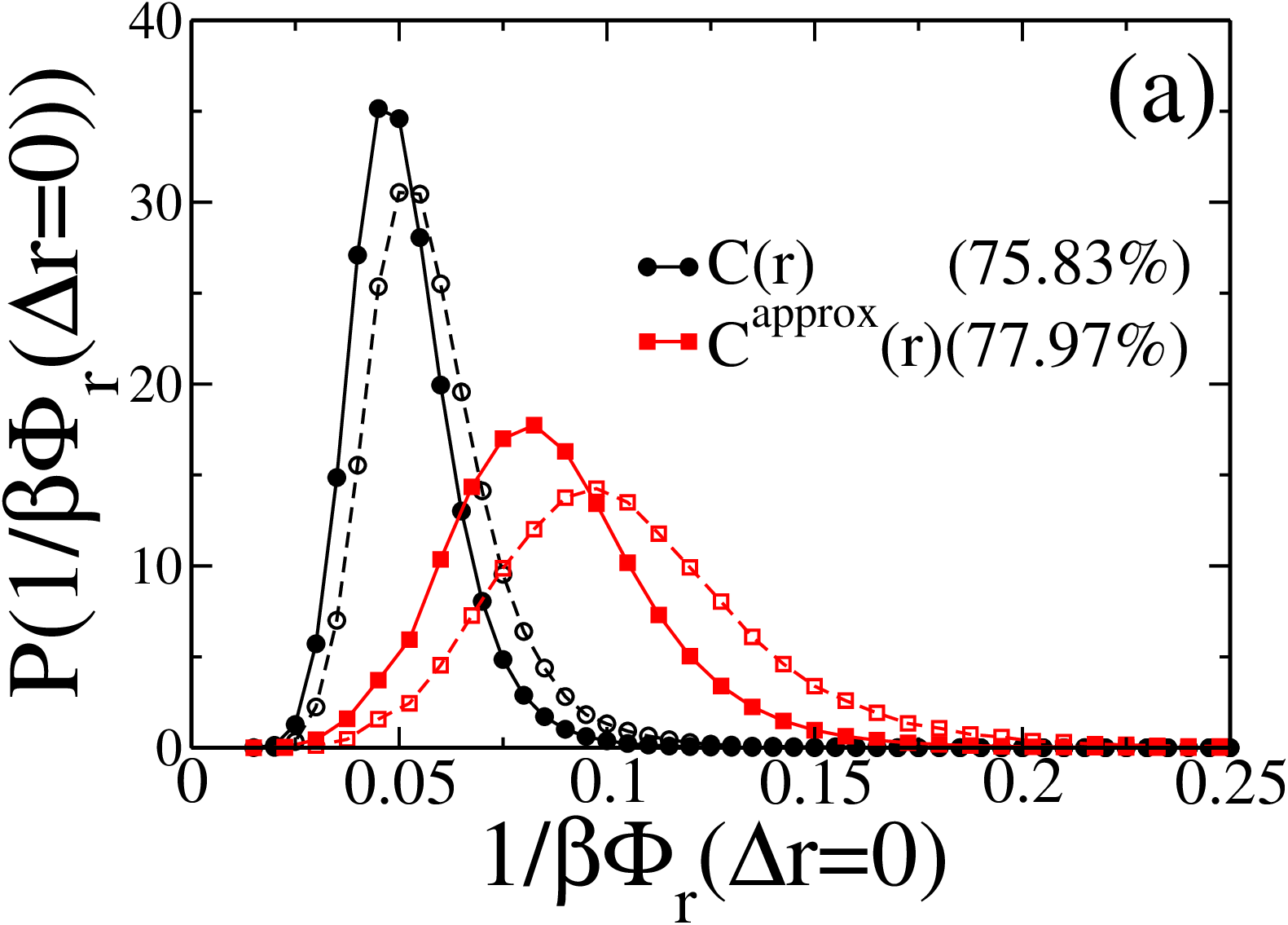}
\end{subfigure}
\begin{subfigure}{0.23\textwidth}
\includegraphics[width=1.0\linewidth]{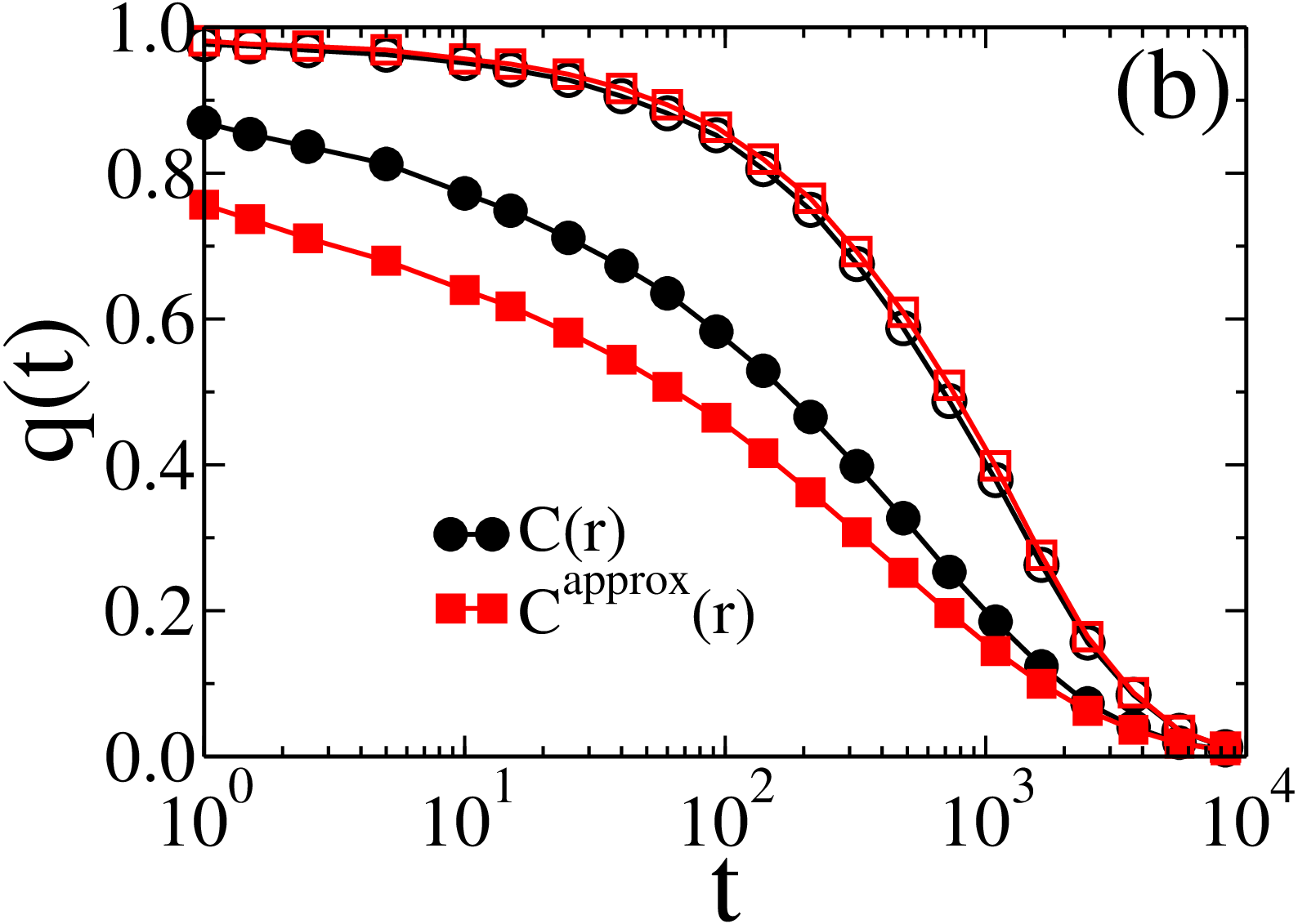}
\end{subfigure}
\vskip\baselineskip
\begin{subfigure}{0.22\textwidth}
\includegraphics[width=1.0\linewidth]{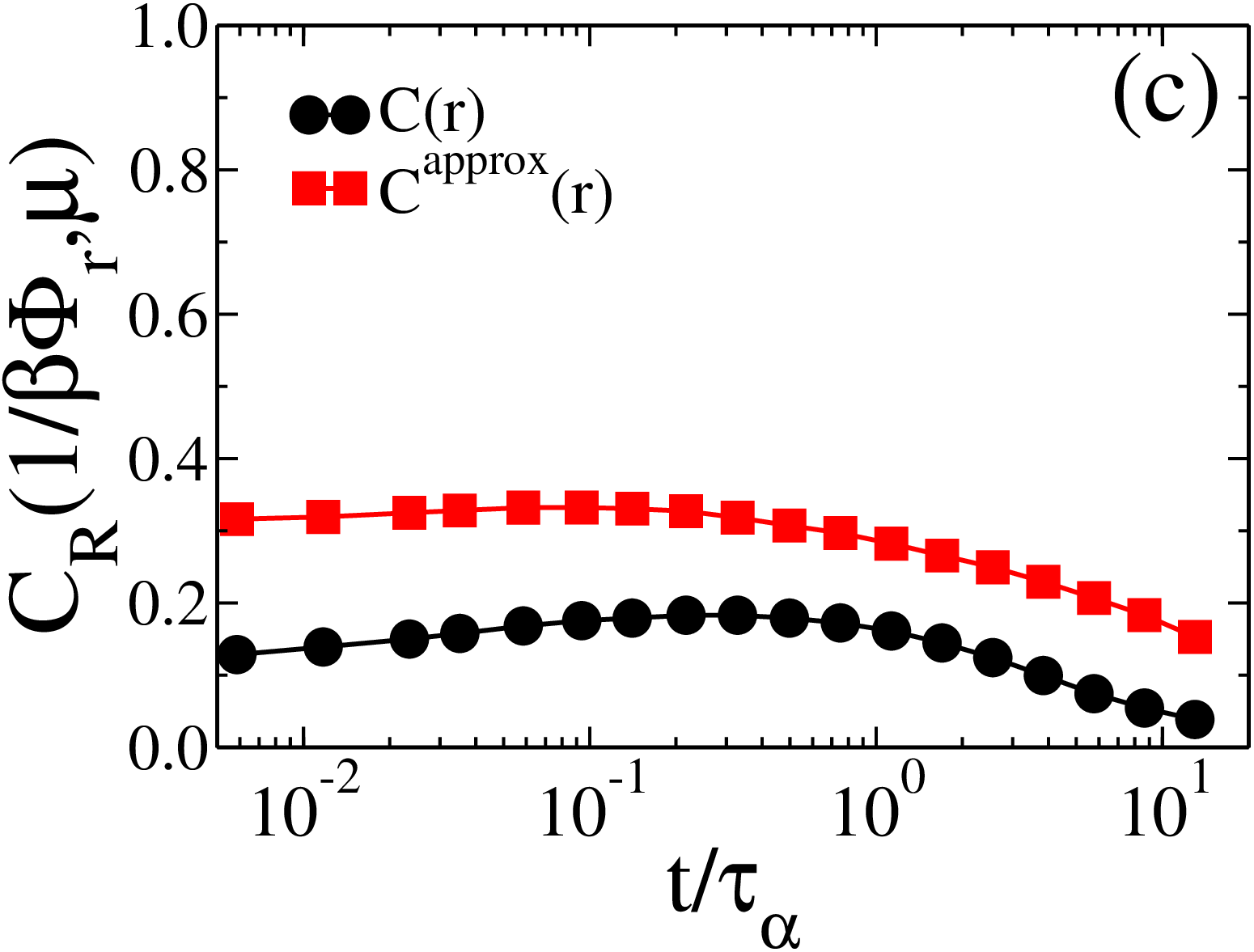}
\end{subfigure}

\caption{\textit{Different analysis using $C_{uv}(r)$ and $C_{uv}^{approx}(r)$. \textbf{(a)} Distribution of microscopic SOP ($1/\beta\Phi_{r}$) for rearranging particles, solid lines correspond to the distribution of all particles, and dashed lines are the distribution of fast particles. \textbf{(b)} Overlap function analysis to differentiate hard and soft particles. Open symbols are for the hard and filled symbols for soft particles. \textbf{(c)} Spearman rank correlation between SOP and $\mu$.}}
\label{Fig_approx}
\end{figure}

As discussed in the main text, we encounter unphysical values of microscopic caging potential, $\beta\Phi_{r}(\Delta r=0)$ when we use the direct correlation function given by Eq.\ref{direct}. This error is higher at high temperatures and also when the potential grows sharply. We plot the distribution of SOP of the fast particles. The percentage of fast particles whose SOP values are higher than the average value of SOP remains similar in both cases, being marginally higher when we use $C^{approx}(r)$(Fig. \ref{Fig_approx}(a)). Furthermore, we assess the ability of the SOP to differentiate between hard and soft particles by analyzing the overlap function (Fig. \ref{Fig_approx}(b)) of a few hardest and softest particles. Interestingly, we observe comparable predictive power, and again we find that $C^{approx}(r)$ performs better than $C(r)$. We also show that the Spearman rank correlation between local SOP and mobility is higher when we replace the direct correlation function with its approximate value (Fig. \ref{Fig_approx}(c)).\\

{\bf ACKNOWLEDGMENT}\\
M.~S. and S.~M.~B. thanks SERB for the funding. S.~M.~B. thanks Hajime Tanaka and Rajesh Murarka for discussions. M.~S. thanks Palak Patel for the discussions and critical reading of the manuscript.\\[4mm]

{\bf AVAILABILITY OF DATA}\\
The data that support the findings of this study are available from the corresponding author upon reasonable request.\\[3mm]

\end{document}